\documentclass[traditabstract]{aa}  

\usepackage{hhline}
\usepackage{booktabs}
\usepackage{bm}

\usepackage[table, x11names]{xcolor}
\usepackage{array, booktabs, boldline} %
\usepackage{cellspace}

\usepackage{soul} 
\usepackage{graphicx}
\usepackage{verbatimbox}
\usepackage{longtable,lscape}
\usepackage{rotating}
\usepackage[draft]{hyperref}
\hypersetup{colorlinks=true, linkcolor=red, citecolor=blue, filecolor=magenta, urlcolor=blue}
\usepackage{amssymb}
\usepackage{txfonts}

\usepackage{fixltx2e}

\usepackage{adjustbox}
\usepackage{amsmath}
\usepackage{relsize}
\usepackage{cleveref}
\usepackage{marginnote}
\usepackage{threeparttable}
\usepackage{comment}
\usepackage{makecell}
\usepackage{float}

\newcommand{\myemail}{antonello.calabro@inaf.it}

\definecolor{jf}{rgb}{0,0.75,0.5}

\usepackage{booktabs,tabularx}

\begin{document}


\title{The VANDELS survey: the relation between UV continuum slope and stellar metallicity in star-forming galaxies at $z\sim3$}

\author{A. Calabr{\`o}\inst{1} 
\and M. Castellano\inst{1} 
\and L. Pentericci\inst{1}
\and F. Fontanot\inst{2,3}
\and N. Menci\inst{1}
\and F. Cullen\inst{4}
\and R. McLure\inst{4}
\and M. Bolzonella\inst{5}
\and A. Cimatti\inst{6}
\and F. Marchi\inst{1}
\and M. Talia\inst{5,6}
\and R. Amor\'in\inst{7,8}
\and G. Cresci\inst{9}
\and G. De Lucia\inst{2}
\and J. Fynbo\inst{10}
\and A. Fontana\inst{1}
\and M. Franco\inst{11}
\and N. P. Hathi\inst{12}
\and P. Hibon\inst{13}
\and M. Hirschmann\inst{14}
\and F. Mannucci\inst{9}
\and P. Santini\inst{1}
\and A. Saxena\inst{15,1}
\and D. Schaerer\inst{16,17}
\and L. Xie\inst{18}
\and G. Zamorani\inst{5}
}

\institute{INAF - Osservatorio Astronomico di Roma, via di Frascati 33, 00078, Monte Porzio Catone, Italy (\myemail)
\and  INAF - Astronomical Observatory of Trieste, via G.B. Tiepolo 11, I-34143 Trieste, Italy
\and IFPU - Institute for Fundamental Physics of the Universe, via Beirut 2, 34151, Trieste, Italy
\and SUPA, Institute for Astronomy, University of Edinburgh, Royal Observatory, Edinburgh EH9 3HJ 
\and INAF - Osservatorio Astronomico di Bologna, via P. Gobetti 93/3, I-40129, Bologna, Italy
\and University of Bologna, Department of Physics and Astronomy (DIFA) Via Gobetti 93/2- 40129, Bologna, Italy
\and Instituto de Investigaci\'on Multidisciplinar en Ciencia y Tecnolog\'ia, Universidad de La Serena, Ra\'ul Bitr\'an 1305, La Serena, Chile
\and Departamento de F\'isica y Astronom\'ia, Universidad de La Serena, Av. Juan Cisternas 1200 Norte, La Serena, Chile
\and INAF - Osservatorio Astrofisico di Arcetri, Largo E. Fermi 5, I-50125, Firenze, Italy
\and Cosmic DAWN Center, Niels Bohr Institute, University of Copenhagen, Juliane Maries Vej 30, 2100 Copenhagen \O, Denmark
\and Centre for Astrophysics Research, University of Hertfordshire, Hatfield, AL10 9AB, UK
\and Space Telescope Science Institute, 3700 San Martin Drive, Baltimore, MD 21218, USA
\and European Southern Observatory (ESO) Chile
\and DARK, Niels Bohr Institute, University of Copenhagen, Lyngbyvej 2, 2100 Copenhagen, Denmark
\and Department of Physics and Astronomy, University College London, Gower Street, London WC1E 6BT, UK
\and Observatoire de Gen\`eve, Universit\'e de Gen\`eve, 51 Ch. des Maillettes, 1290 Versoix, Switzerland
\and CNRS, IRAP, 14 Avenue E. Belin, 31400 Toulouse, France
\and Tianjin Astrophysics Center, Tianjin Normal University, Binshuixidao 393, 300384, Tianjin, China
}
\date{Received XXX}

\abstract 
{
The estimate of stellar metallicities (Z$_\ast$) of high-z galaxies are of paramount importance in order to understand the complexity of dust effects and the reciprocal interrelations among stellar mass, dust attenuation, stellar age and metallicity. Benefiting from uniquely deep FUV spectra of $>500$ star-forming galaxies at redshifts $2<z<5$ extracted from the VANDELS survey and stacked in bins of stellar mass and UV continuum slope ($\beta$), we estimate their stellar metallicities Z$_\ast$ from stellar photospheric absorption features at $1501$ and $1719$ \AA, which are calibrated with Starburst99 models and are largely unaffected by stellar age, dust, IMF, nebular continuum or interstellar absorption. 
Comparing them to photometric based spectral slopes in the range $1250$-$1750$ \AA, we find that the stellar metallicity increases by $\sim0.5$ dex from $\beta \sim -2$ to $\beta \sim -1$ ($1 \lesssim$ A$_{1600}$ $\lesssim 3.2$), and a dependence with $\beta$ holds at fixed UV absolute luminosity M$_{UV}$ and stellar mass up to $\sim10^{9.65}$ M$_\odot$. As a result, the metallicity is a fundamental ingredient for properly rescaling dust corrections based on M$_{UV}$ and M$_\ast$.
Using the same absorption features, we analyze the mass-metallicity relation (MZR), and find it is consistent with the previous VANDELS estimation based on a global fit of the FUV spectra. Similarly, we do not find a significant evolution between $z\sim2$ and $z\sim3.5$. 
Finally, the slopes of our MZR and Z$_\ast$-$\beta$ relation are in agreement with the predictions of well-studied semi-analytic models of galaxy formation (SAM), while some tensions with observations remain as to the absolute metallicity normalization.  
The relation between UV slope and stellar metallicity is fundamental for the exploitation of large volume surveys with next generation telescopes and for the physical characterization of galaxies in the first billion years of our Universe.
}


\titlerunning{\footnotesize $\beta$-Z$_\ast$ relation of star-forming galaxies in VANDELS}
\authorrunning{A.Calabr\`o et al.}
 \maketitle
\section{Introduction}\label{introduction}

Metals and dust are the main final products of stellar evolution, hence they are key to understanding the history of star-formation (SF) in the first $1$-$3$ billion years since the birth of our Universe, during which the gas composition, the dust production mechanisms and stellar properties were radically different from today \citep{maiolino04,dayal12}. 
Indeed, feedback mechanisms from stars and AGNs not only can stop SF, but significantly affect the metal content of galaxies and new generations of stars. At the same time, this picture is complicated by the presence of dust, which may play a relevant role in obscuring even the most active star-froming galaxies, and thus providing a biased view of the intrinsic properties of high-$z$ sources. 
Investigating how these quantities are related to each other and with additional properties, including the stellar mass (M$_\ast$) and the star-formation rate (SFR), is thus of paramount importance to constrain formation models of galaxy formation and evolution.

In previous years, thanks to the relative ease of performing observations and data reduction at optical wavelengths, impressive imaging campaigns have discovered large samples of (rest-frame) UV bright star-forming galaxies in the redshift range between $2$ and $4$ \citep{steidel95,giavalisco96,giavalisco02,hathi10,parsa16}.
These efforts have begun to push the study of the UV luminosity function and SFR density up to the first $3$ Gyrs of the Universe \citep[e.g.,][]{bouwens15,finkelstein15,oesch18}. It was also possible to correct these quantities for the influence of dust when it became clear that dust attenuation is primarily responsible for the slope of the UV continuum (also called $\beta$) \citep{meurer99,steidel99,shapley03,pannella09}, 
which can be measured from narrow and broad photometric bands \citep[e.g.,][]{bouwens09,rogers13,castellano12,castellano14,hathi13,pilo19}. 
Currently, UV-based, dust-corrected SFRs reach by far higher sensitivity levels and statistics than any other alternative tracer at other wavelengths. More recently, some works \citep[e.g.,][]{pannella15,mclure18a} have shown that $\beta$ values correlate with the stellar masses of the galaxies, which can then be used as indirect probes of attenuation, even though the mass derivation usually requires more information on the SED distribution of the objects at longer wavelengths.

Spectroscopy provides an alternative, powerful, method to constrain galaxy evolution through measurements of accurate spectroscopic redshifts, kinematics, gaseous and stellar metallicity. 
For example, \citet{fanelli88,fanelli92} proposed a series of UV absorption lines as useful tracers of the physical properties of young stellar populations, including, but not limited to, metal content, stellar wind strength, and IMF. 
\citet{rix04} and \citet{leitherer11} first introduced three absorption indexes around $1370$, $1425$ and $1978$ \AA\ to study a sample of star-forming galaxies at redshift $\sim1$. These features, which are mostly blends of multiple elements, were found to depend only on metallicity, according to Starburst99 stellar models \citep{leitherer99}. 
It became immediately clear that using these rest-frame UV features for high-redshift observations, where they fall in the optical or near-infrared (NIR) range, could provide a lot of insight about the nature of pristine, young galaxies. 

To this aim, a series of spectroscopic campaigns started in $2014$-$2015$ \citep[e.g., VUDS,][]{lefevre15}, targeting thousand of star-forming galaxies at redshifts $2$-$4$ to study their properties. \citet{chisholm19} measured the stellar metallicity of $19$ galaxies at $z \sim 2$, observed through the Magellan Spectrograph. Ultraviolet and optical rest-frame spectra of SF galaxies at $z \sim 2.4$ were also obtained from Keck/LRIS and Keck/MOSFIRE by \citet{steidel16} and \citet{topping20}, who investigated the relation between the stellar and gas-phase metallicity.
Along the same direction, the ESO-VANDELS spectroscopic survey \citep{pentericci18,mclure18b} has continued to dig deep into this cosmic epoch, and currently represents the state-of-the-art with respect to number of targeted galaxies and depth reached. In fact, VANDELS observed from $2015$ to $2018$ more than two thousand galaxies in the rest-frame UV down to a limiting magnitude of $i_{AB}\simeq27.5$ (at $5\sigma$), with integration times ranging from $20$ to $80$ hours per source, ensuring enough SNR of the continuum for the derivation of reliable constraints on stellar mass, attenuation, SFR, and stellar metallicity.
Most importantly, these performances enabled the determination, for the first time at redshift $>2.5$, of the stellar metallicity from stacked spectra of galaxies in bins of stellar masses or Ly$\alpha$ equivalent width, by fitting stellar population synthesis templates to their entire FUV emission \citep{cullen19,cullen20}. This has shed light on the chemical evolution of intermediate-high mass ($8.5<$ log$_{10}$ M$_\ast$ $<10.2$) systems before the peak of cosmic SF activity at $z\sim2$ \citep{madau14}.

Despite this rapid progress, the growth of galaxies and the increase of SFR density in the early Universe are far from being completely understood. The main limitation in the analysis is not connected with data availablity, but with the systematic uncertainties involved in the determination of key physical quantities, and related to the known degeneracies involved in the estimation of several parameters. 
Most notably, the SED fitting technique is subject to the attenuation, age, metallicity, IMF degeneracy. 
The stellar mass and $\beta$ slope are usually taken as proxies for predicting the level of dust attenuation, but these correlations have a large scatter, which leaves dust corrections largely inaccurate, especially when applied to individual objects. Moreover, UV slopes can also be affected by other quantities than dust, namely the IMF, the stellar age, the nebular continuum, and the metallicity \citep{bouwens09,castellano14,raiter10}. In particular, the latter could have a crucial role to solve part of the degeneracies still affecting these scaling relations.

For deriving stellar metallicities, a whole spectral fitting could still be influenced by complex dependencies on the age and IMF, which is typical for the majority of absorption complexes in the FUV. In addition, it needs a good quality spectrum over a large wavelength range.
In order to reduce potential biases, single absorption lines provide an alternative method to measure the metallicity.  
Needing only specific indexes $10$-$20$ \AA\ wide, this method can work on a limited portion of the FUV spectral range, typically $\sim100$ \AA, required for a good estimate of the underlying UV continuum. In addition, this estimate is also independent on dust extinction and in most cases it is insensitive to stellar age and IMF, at least for slopes $\alpha$ close to Salpeter (within 0.2) and ages higher than $\sim50$ Myr. The reason of this behavior is that the depth of photospheric absorption features in the far-UV depends on the relative abundance of O and B stars. After an initial period of the same order of the average lifetime of these stars, if the SFR is constant, the same number of young stars forms, and their contribution to the UV spectrum does not change on average over time.

However, several studies conducted in the past $20$ years have generated some uncertainties on the best absorption lines for measuring the metallicity (i.e., those least affected by age or IMF variations) and on the correct calibration functions to adopt. Moreover, many lines were found to be strongly contaminated by ISM absorption, thus are not reliable tracers of the metallicity in stars.
In addition, most of them were tested on a limited number of objects, with various FWHM resolution data (from $0.25$ to $3.8$ \AA\ rest-frame) and redshift (from $0$ to $\sim 3$). 
For example, \citet{sommariva12} found that the $1978$ \AA\ index is quite sensitive to the IMF assumptions, but defined three additional indexes near $1460$, $1501$ and $1533$ \AA, independent on age and IMF, from which they measured the stellar metallicity of five star-forming galaxies at redshift $\sim3$. 
While the $1460$ \AA\ feature is produced by NiII and the $1533$ \AA\ line by SiII, the absorption region at $1501$ \AA\ arises in the photosphere of young, hot stars and is due to the ionized SV species \citep{pettini99,quider09}.
\citet{leitherer11} and \citet{faisst16} adopted other metal-sensitive indexes near $1400$ and $1550$ \AA\ to study the metal content of a sample of local starbursts and star-forming galaxies at $z\sim5$, respectively. In particular, the CIV absorption feature around $1550$ \AA\ has a strong wind component, indicated by its P-Cygni profile, whose strength is known to correlate with metallicity of the parent stars \citep{castor75,walborn95}. 
On the other hand, winds from hot stars also contribute to the absorptions at $1300$ and $1400$ \AA, which are due to SiIII and SiIV, respectively. However, their metallicity is more representative of the ISM component of the galaxies: even though these lines are influenced by stellar photospheric absorption, they are mainly affected by interstellar absorption and, in part for the second index, by nebular emission. 
More recently, thanks to ISM and radiative tranfer models, \citet{vidalgarcia17} studied the influence of ISM absorption and emission for most of the commonly adopted stellar photospheric indexes in the literature, finding that the $1425$ \AA\ line and the $\sim1719$ \AA\ complexes \citep[already studied in][]{fanelli92} are among the cleanest and least contaminated 
stellar metallicity tracers up to at least solar metallicity values. In particular, the latter index is a blend of medium and highly ionized species, including NIV ($1718.6$ \AA), SiIV ($1722.5$, $1727.4$ \AA), and multiple transitions of AlII and FeIV ranging from $1705$ to $1729$ \AA. 

In this paper we revisit most of the absorption indexes that have been previously adopted in the literature. Comparing the predictions of multiple stellar models, we infer new calibrations for VANDELS-like spectra and measure the stellar metallicity of high-redshift galaxies at $2<z<5$ from a combination of two robust UV absorption lines, located at $1501$ and $1719$ \AA. With these in hand, we explore how the metallicity is related to other properties, including UV slope, UV magnitude and stellar mass, and whether it can remove the degeneracies still affecting the scaling relations involving such quantities.

The paper is organized as follows. In Section~2 we describe the VANDELS spectral observations and the procedures adopted to measure the UV slope, the stellar mass, and the stellar metallicity. We conclude this part by illustrating the final sample selection for this work. 
In Section~3 we present our results. First we explore the mass-metallicity relation from two UV absorption line metallicity tracers, and its evolution with redshift. Then we investigate the role of stellar metallicity in the UV magnitude-$\beta$ and stellar mass-$\beta$ relations, and assess the dependence between $\beta$ slope and metallicity.
Finally, in Section~4 we discuss our results and compare them to semi-analytic models of galaxy evolution. A summary with conclusions is the content of the fifth Section, while an appendix with additional material is included in the last part of the paper.
In our analysis, we adopt AB magnitudes and \citet{chabrier03} initial mass function (IMF) for deriving stellar masses, star-formation rates, and UV absolute magnitudes. Throughout this work, unless otherwise stated, we assume a cosmology with $H_{0}=70$ $\rm km s^{-1}Mpc^{-1}$, $\Omega_{\rm m} = 0.3$, $\Omega_\Lambda = 0.7$ and the most recent estimation of the solar metallicity Z$_\odot$ $=0.0142$ \citep{asplund09}. We also assume by convention a positive equivalent width (EW) for absorption lines and a negative EW for lines in emission.

\section{Methodology}\label{methodology}

In this section we describe VANDELS observations, the spectral reduction and calibration. Then we illustrate in detail the derivation of the two key physical quantities of this work: the UV continuum slope from photometric data, and the stellar metallicity from rest-frame UV spectra. Finally, we specify the sample selection adopted in our analysis. 

\subsection{Spectral observations and reduction}\label{observations}

The galaxies analyzed in this study are selected from the ESO-VANDELS project (ESO Large Program ID 194.A- 2003(EK), P.I. L.Pentericci and R.McLure) \footnote{Link to VANDELS project: http://vandels.inaf.it}. VANDELS is, to date, the deepest optical spectroscopic survey of high redshift galaxies. We refer to the two introductory papers by \citet{mclure18b} and \citet{pentericci18} for all the details concerning the observations and data reduction, and highlight here only the main characteristics.
The survey targeted $\sim2100$ galaxies at redshift $z \geq 1$ in an area of the sky of $0.2$ deg$^2$ in total, in the UDS (Ultra Deep Survey) and CDFS (Chandra Deep Field South) fields around the CANDELS region \citep{grogin11,koekemoer11}. The interesting targets for our goals are: (1) bright star-forming galaxies (SFG) with photometric redshift ranging $2.5<z_{phot}<5.5$ and magnitude limit $i_{AB} < 25$, and (2) lyman-break galaxies (LBG) in the range $3<z_{phot}<5.5$, which have fainter magnitudes and lower SNR compared to SFGs. The initial magnitude limits were H$<27$ and $i_{AB}<27.5$ in this case. All targeted galaxies have specific SFRs (SSFR) higher than $0.1$ Gyr$^{-1}$, even though the majority of them have SSFR $>0.4$ Gyr$^{-1}$ and SFRs higher than $2.5$ M$_\odot$/yr.

The observations were performed with the VIMOS multi-object spectrograph mounted at the ESO-VLT, which delivers high-quality spectra in the wavelength range $4900 \AA < \lambda < 9800 \AA$ with an average resolving power R$=580$, corresponding to an average spectral resolution in rest-frame of $\simeq 2.8$ \AA. The VIMOS spectra were reduced in a fully automatic way with the EASYLIFE pipeline \citep{garilli12}.
This procedure yields fully wavelength and flux calibrated two and one dimensional spectra, corrected for atmospheric and galactic extinction, and normalized to the i-band photometry available for all targets. As described in \citet{pentericci18}, since an artificial flux loss was observed in the extreme blue end of the spectra ($\lambda < 5600$ \AA) when compared to broad-band photometry, an empirical correction estimated in a statistical way was applied in post-processing to ensure the correct flux density shape at lower wavelengths. This correction is in all cases no larger than $10$-$20\%$ of the original flux density.  

After the reduction process, the VANDELS team was in charge for measuring spectroscopic redshifts $z_{spec}$ for all the observed targets. The derivation of $z_{spec}$ was made with the help of the EZ software package \citep{garilli10}, which cross-correlates each spectrum with a subset of reference templates derived from previous VIMOS observations, representative of a large variety of stellar and galaxy types. The measurements were supervised by two independent team members and then further checked by the two co-PI until reaching a final agreement. A quality flag (from 0 to 9) was also assigned to each measurement, representing the probability of the redshift to be correct. Spectra with flags $3$ or $4$ are the most reliable, with $95\%$ and $100\%$ probability to be correct, respectively. In these cases, multiple emission or absorption lines could be typically recognized in a moderate S/N continuum. The median accuracy of spectroscopic redshift determinations is $0.0005$ ($\sim 150$ km/s). 

As first step of this work, we preselected $872$ galaxies from the parent catalog in VANDELS by requiring a secure spectroscopic redshift (flags 3 or 4) between $2$ and $5$. The former is slightly below the lower limit in the original selection, and includes some objects for which the photometric estimate ($z_{phot}$) was slightly overestimated. The latter condition was set to have good quality spectra: above $z=5$, galaxies have too noisy spectra to give a statistically significant contribution to our analysis, and none of them pass further constraints that will be introduced in the following sections to ensure good quality measurements of the parameters we are interested in this work. 
We also specify that bright sources in the infrared, which could be starburst (off-MS) systems, have been excluded from our selection. Our sample thus contains systems, either preselected as star-forming galaxies (based on specific SFR $>0.1$ Gyrs$^{-1}$) or Lyman-break galaxies, which are fairly representative of the Main Sequence of star-formation \citep[see ][]{mclure18b}.

\subsection{Stacking procedure}\label{stacks}

In this paper we derive the stellar metallicity from two photospheric absorption features of young O-B stars, located in the far-UV spectral range at $1501$ and $1719$ \AA, as described in the introduction. These indicators are typically very faint, with equivalent widths (EW) lower than $8$ \AA\ over a wavelength range of $\sim 10$ \AA, depending on the index, thus they require a high signal-to-noise continuum to be properly constrained, as we will quantify below. For this reason, prior to the metallicity measurements, we stacked our spectra in multiple bins of different physical quantities (i.e., stellar mass, UV slope, UV absolute magnitude M$_\text{UV}$, and redshift). 
In a representative example in Fig. \ref{Zerror_SNR} in the Appendix, we show that the 1$\sigma$ error on the equivalent width (hence on metallicity) is tightly correlated and strongly decreasing with the SNR of the underlying continuum. Ideally, a SNR of at least $\sim25$-$30$ is required to measure Z with an uncertainty of the order of $0.3$-$0.4$ dex in a single index, which improves to $0.1$-$0.2$ dex by combining multiple indicators. Throughout the rest of this work, our bins will be constructed to ensure high enough SNR in each stacked spectrum, $>30$ if possible, and $>20$ in all cases. This was found to be the best compromise between minimization of the uncertainty on the final metallicity estimate and the number of bins needed to test our relations.  

We illustrate here the general stacking procedure adopted in this paper. 
First, we converted all the spectra to rest-frame according to their spectroscopic redshifts estimated by the VANDELS collaboration. We normalized them using the median flux estimated in the range $1250<\lambda<2000$ \AA, and then resampled the spectra to a common wavelength grid of $0.4$ \AA\ per pixel, similar to the sampling of Starburst99 stellar models, and which corresponds to nearly half of the wavelength sampling of individual galaxies in VANDELS. This way, we are less prone to introduce systematic biases in the calibration functions compared to resampling simultaneously both the original S99 models to a coarser grid and the observed spectra. 
Then, we built composite spectra of all the objects falling in the same bin by taking the median flux at each dispersion point, while we calculated the noise from $500$ simulations by taking each time a random $80\%$ of the spectra in the bin, computing the median of individual flux values, and finally deriving the standard deviation of all the $500$ realizations. We did not perform the bootstrap resampling with replacement (requiring for the subsets the same size of the original sample) as this would be more affected by peculiar spectra, which could enter many times in the calculation. In addition, we did not derive the composite flux with an error-weigthed method, as this would bias the result toward lower redshift or more star-forming objects, which have typically a higher SNR. We also remark that a coarser spectral resampling in the stacking procedure (e.g., $1$\AA\ /pixel) does not alter the results presented in the following sections and the uncertainties associated to the measured equivalent widths. 
Finally, as mentioned in Section \ref{observations}, the redshift determinations in VANDELS are mainly based on the presence of UV rest-frame emission or absorption lines, which are produced by different components of the galaxies (e.g., stars, ISM, gas inflows/outflows), possibly in relative motion among them. 
It is thus useful to check our stacking procedure by using a common reference redshift for the galaxies, such as the systemic redshift (defined as the redshift of the bulk of the stars) as traced by the CIII]$\lambda\lambda 1907$-$1909$ \AA\ emission line doublet \citep{shapley03}. Thanks to the relative brightness of this line, we identified a subset of $150$ CIII] emitters by visual inspection of both the 2D and 1D spectra (SNR $>7$). For this sample, we found that adopting either the VANDELS released or the systemic redshift in the stacking procedure yields fully consistent metallicity measurements, thus would not modify the results of this work.

\subsection{Photometry in VANDELS fields and stellar masses from SED fitting}\label{photometry}

\begin{table}[t!]
    \footnotesize
    \begin{tabular}{|l|l|}
    \hlineB{2}
    field & bands used \\
    \hlineB{2}
    CDFS-HST & \makecell[l]{F435W, F606W, F775W, F814W, F850LP,\\F098M, F105W, F125W, F140W, F160W,\\ VIMOS-R, ISAAC-Ks, HAWKI-Ks,\\ IRAC-ch1, IRAC-ch2} \\ 
    \hlineB{2}  
    CDFS-GROUND & \makecell[l]{B-WFI, F606W, F850LP, subaru-IA484,\\ subaru-IA527, subaru-IA598, subaru-IA624,\\ subaru-IA651, VIMOS-R, subaru-IA679,\\ subaru-IA738, subaru-IA767, VISTA-Z,\\ VISTA-Y, VISTA-J, VISTA-H, VISTA-K, \\IRAC-ch1, IRAC-ch2} \\
    \hlineB{2}
    UDS-HST & \makecell[l]{subaru-B, F606W, subaru-V, subaru-i, F814W,\\ subaru-z, HAWKI-Y, WFCAM-J, F125W,\\ F160W, wfcam-H, wfcam-K, HAWKI-Ks,\\ IRAC-ch1, IRAC-ch2} \\
    \hlineB{2}
    UDS-GROUND &  \makecell[l]{subaru-B, subaru-V, subaru-i, subaru-z,\\ subaru-znew, VISTA-Y, wfcam-J,\\ wfcam-H, wfcam-K, IRAC-ch1, IRAC-ch2} \\
    \hlineB{2}
    \end{tabular}
\caption{\small Photometric bands available in the four fields covered by the VANDELS survey, and used for the estimation of stellar masses, SFR, and UV slope.}\label{Tablebands}
\end{table}

All targets in VANDELS have either space-based or ground-based photometric data available. The pointings in the UDS and CDFS fields are centered on the area covered by the CANDELS survey \citep{grogin11,koekemoer11}. For these regions, we have deep optical+near-IR (ACS + WFC3/IR) HST images, Spitzer images, and H-band selected, PSF-homogenized photometric catalogs assembled by \citet{galametz13,guo13}, including total magnitudes in 6 and 10 space-based broad-band filters, in the UDS and CDFS fields respectively. In the same area, also photometric data from ground are available, so that we are able to cover the full range between band B ($\lambda_{cen} \simeq 4450$ \AA) and IRAC channel 2 ($\lambda_{cen} \simeq 4.5 \mu m$). The limiting magnitude of this dataset is H$_{AB} < 27.05$ ($5\sigma$).

On the other hand, $\sim50\%$ of the total VANDELS area (both CDFS and UDS) falls outside of CANDELS. In these regions, while some HST filters are still present, most of the optical+near-IR imaging is performed with ground-based facilities, including Subaru, CFHT, UKIRT, VISTA, and VLT. Photometric catalogs of H-band detected sources (H$_{AB} < 27.05$) were produced by the VANDELS team using the SExtractor tool v2.8.6 \citep{bertin96}, providing PSF-homogeneized photometry and total magnitudes in $13$ and $11$ broadband filters in UDS and CDFS, respectively, ranging from $0.3$ to $8.0\ \mu m$. 
In Table \ref{Tablebands} we show the photometric bands that were used in all the fields covered by the VANDELS survey. More information on the observing campaigns, instruments adopted, and depth of the survey can be found in Table~1 of \citet{mclure18b}. 

Using all the available photometry ranging from band U to IRAC channel 2, the stellar masses are derived through the SED fitting technique as described in \citet{mclure18b}. This fit adopts \citet{bruzual03} stellar population templates with solar metallicity, a Chabrier IMF, declining $\tau$-model star-formation histories (SFH) with $\tau$ ranging $0.3 < \tau < 10.0$ Gyr and ages $\geq 50$ Myr. The dust is modeled with a \citet{calzetti00} attenuation law, with A$_V$ values in the range $0 <$ A$_{V}$ $< 3.0$, while the effect of the inter-galactic medium (IGM) transmission is taken into account following \citet{inoue14}. Similarly, the UV-based SFRs are derived from the best-fit UV rest-frame absolute luminosity following \citet{madau14}, and then dust corrected using A$_{V}$ estimated from the same fit.

\subsection{Beta slope and UV absolute magnitude estimations}\label{beta}

\begin{figure*}[h!]
    \centering
    \includegraphics[angle=0,width=\linewidth,trim={0cm 0cm 0cm 0cm},clip]{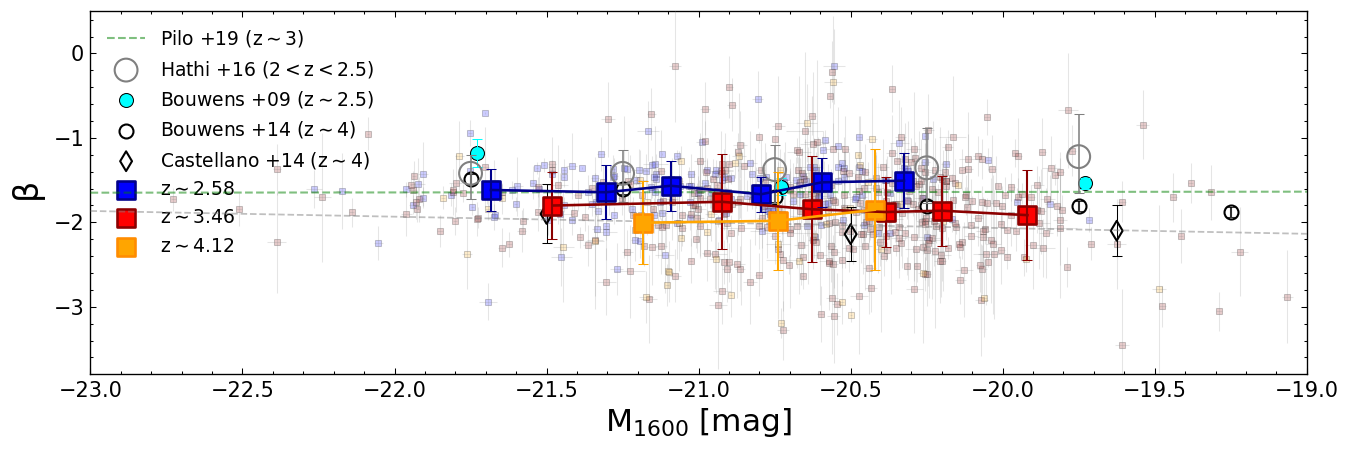} 
    \caption{\small UV slope vs M$_{1600}$ for VANDELS galaxies selected in this work with $\sigma_\beta \leq 1$ and at least three photometric bands available in the range $1230$-$2750$ \AA. Our data are color coded according to three different redshift bins ($2<z_{bin 1}<3<z_{bin 2}<4<z_{bin 3}<5$ in blue, dark-red, and orange, respectively), with $z$ estimated from VANDELS spectra. The median $z$ in each bin is reported in the legend. For comparison are shown the M$_{1600}$-$\beta$ relations found by \citet{hathi16} at $2<z<2.5$ (gray empty circles), \citet{bouwens09} at $\sim2.5$ (cyan circles), \citet{bouwens14} at $\sim4$ (black empty circles), and \citet{castellano14} for LBGs at $z \sim 4$ (black empty diamonds and light-gray dashed best-fit line). The relation found by \citet{pilo19} for LBGs at $z\sim3$ is displayed as a light-green dashed line. 
    }\label{MUV-beta}
\end{figure*}

In this work, we assume that the UV-continuum emission of each galaxy can be approximated with a power-law of the form $f(\lambda) \propto \lambda^\beta$ \citep{meurer99,calzetti94}. To estimate the exponent $\beta$, we first converted all the observed (total) AB magnitudes into flux densities $f_\lambda$, and removed all photometric bands whose bandwidths are outside the $1230$-$2750$ \AA\ rest-frame wavelength range, to exclude any contamination from the Ly$\alpha$ line, while the redward limit is the same adopted in \citet{pilo19}. We note that the redward limit is slightly higher than in the original \citet{calzetti94} definition. This ensures more statistics for our analysis, while it does not introduce systematic biases to the results: we note indeed that the central wavelength of the reddest bandpass typically does not lie outside of $2600$ \AA. 
When multiple photometric bands with similar pivot wavelengths were available, we determined the weighted average of their fluxes in order to provide a more uniform, evenly sampled coverage of the wavelength space.
Then we fitted a linear relation between log($\lambda$) and log($f_\lambda$) by using an orthogonal distance regression (ODR) technique \citep{boggs92}.
From the best-fit relation and the spectroscopic redshift of each galaxy, we also estimated the UV absolute magnitude M$_{UV}$ at $1600$ \AA, M$_{1600}$.

\subsubsection{Systematics and uncertainties}\label{uncertainties}

In the measurements of $\beta$, for each galaxy we could use from a minimum of two to a maximum of six bands (four on average) from the list of Table \ref{Tablebands}. 
We proved the stability of our values by removing for each galaxy one or two random photometric bands from the initial dataset (keeping at least three bands), and computing again the slope with the same ODR fitting procedure. This yields determinations that are in qualitative agreement with the original values based on the full available dataset, and do not have systematic discrepancies, indicating that our results are not driven by some specific bands adopted in the fit, and are stable against the exact number of photometric points that are used in the fit. We note that the wavelength range for fitting the UV slope does not contain strong emission lines that can significantly affect the photometry, as those bands possibly contaminated by the Ly$\alpha$ line have always been excluded at the beginning. 

On the other hand, since our galaxies are located in different fields for which a heterogeneous set of photometric filters is available, we checked for the presence of systematic differences among $\beta$ determinations in the four different VANDELS fields.  
We found that our results are generally in agreement, except for galaxies in the CDFS-GROUND field, which show a higher UV-slope at fixed stellar mass or selection magnitude. 
We also note that CDFS-GROUND data have a lower quality than in UDS-GROUND, and most of the discrepant objects have a low SNR, with beta measurements just based on two bands. 
We found that applying a cut to the $\beta$ uncertainty ($\sigma_\beta<1$), and requiring at least three data points to perform the fit, removes these outliers and restores the consistency of $\beta$ distribution among all VANDELS fields.
We also remark that these offsets do not affect the derivation of stellar masses in CDFS-GROUND, as these are based on fitting the entire SED up to IRAC channel 2, and they are most sensitive to the optical rest-frame range rather than the UV. 

We compared our $\beta$ estimations to those obtained from the best-fit photometric SED (see Section \ref{photometry}), while we leave for the Appendix \ref{appendix0} a discussion about the UV slope inferred from VANDELS spectra. 
As far as the first are concerned, we found a systematic difference with the SED-based estimates of $\sim - 0.2$ (i.e., $10\%$ of $\beta$ measurements), which is likely related to the different method adopted and to the different treatment of photometric bands at the left- and right-most extremes of the wavelength range $1230$-$2750$ \AA. In particular, if we require that only the central pivot wavelength should not exceed those limits (instead of the whole bandwidths), we obtain UV-slopes $\sim0.15$ flatter, more in agreement with SED-fitting based values. However, we remark that this difference is below the typical uncertainty of the $\beta$ estimations for our galaxies ( ($1\sigma_{\beta}$)$_\text{median}=0.23$).
Overall, compared to $\beta$ and M$_{UV}$ estimations based on the best-fit photometric SED, an advantage of our procedure is that it is not model dependent. 

\subsection{The UV absolute magnitude - $\beta$ relation}

In Fig. \ref{MUV-beta} we show the distribution of beta slopes as a function of M$_{UV}$, which is often taken as a reference to dust-correct the luminosity function. 
Our entire sample has a median $\beta$ slope of $-1.76$ ($1 \sigma$ dispersion of $0.54$) and M$_{UV}$ of $-20.62$ ($\sigma=0.57$). A best-fit linear relation can be written explicitly as $\beta =$ ($-0.07 \pm 0.03$) $\times$ M$_{UV}$ $-$ ($3.18 \pm 0.7 $), indicating a slight increase of $\beta$ for bright objects, at $2\sigma$ level significance. 
A similar slope to our analysis was also found by \citet{bouwens09,bouwens14} for U- and B-dropouts at redshift $\sim2.5$ and $4$, and by \citet{castellano12} for LBGs at redshift $\sim4$.

Dividing the sample into three redshift bins, we also find a $\beta$ evolution in our redshift range, increasing on average from $-1.98$ at $z\sim4.1$ to $-1.59$ at $z\sim2.6$, in agreement with the strong evolution in $\beta$ expected in this cosmic epoch \citep[e.g.,][]{pannella15}, and with results found by other works at similar or slightly different redshifts. First, photometric based UV slopes derived by \citet{hathi16} for star-forming galaxies in VUDS at redshifts $2 < z < 2.5$ are slightly redder (by $\sim 0.15$) than our estimates at $z_\text{median}=2.58$, across the same UV magnitude range $-22<M_{UV}<-20$. 
Then, our objects follow approximately the same distribution expected for star-forming galaxies at redshift $\sim 3$: the median values of $\beta$ calculated in our first two subsets around $z\sim 3$ (z$_\text{median}=2.58$ and $3.46$) are indeed largely consistent, over all the range of UV magnitudes, with $z \sim 3$ LBGs in the COSMOS field presented by \citet{pilo19}. 
The last bin at the highest redshift ($4<z<5$) was created to compare with the work of \citet{castellano12}, which adopts a similar derivation of the UV slopes, and whose results are in agreement within the errors with our findings.

\subsection{Metallicity calibrations from UV absorption indexes in Starburst99 models}\label{calibrations}

\begin{table}[t!]
    \centering
    \footnotesize
    \begin{tabular}{|c|c|c|c|c|c|c|}
    \hlineB{2}
    index & $\lambda_1$ ($\AA$) & $\lambda_2$ ($\AA$) \\
    \hlineB{2}
    1370 & 1360 & 1380 \\ 
    \hlineB{2}  
    1400 & 1385 & 1410 \\
    \hlineB{2}  
    1425 & 1413 & 1435 \\
    \hlineB{2}
    1460 & 1450 & 1470 \\
    \hlineB{2}
    1501 & 1496 & 1506 \\
    \hlineB{2}
    1533 & 1530 & 1537 \\ 
    \hlineB{2}  
    1550 & 1530 & 1560 \\
    \hlineB{2}
    1719 & 1705 & 1729 \\
    \hlineB{2}
    1853 & 1838 & 1858 \\
    \hlineB{2}
    1978 & 1935 & 2020 \\
    \hlineB{2}
    \end{tabular}
\caption{\small Absorption complexes analyzed in this work. The second and third columns correspond to $\lambda_1$ and $\lambda_2$ in Equation \ref{eqEW_formula}.}\label{table_features}
\end{table}

\begin{figure}[t!]
    \centering
    \includegraphics[angle=0,width=0.495\linewidth,trim={0.7cm 0.cm 0.3cm 0.3cm},clip]{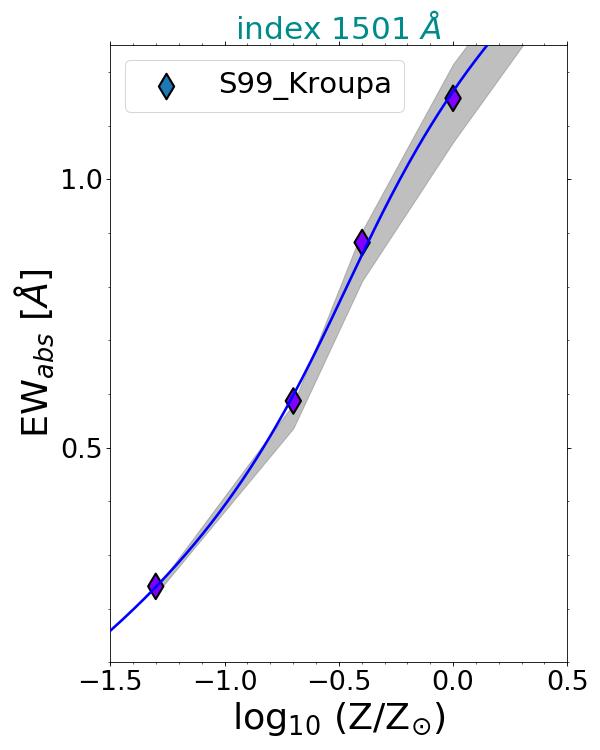}
    \includegraphics[angle=0,width=0.495\linewidth,trim={0.7cm 0.cm 0.3cm 0.3cm},clip]{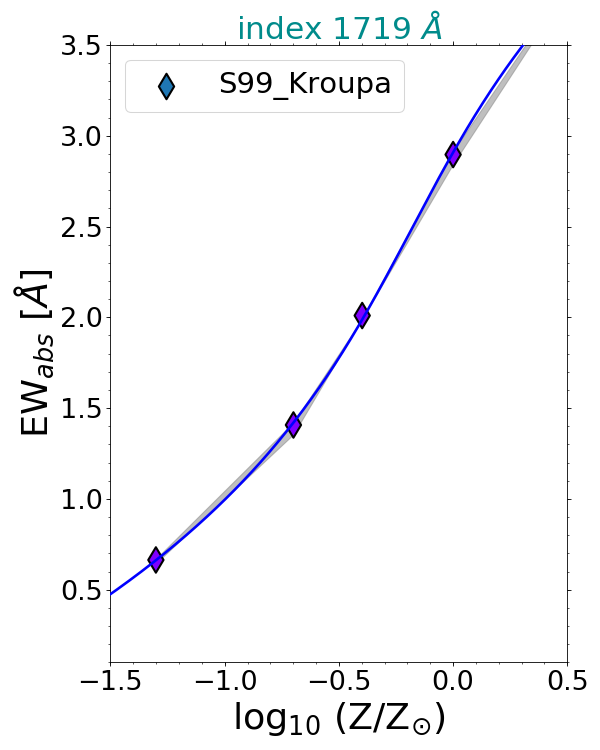}
    \caption{Diagrams showing the dependence between equivalent width (EW) and metallicity for the $1501$ and $1719$ \AA\ absorption line indexes that we adopt in this paper according to Starburst99 models. 
    All the data points are derived assuming a constant SFH for $100$ Myr and Kroupa IMF. The gray shaded region around the main relations represents the variation of EW with the IMF (Kroupa-Salpeter) and with the ages of the stellar population chosen ($50$ Myr - $2$ Gyr), as described in the text. The final metallicity calibrations are based on a third order polynomial fit to the data points, and are shown with a blue continuous line in each panel. Their explicit forms are given in the text in Equations \ref{calib1719} and \ref{calibothers}.}\label{calibration_figure1}
\end{figure}

Absorption lines in the UV rest-frame spectra carry important physical information about the properties of the host galaxies \citep{fanelli88,fanelli92}. While some of them are produced in the interstellar medium (ISM) of the galaxy itself, others are instead produced by chemical elements in the photospheres of hot, young, O and B stars, or in stellar winds generated by their radiation pressure. The latter two cases are extremely interesting as they can be used as stellar metallicity diagnostics. These include the indexes at $1501$ and $1719$ \AA\ that are adopted in this paper.

In order to derive a proper conversion between absorption strength and Z$_\ast$, we used Starburst99 WM Basic models \citep{leitherer10,leitherer11} to remain consistent with the previous VANDELS work on the stellar metallicity by \citet{cullen19}. Moreover, among all currently available stellar templates, they offer the highest native spectral resolution across a wide range of wavelengths ($0.4$ \AA\ in the range $900<\lambda <3000$ \AA), offering the possibility to test the metallicity calibration functions also for significantly higher resolutions than VANDELS. These models have also been widely tested in many studies on faint photospheric absorption lines \citep[e.g.,][]{rix04,sommariva12,leitherer11}. However, for completeness, a comparison with BPASS models \citep{eldridge17} is also included in the Appendix \ref{appendix3}.

We produced far-UV S99 spectra assuming a continuous star-formation history (SFH) and a grid with different stellar ages ($50$, $100$, $150$, $200$, $500$ Myr, and $1$, $1.5$, $2$ Gyr), metallicities ($0.05$, $0.2$, $0.4$, $1$ and $2.5$ times solar), and the two IMFs available in the simulation (i.e., Salpeter and Kroupa \footnote{We note that, even though a Chabrier IMF was used in our SED fitting, the Kroupa and Chabrier IMFs yield very similar results for the stellar masses and SFRs of our galaxies.}). The lower limit of $50$ Myr is also the lowest stellar age adopted in the SED fitting, while the upper bound of $2$ Gyr corresponds approximately to the age of Universe at the median redshift of our sample.

The models, which have the same wavelength sampling of VANDELS stacked spectra (see Section \ref{stacks}), were smoothed with a gaussian kernel with $\sigma=1.3$ \AA\ to match the VANDELS average resolution.
Afterwards, for each model, we measured the equivalent widths (EW) of absorption features according to the following definition :
\begin{equation}\label{eqEW_formula}
{\rm EW_i} \equiv \int_{\lambda_1}^{\lambda_2} \left( \frac{f(\lambda)-f_{\rm cont}(\lambda)}{f_{\rm cont}(\lambda)} \right) d\lambda,
\end{equation}
\noindent
where $f(\lambda)$ is the flux density spectrum across the absorption/emission feature, $f_{\rm cont}(\lambda)$ is the continuum (both in erg/s/cm$^2$/A), while $\lambda_1$ and $\lambda_2$ are the starting and ending wavelengths of the features. The list of features analyzed in this work, including the corresponding $\lambda_1$ and $\lambda_2$, are shown in Table \ref{table_features}.

\begin{figure}[t!]
    \centering
    \includegraphics[angle=0,width=\linewidth,trim={0.1cm 0.2cm 0.3cm 0.3cm},clip]{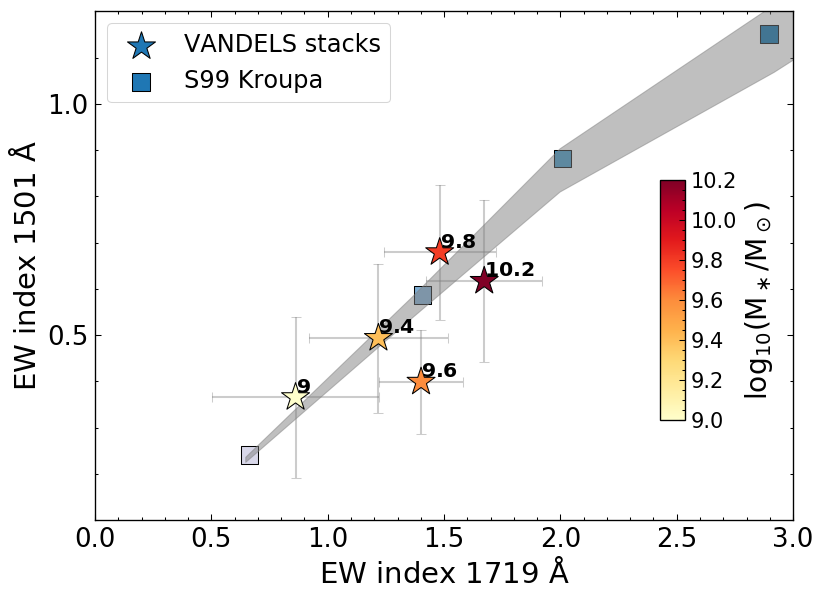}
    \caption{Comparison between the equivalent widths of the $1719$ and $1501$ \AA\ absorption indicies predicted by S99 models (big squares color coded by metallicity). We show in each diagram the EWs measured in five VANDELS stacks (big stars), obtained with the same procedure adopted for synthetic templates. The stacks used here were produced at different stellar mass bins (see Section \ref{MZR_results} and Table \ref{table_MZR}), whose median stellar masses (in log$_{10}$ (M$_\ast$/M$_\odot$)) are highlighted in black above each point. The gray shaded area has the same meaning as in Fig. \ref{calibration_figure1}. 
    }\label{calibration_figure2}
\end{figure}

The EW quantifies the relative absorption strength of the line with respect to the underlying UV continuum, hence it is critical to specify the calculation of $f_{\rm cont}(\lambda)$ in Eq. \ref{eqEW_formula}. Given the relatively low resolution of VANDELS spectra, we cannot identify the 'real', unobscured continuum level, as there are basically no spectral regions free of absorption. To overcome this problem, we adopt the so-called 'pseudo-continuum', which is defined by ranges relatively free of strong absorption or emission lines, as introduced by \citet{rix04}. Therefore, for each index, we calculated $f_{\rm cont} (\lambda)$ in Eq. \ref{eqEW_formula} as the linear interpolation of the error-weigthed average flux density in the closest blueward and redward pseudocontinuum windows defined in Table~3 of \citet{rix04}. Since the original pseudocontinuum widths of $\sim3$-$4$ \AA\ barely correspond to one resolution element in VANDELS spectra, we also increased them by $\pm3$ \AA, yielding a total width of $\sim10$ \AA. This choice also produces more stable measurements, less affected by noise. However, we remark that it does not affect the final results, providing that we use the same definition for both the calibrations and the observations.

In order to mitigate even more noise effects in observations, we estimated the EW of absorption lines in VANDELS stacks using Eq. \ref{eqEW_formula} and taking the median value from $1000$ Monte Carlo realizations, generated by perturbing the flux at each wavelength according to the noise spectrum. This procedure allows a robust estimate of the $1\sigma$ uncertainty of the associated EW as the standard deviation of those different realizations. Moreover, we find that our results are not significantly affected if we just perform a single estimate of the pseudo-continuum level and of the EW of absorption indexes, even though we do not have an associated error for the observation in this case.

We remind that another approach to calculate the pseudocontinuum spectrum is based on fitting a spline to all the \citet{rix04} windows simultaneously, as done by \citet{sommariva12}. However, we found that the two approaches yield very similar calibration functions and thus fully consistent results. In the rest of this paper, we use only the local fit explained above, because it has the clear advantage of being applicable even when the full far-UV spectrum is not available or any of the pseudocontinuum windows has to be discarded because of contamination from sky line residuals in individual observed spectra. 

In Fig. \ref{calibration_figure1} we show the main results of this analysis for the two absorption indexes adopted in this study. 
First, the EW$_{1719}$ spans a large dynamic range of $\sim2.5$ \AA\ (from $0.2$ to $2.7$) from $1\%$ Z$_\odot$ to solar metallicity, making it relatively easy to constrain $Z$ within $\pm0.1$-$0.15$ dex uncertainty with the SNR imposed on our VANDELS stacks (see Fig. \ref{Zerror_SNR}). A third order polynomial fit yields the following calibration :
\begin{equation}\label{calib1719}
\mathrm{log}_{10}(Z/Z_\odot)= 0.06\ EW_{1719}^3 - 0.44\ EW_{1719}^2 + 1.51\ EW_{1719} - 2.12
\end{equation}
Overall, we confirm that the EW of the $1719$ \AA\ metallicity tracer is largely independent on the IMF chosen and stellar age (Fig. \ref{calibration_figure1}-left). Since it is basically uncontaminated by ISM absorption even at higher (solar) metallicities, according to \citet{vidalgarcia17}, we take this index as reference as it should be the most reliable. We also notice that our definition for this index is slightly different from the original version, as this allows to fully include also the leftmost blend generated by FeIV absorptions between $1709$ and $1712$ \AA\ (visible later in Fig. \ref{fitlineprofile}).

Applying the same above procedure to the index located at $1501$ \AA\ yields the following metallicity calibration inferred from a third order polynomial fit (Fig. \ref{calibration_figure1}-right):
\begin{equation}\label{calibothers}
\begin{split}
\mathrm{log}_{10}(Z/Z_\odot)= 1.24\ EW_{1501}^3 - 2.97\ EW_{1501}^2 + 3.48\ EW_{1501} - 1.98
\end{split}
\end{equation}
The $1501$ \AA\ index was first proposed by \citet{sommariva12} as a very promising metallicity tracer. However, because it is narrower, it also has the lowest spread in general, showing EWs $\leq 1$ \AA\ for all metallicities below solar.

In Fig. \ref{calibration_figure2} we display the EWs of the $1719$ and $1501$ indexes obtained from $5$ stacks in M$_\ast$ bins constructed to analyze the mass-metallicity relation (see later in Section \ref{MZR_results}), as they have the highest SNR among all the stacked spectra derived in this work. This figure indicates that the $1501$ tracer not only shows EWs that are correlated to those of the $1719$ index, but that the two EWs are qualitatively in agreement with the predictions of Starburst99 models. In other words, it means that calibrations built from the same S99 templates yield very consistent metallicities from both the $1501$ and $1719$ \AA\ indexes.
We also caution that the two calibrations derived in this section could be safely applied in the metallicity regime spanned by the VANDELS data \citep{cullen19}, but we cannot verify whether significant ISM absorption would affect the lines at higher $Z$, and whether the models (at the VANDELS resolution) still reproduce simultaneously the two EWs outside of our range.

As far as the remaining UV absorption lines are concerned, we find two different situations. 
On the one hand, the EWs of absorption indexes located at $1370$, $1425$, $1460$, $1533$ and $1978$ \AA\ are weakly (or not at all) correlated with the EW of the $1719$ index, hence with the metallicity. The majority of these features measured in the stacked spectra are also very faint and would yield unrealistically low metallicities with very large uncertainties at our SNR and spectral resolution. As a result, they are unusable to constrain the chemical abundances from VANDELS spectra.
On the other hand, the absorption lines at $1400$, $1550$ and $1853$ \AA, even though they are correlated with the EW of the $1719$ index, are systematically deeper than predicted by S99 models, indicating they are contaminated by ISM absorption at various degrees also at sub-solar metallicities. Since our work is based on the stellar metallicity and a modeling of the ISM is beyond our goals, we exclude them from the subsequent analysis. 

We refer the reader to the Appendix \ref{appendix3} for a more detailed discussion of the EWs and calibrations obtained with all the indexes listed in Table \ref{table_features}. 
Hereafter, we focus exclusively on the two aforementioned metallicity indicators at $1719$ and $1501$ \AA, displayed in Fig. \ref{calibration_figure1} and \ref{calibration_figure2}. In order to define a unique, representative metallicity for the stacks derived in this paper, we will consider the average metal abundance obtained from the EWs of those two indexes separately, and we will use the error propagation to determine the final uncertainty on each estimation.

Finally, in Appendix \ref{IMF_variations} we also explore for all the lines the effect of different IMFs than those adopted in this Section. While modifying the IMF upper mass cutoff (up to 300 M$_\odot$/yr) and the slope $\alpha$ of the high-mass end (within $\sim 0.2$ of the Salpeter value $\alpha=-2.35$) do not produce significant variations of the metallicity calibrations for the $1719$ and $1501$ absorption indexes, choosing even flatter slopes (e.g., $\alpha \lesssim -2$) would yield systematically higher metallicity values. However, typical Main Sequence star-forming galaxies at $z \sim 3$ with intermediate stellar masses (median M$_\ast$ $= 10^{9.7}$ M$_\odot$) are not expected to have IMFs extremely different from Salpeter \citep[e.g.,][]{elmegreen06,bouwens12}. %
According to \citet{fontanot18}, flatter and top-heavy IMFs can be found for galaxies more massive and star-forming compared to our VANDELS selection. Finally, we remark that a more detailed analysis of a variable IMF on the MZR and M$_\ast$-$\beta$ relation is clearly beyond the aim of this paper. 

\subsection{Final sample selection}\label{selection}

\begin{figure}[t!]
    \centering
    \includegraphics[angle=0,width=1\linewidth,trim={0cm 0.1cm 0cm 0cm},clip]{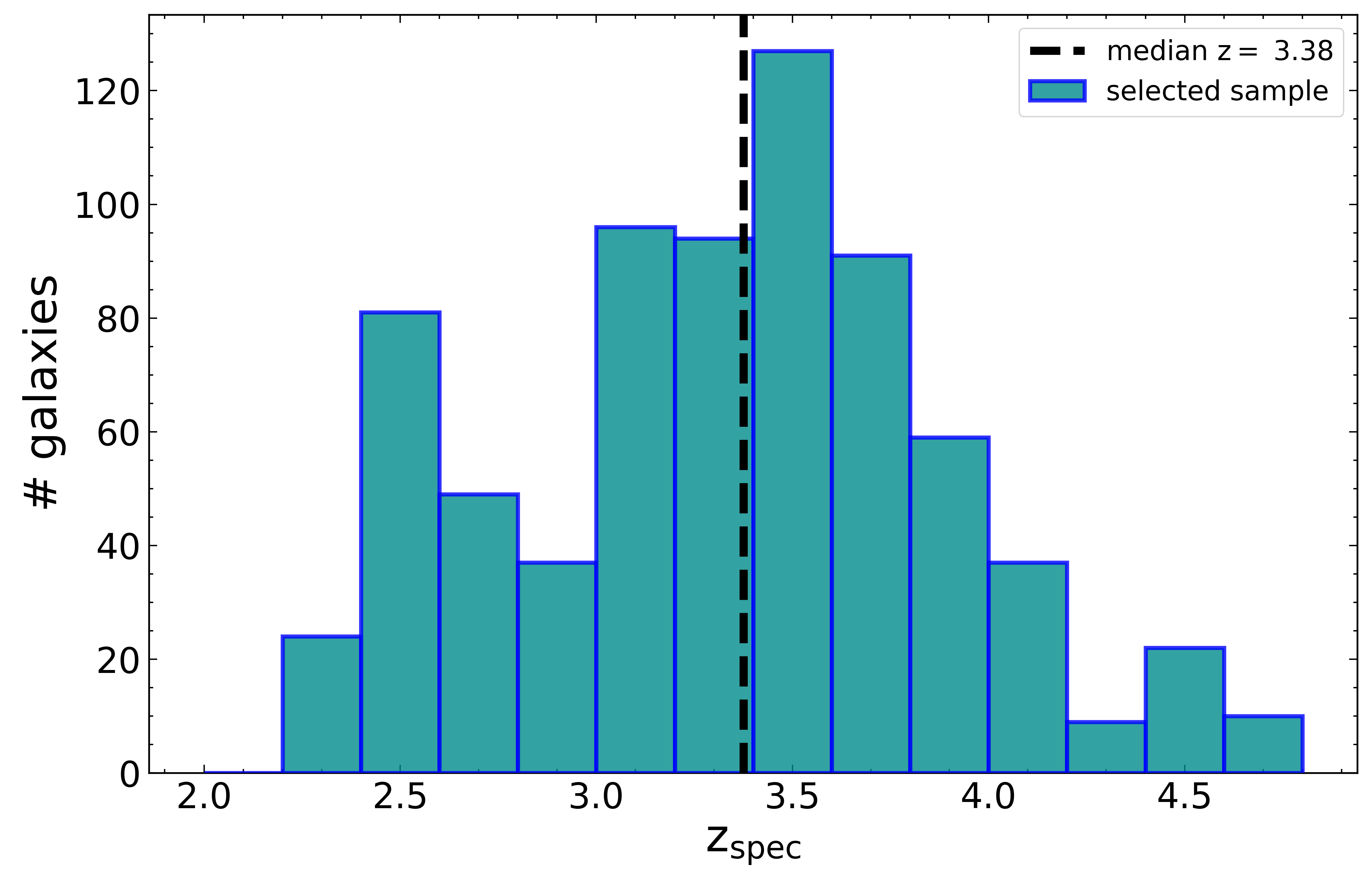}
    \caption{\small Diagram showing the distribution of spectroscopic redshifts of VANDELS galaxies selected for this work. The vertical dashed line indicates the median redshift of the sample ($z_{med}=3.38$).
    }\label{redshift_histo}
\end{figure}

We include additional constraints to our sample selection criteria in order to exclude objects where UV slopes and/or metallicity indicators are not reliable.
From the sample selected at the end of Section \ref{observations}, we thus selected spectra free from bad sky subtraction residuals, noise spikes, or reduction problems in the spectral windows used to estimate the pseudo-continuum level and the EW of the $1501$ and $1719$ metallicity tracers. This yields $732$ galaxies (called subset $\#1$), of which $372$ are in the CDFS field. We will use this subset to study the mass-metallicity relation (MZR) and its evolution with redshift in the following section. The histogram distribution of spectroscopic redshifts for this selected sample is shown in Fig. \ref{redshift_histo}. 

Afterwards, we built a second smaller subset requiring in addition a reliable estimate of the UV slope, with $\sigma_\beta < 1$ and at least three photometric bands used in the fit, as already discussed in Section \ref{uncertainties}. We end up in this case with $576$ galaxies, named subset $\#2$ (all of them are included in the first subset), that we consider for all the remaining diagrams. In this sample, $254$ objects are located in the CDFS field.
We remark that adopting this second galaxy subset for the mass-metallicity relation, the results would not be significantly altered, however we prefer to maintain the larger sample in order to have a better statistics to study the evolution in redshift of that relation. 
We also note that these criteria do not introduce biases in the stellar mass and beta distributions of the original sample, hence the selected galaxies are still representative of the star-forming Main Sequence.

\section{Results}\label{results}

\begin{figure*}[t!]
    \centering
    \includegraphics[angle=0,width=0.49\linewidth,trim={0cm 0cm 0cm 0.1cm},clip]{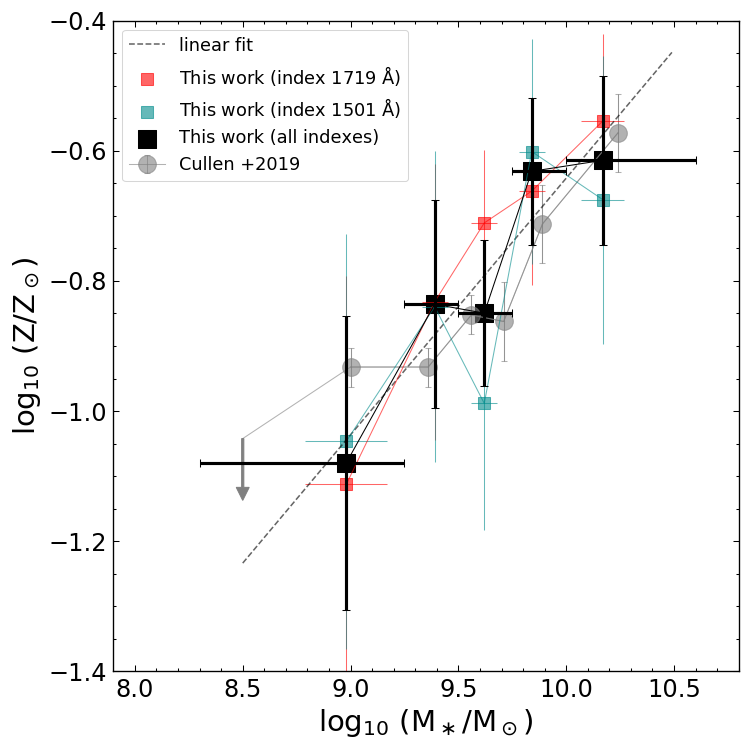}
    \includegraphics[angle=0,width=0.49\linewidth,trim={0cm 0cm 0cm 0.1cm},clip]{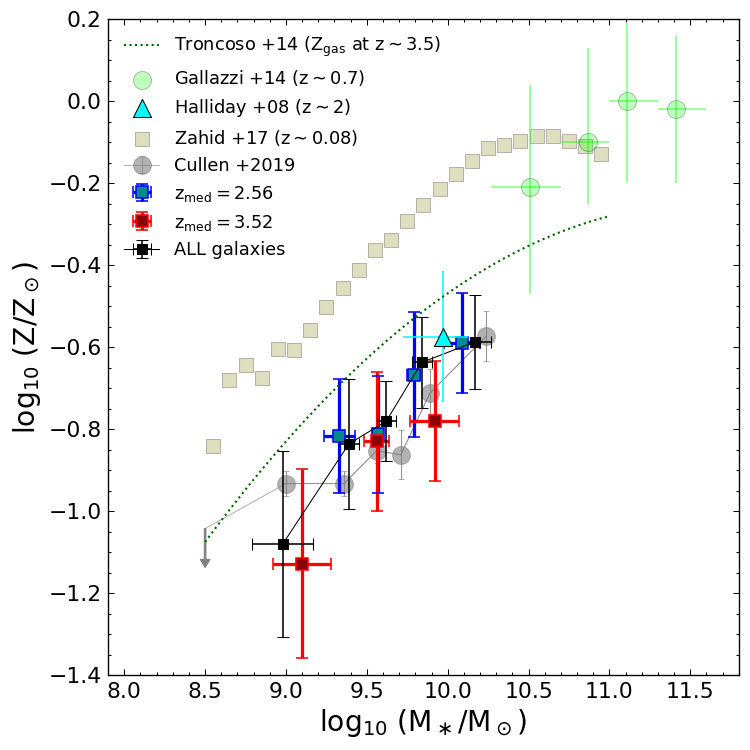}
    \caption{\footnotesize \textit{Left}: Mass-metallicity relation of star-forming galaxies in VANDELS, with median stacks in bins of stellar mass (black squares). The metallicities derived from single indexes (i.e., in order, $1550$, $1719$ and $1501$ \AA) are drawn with pale colored squares in blue, red and dark-cyan, respectively. The linear fit to the median metallicities in the range $10^{8.5}<$ M$_\ast$ $<10^{10.5}$ M$_\odot$ is highlighted with a dark-cyan dashed line. \textit{Right}: Mass-metallicity relation followed by VANDELS star-forming galaxies at redshift $<3$ (dark-cyan squares) and $>3$ (dark-red squares). The median redshifts of the two subsets are $2.58$ and $3.54$, respectively. The relation for all the sample is overplotted in black. In both diagrams, the relation by \citet{cullen19}, calculated at same median redshift of our work (and of the VANDELS survey), is displayed with gray filled circles, while in the bottom panel we also include the MZR found by \citet{zahid17} at $z\sim0.08$ from the SDSS survey (olive squares) and by \citet{gallazzi14} for massive star-forming galaxies (M$_\ast$ $>10^{10.2}$ M$_\odot$) at $z\sim0.7$. The linear fit to the MZR in the lower and higher redshift bin is drawn with a shaded dashed line in the corresponding color. The representative M$_\ast$ and metallicity estimated from the spectral stack of $75$ star-forming galaxies at redshift $\sim2$ from \citet{halliday08} is shown with a cyan square and error bars. Finally, the stellar mass - gas-phase metallicity relation for star-forming galaxies at $z\sim3.5$ \citep{troncoso14} is drawn with a green dashed line.
    }\label{MZR_figure}
\end{figure*}


\begin{table*}[ht]
  \caption{Summary of the best-fit relations parameters. Each panel of the table shows for each bin the stellar mass range, the number of galaxies falling in the bin, the average SNR of the stack, the median stellar mass of the galaxies, and the stellar metallicity computed from the stack using the $1501$ and $1719$ indexes (as explained in the text).}
  \label{summary_relations}

  \medskip

  \begin{tabularx}{\linewidth}{ X X X X X X X }
    \multicolumn{7}{l}{Panel A: MZR from UV absorption lines ($1501$ and $1719$ \AA) of VANDELS selected galaxies in this work (Fig. \ref{MZR_figure}-\textit{left}).} \\
    \toprule
    \multicolumn{1}{l}{bin $\#$} &  \multicolumn{2}{l}{Stellar mass range}  &  \multicolumn{1}{l}{$\#$ galaxies} & \multicolumn{1}{l}{SNR} & \multicolumn{1}{l}{$log_{10}$(M$_\ast$/M$_\odot$)$_{med}$} & \multicolumn{1}{l}{$log_{10}$(Z/Z$_\odot$)} \\
    \midrule

    \multicolumn{1}{l}{1} &  \multicolumn{2}{l}{$8.3<log_{10}$(M$_\ast$/M$_\odot$)$<9.25$} & \multicolumn{1}{l}{207} & \multicolumn{1}{l}{30.8} & \multicolumn{1}{l}{9.0 $\pm$ 0.2} & \multicolumn{1}{l}{-1.08 $\pm$ 0.22} \\
    
    \multicolumn{1}{l}{2} &  \multicolumn{2}{l}{$9.25<log_{10}$(M$_\ast$/M$_\odot$)$<9.5$} & \multicolumn{1}{l}{147} & \multicolumn{1}{l}{34.8} & \multicolumn{1}{l}{9.39 $\pm$ 0.06} & \multicolumn{1}{l}{-0.84 $\pm$ 0.16} \\
    
    \multicolumn{1}{l}{3} &  \multicolumn{2}{l}{$9.5<log_{10}$(M$_\ast$/M$_\odot$)$<9.75$} & \multicolumn{1}{l}{184} & \multicolumn{1}{l}{49.8} & \multicolumn{1}{l}{9.62 $\pm$ 0.06} & \multicolumn{1}{l}{-0.85 $\pm$ 0.11} \\
    
    \multicolumn{1}{l}{4} &  \multicolumn{2}{l}{$9.75<log_{10}$(M$_\ast$/M$_\odot$)$<10$} & \multicolumn{1}{l}{110} & \multicolumn{1}{l}{39.8} & \multicolumn{1}{l}{9.84 $\pm$ 0.06} & \multicolumn{1}{l}{-0.63 $\pm$ 0.11} \\
    
    \multicolumn{1}{l}{5} &  \multicolumn{2}{l}{$10<log_{10}$(M$_\ast$/M$_\odot$)$<10.6$} & \multicolumn{1}{l}{83} & \multicolumn{1}{l}{32.4} & \multicolumn{1}{l}{10.17 $\pm$ 0.1} & \multicolumn{1}{l}{-0.61 $\pm$ 0.13} \\
    
    \bottomrule
  \end{tabularx}\label{table_MZR}
  
  \bigskip

  \begin{tabularx}{\linewidth}{ X X X X X X X }
    \multicolumn{7}{l}{Panel B: Redshift dependence of the MZR (Fig. \ref{MZR_figure}-\textit{right}).} \\
    \toprule
    \multicolumn{1}{l}{bin $\#$} &  \multicolumn{2}{l}{Stellar mass range}  &  \multicolumn{1}{l}{$\#$ galaxies} & \multicolumn{1}{l}{SNR} & \multicolumn{1}{l}{$log_{10}$(M$_\ast$/M$_\odot$)$_{med}$} & \multicolumn{1}{l}{$log_{10}$(Z/Z$_\odot$)} \\
    \midrule
    \multicolumn{1}{l}{} &  \multicolumn{2}{l}{$2<z<3$} & \multicolumn{1}{l}{} & \multicolumn{1}{l}{} & \multicolumn{1}{l}{} & \multicolumn{1}{l}{} \\
    \midrule
    \multicolumn{1}{l}{1} & \multicolumn{2}{l}{$8.8<log_{10}$(M$_\ast$/M$_\odot$)$<9.5$} & \multicolumn{1}{l}{50} & \multicolumn{1}{l}{33.3} & \multicolumn{1}{l}{9.33 $\pm$ 0.10} & \multicolumn{1}{l}{-0.82 $\pm$ 0.14} \\
    \multicolumn{1}{l}{2} & \multicolumn{2}{l}{$9.5<log_{10}$(M$_\ast$/M$_\odot$)$<9.7$} & \multicolumn{1}{l}{50} & \multicolumn{1}{l}{36} & \multicolumn{1}{l}{9.57 $\pm$ 0.03} & \multicolumn{1}{l}{-0.81 $\pm$ 0.14} \\
    \multicolumn{1}{l}{3} & \multicolumn{2}{l}{$9.7<log_{10}$(M$_\ast$/M$_\odot$)$<9.9$} & \multicolumn{1}{l}{40} & \multicolumn{1}{l}{31.3} & \multicolumn{1}{l}{9.79 $\pm$ 0.04} & \multicolumn{1}{l}{-0.67 $\pm$ 0.15} \\
    \multicolumn{1}{l}{4} & \multicolumn{2}{l}{$9.9<log_{10}$(M$_\ast$/M$_\odot$)$<10.6$} & \multicolumn{1}{l}{48} & \multicolumn{1}{l}{33.6} & \multicolumn{1}{l}{10.09 $\pm$ 0.11} & \multicolumn{1}{l}{-0.59 $\pm$ 0.12} \\
    \midrule
    \multicolumn{1}{l}{} &  \multicolumn{2}{l}{$3<z<5$}  & \multicolumn{1}{l}{} & \multicolumn{1}{l}{} & \multicolumn{1}{l}{} & \multicolumn{1}{l}{} \\
    \midrule
    \multicolumn{1}{l}{5} & \multicolumn{2}{l}{$8.3<log_{10}$(M$_\ast$/M$_\odot$)$<9.4$} & \multicolumn{1}{l}{247} & \multicolumn{1}{l}{27.5} & \multicolumn{1}{l}{9.1 $\pm$ 0.18} & \multicolumn{1}{l}{-1.13 $\pm$ 0.23} \\
    \multicolumn{1}{l}{6} & \multicolumn{2}{l}{$9.4<log_{10}$(M$_\ast$/M$_\odot$)$<9.7$} & \multicolumn{1}{l}{148} & \multicolumn{1}{l}{30.7} & \multicolumn{1}{l}{9.56 $\pm$ 0.08} & \multicolumn{1}{l}{-0.83 $\pm$ 0.17} \\
    \multicolumn{1}{l}{7} & \multicolumn{2}{l}{$9.7<log_{10}$(M$_\ast$/M$_\odot$)$<10.6$} & \multicolumn{1}{l}{135} & \multicolumn{1}{l}{31.7} & \multicolumn{1}{l}{9.92 $\pm$ 0.15} & \multicolumn{1}{l}{-0.78 $\pm$ 0.15} \\
    
    \bottomrule
  \end{tabularx}\label{table_MZR_redshift}

  \bigskip

  \begin{tabularx}{\linewidth}{ X X X X X X X }
    \multicolumn{7}{l}{Panel C: $\beta$ vs Z$_\ast$ relation (Fig. \ref{beta_metallicity}).} \\
    \toprule
    \multicolumn{1}{l}{bin $\#$} &  \multicolumn{2}{l}{UV slope range}  &  \multicolumn{1}{l}{$\#$ galaxies} & \multicolumn{1}{l}{SNR} &  \multicolumn{1}{l}{$\beta$ $_{med}$} & \multicolumn{1}{l}{$log_{10}$(Z/Z$_\odot$)} \\
    \midrule
    \multicolumn{1}{l}{1} & \multicolumn{2}{l}{$-2.5<\beta<-1.7$} & \multicolumn{1}{l}{308} & \multicolumn{1}{l}{52} & \multicolumn{1}{l}{-1.98 $\pm$ 0.18} & \multicolumn{1}{l}{-0.93 $\pm$ 0.10} \\ 
    \multicolumn{1}{l}{2} & \multicolumn{2}{l}{$-1.7<\beta<-1.3$} & \multicolumn{1}{l}{200} & \multicolumn{1}{l}{53} & \multicolumn{1}{l}{-1.51 $\pm$ 0.10} & \multicolumn{1}{l}{-0.72 $\pm$ 0.07} \\ 
    \multicolumn{1}{l}{3} & \multicolumn{2}{l}{$-1.3<\beta<-0.5$} & \multicolumn{1}{l}{65} & \multicolumn{1}{l}{26} & \multicolumn{1}{l}{-1.10 $\pm$ 0.11} & \multicolumn{1}{l}{-0.45 $\pm$ 0.16} \\ 
    \bottomrule
  \end{tabularx}\label{table_betaZ}

\end{table*}\label{table2}

In this Section we explore how the stellar metallicity and UV continuum slope are related to each other and to other galaxy properties. We then assess the role of stellar metallicity in the estimation of dust attenuations for galaxies with different stellar masses and UV luminosities. 
As a first step, we take advantage of our new metallicity measurements to investigate the mass-metallicity relation from UV absorption lines, and compare the result with the relation presented in \citet{cullen19}, derived from fitting S99 models to the entire VANDELS FUV spectral range.

\subsection{Mass-metallicity relation from UV absorption indexes}\label{MZR_results}

The mass-metallicity relation (MZR) is a powerful diagnostic of the chemical evolution history of galaxies, with its shape and normalization providing important constraints on the star-formation history, feedback processes, gas inflows and outflows \citep{maiolino19}. 

In Fig. \ref{MZR_figure}-\textit{left} we show the stellar mass - stellar metallicity relation (MZR) derived for our VANDELS subset $\#1$ in the redshift range $2<z<5$ from UV absorption tracers, as explained in Section \ref{calibrations}.
The data points represent the median stellar masses of galaxies residing in the same bin, and the chemical abundances from the corresponding stacks. We chose a stellar mass bin width of $0.25$ dex (larger at the borders), as a compromise between the highest possible number of bins and a minimum SNR ($=30$) required for each stacked spectrum to derive accurate metallicities. 
The final values of Z and errors (represented as black squares and vertical bars) were derived by averaging the estimates obtained from the two absorption indexes at $1501$ and $1719$ \AA. In Fig. \ref{MZR_figure}-\textit{top}, we also report the corresponding metal estimates for each individual index with pale dark-cyan and red squares, respectively. For comparison purposes, we also include the MZR of \citet{cullen19} as gray circles, which is derived from the DR2\footnote{http://vandels.inaf.it/dr2.html} release of VANDELS with a similar selection to our ($2 < z < 5.0$ and $z_\text{average} = 3.5$).

At a first analysis, we note that the MZR built from our two absorption indicators, either taking the averages or the single values separately, is consistent within $1\sigma$ to that derived by \citet{cullen19}. The metallicity rises by $\sim0.5$ dex, from log$_{10}$ (Z$_\ast$/Z$_\odot$) $\simeq -1.1$ to $\simeq -0.6$, in the range of stellar masses between log$_{10}$ (M$_\ast$/M$_\odot$) $\simeq 9$ and $\simeq 10.2$. 
The increasing trend of metallicity in this mass range can be approximated by a linear function (dark-cyan dashed line), whose best-fit coefficients are displayed in Table \ref{fit_MZR}. 
We notice that our relation is sampled by a lower number of points in the low mass range compared to \citet{cullen19}, as galaxies in this regime are generally fainter and larger bins are necessary to reach the required SNR. Nevertheless, the upper limit at M$_\ast \simeq 10^{8.5}$ M$_\odot$ established by \citet{cullen19} and the metallicity of our lowest mass bin suggest that the same decreasing trend may also continue to stellar masses substantially lower than 10$^9$ M$_\odot$. 
In Table \ref{summary_relations} we present the definition of the bins used to build the MZR and the other relations studied in this work. It also includes for each bin the number of galaxies considered, the SNR reached in the stacked spectra, and the median properties of the corresponding subsets.

\subsection{Redshift dependence of the MZR}\label{redshift_evolution}

\begin{table}[t!]
  \label{fit_MZR}
  \caption{Linear fit of the MZR in two bins of redshift and in the whole sample, as explained in the text and shown in Fig. \ref{MZR_figure}. The relations that we fit are of the form log$_{10}$ (Z/Z$_\odot$) $=$ $m$ $\times$ (log$_{10}$ (M$_\ast$/M$_\odot$) $-10$) $+q$, where the coefficients are listed below.}

  \medskip

  \begin{tabularx}{\linewidth}{ X X X X}
    \toprule
    \multicolumn{2}{l}{redshift range} &  \multicolumn{1}{l}{$m$}  &  \multicolumn{1}{l}{$q$} \\
    \midrule

    \multicolumn{2}{l}{$2<z<3$} &  \multicolumn{1}{l}{$0.33 \pm 0.06$}  &  \multicolumn{1}{l}{$-0.62 \pm 0.02$} \\
    \multicolumn{2}{l}{$3<z<5$} &  \multicolumn{1}{l}{$0.39 \pm 0.15$}  &  \multicolumn{1}{l}{$-0.72 \pm 0.07$} \\
    \multicolumn{2}{l}{$all$} &  \multicolumn{1}{l}{$0.39 \pm 0.10$}  &  \multicolumn{1}{l}{$-0.64 \pm 0.04$} \\
    \bottomrule
  \end{tabularx}
\end{table}

We analyze in this section the mass-metallicity relation as a function of redshift. To this aim, we divided our sample (subset $\#1$) in two redshift bins with $z$ lower and higher than $3$. Then we measured the metal content from our two indexes, as already done for the global relation, in four (respectively three) bins of stellar masses, as can be seen in the right panel of Fig. \ref{MZR_figure}. 
In the range of M$_\ast$ that is in common between the two subsets (log$_{10}$ (M$_\ast$/M$_\odot$) from $\sim 9.5$ to $\sim 10$), the metallicities of the stacks in the upper redshift bin are systematically lower than at lower redshift by $\sim 0.1$ on average, even though all the estimates are still consistent within $1 \sigma$. 
Secondly, we fitted a linear relation to all the data points belonging to the same redshift bin, finding a normalization difference between the two MZR (at $10^{10}$ M$_\odot$) of $0.125 \pm 0.134$, hence they are consistent within the errors. The coefficient results for the two redshift bins are listed in Table \ref{fit_MZR}. Furthermore, also a series of Monte Carlo simulations, where we perturbed the median Z$_\ast$ and M$_\ast$ of the bins according to the estimated (gaussian-like) uncertainties, yield a $28\%$ probability to obtain our results if there is no redshift evolution of the MZR normalization, hence again the difference is not statistically significant. We note that this result, even though obtained with a different approach, is in agreement with the conclusions of \citet{cullen19}, who also find no clear monothonic decrease of stellar metallicity at fixed mass in the redshift range $2<z<5$.

In Fig. \ref{MZR_figure}-right, we also compare the shape of our relation with other studies at similar or different redshifts. 
First, \citet{halliday08} derived a stellar metallicity from the stacked spectrum of $75$ star-forming galaxies at $z\sim2$ observed with GMASS (Galaxy Mass Assembly ultra-deep Spectroscopic Survey). 
For their estimation, they used the absorption index at $1978$ \AA, which has been shown by subsequent studies to be significantly affected by stellar age and by the choice of the IMF \citep[e.g.,][]{sommariva12}, and also from our work it is not a good indicator (see Appendix \ref{appendix3}). However, they also fitted stellar population models to the far-UV spectra including the $1501$ and $1719$ absorption lines, and they find that observations are better reproduced by models with a metallicity of $0.2$ $\times$ Z$_\odot$, even though the correct value resides between $0.2$ and $0.4$ Z$_\odot$. This suggests that no significant evolution of the MZR can be claimed down to $z\sim 2$, or that the metallicity variation is very mild in the range $2<z<3.5$.

Furthermore, other works were published at significantly lower redshifs than in our study. 
\citet{zahid17} investigated the MZR at redshifts $z\sim0.08$ ($0.027 < z < 0.25$) for star-forming galaxies extracted from the Sloan Digital Sky Survey (SDSS). 
Secondly, \citet{gallazzi14} analyzed a mass-selected sample of $\sim70$ galaxies at redshift $\sim 0.7$, deriving a MZR representative of the whole galaxy population, both star-forming and quiescent. 
While a direct comparison to the latter cannot be performed as we probe systematically lower stellar masses, the dataset of \citet{zahid17} shows that there is a decrease of stellar metallicity by $\sim0.4$ dex between $z\sim0$ and $z\sim3$ in all the mass range explored, from $10^9$ to $10^{10.3}$ M$_\odot$.
Remarkably, the slopes of the mass-metallicity relations found at such different cosmic epochs are very similar. Fitting a linear relation between log$_{10}(M_\ast/M_\odot)=8.5$ and $10.5$ indeed yields slopes that are all in agreement within $1\sigma$. The results of the linear fit in this mass range for our two subsets at lower and higher redshifts are summarized in Table~3. This likewise suggests that, if the linear fit holds up to $M_\ast \simeq 10^{10.5}$ M$_\odot$, the same metallicity offsets might exist down to substantially lower stellar masses than those probed here. 

Even though it is interesting to compare to results obtained at other cosmic epochs, we warn that these studies adopt in general a different procedure for the derivation of the stellar metallicity, which might affect the measured level of normalization. \citet{gallazzi14} infer Z$_\ast$ from the metal-sensitive absorption indices [Mg$_2$Fe] and [MgFe] in the optical spectrum, hence their metallicity might be representative of slightly older stellar populations compared to those probed with far-UV rest-frame absorption lines. 
A similar conclusion holds for the MZR of \citet{zahid17}, who fit stellar population synthesis models to stacked spectra (in the optical range) of star-forming galaxies in bins of stellar mass. However, their relation is also consistent with the gas-phase metallicity obtained from emission lines.  
For comparison purposes, we also show in \ref{MZR_figure}-right the stellar mass - gas-phase metallicity relation at similar redshifts from \citet{troncoso14}, which is approximately $0.2$ dex above our MZR estimated from VANDELS. We refer to \citet{cullen19} for a more detailed discussion of this comparison.

\subsection{UV slope and stellar mass}\label{beta_mass}

\begin{figure}[t!]
    \centering
    \includegraphics[angle=0,width=\linewidth,trim={0.4cm 0cm 0.1cm 0cm},clip]{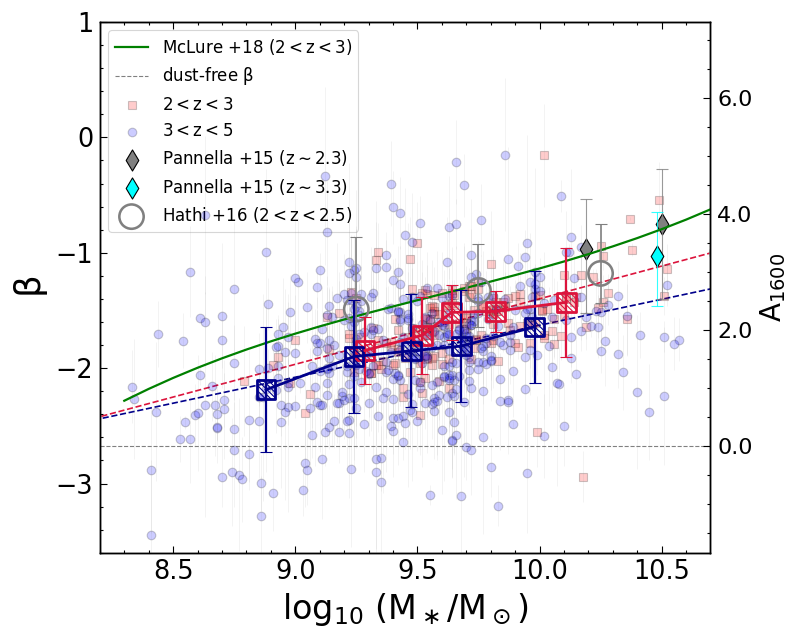} 
    \caption{\small The UV slope as a function of stellar mass in two redshift bins: $2<z<3.5$ and $3.5<z<5$ (red and blue colors, respectively), where $3.5$ is the median redshift of VANDELS. A linear fit to the relations in each redshift bin is also shown with dashed lines with their corresponding colors. The M$_\ast$-$\beta$ relation from \citet{mclure18a} at $z<3$ is shown with a green line for comparison, while gray and cyan diamonds come from \citet{pannella15}. On the left y-axis, A$_{1600}$ is shown using the $\beta$-A$_{1600}$ conversion of \citet{meurer99}. 
    }\label{mass-beta}
\end{figure}

Another way to study the evolution of galaxies in the early Universe is tracing their dust attenuation as a function of stellar mass, metallicity, and redshift. After the formation of the first stars and galaxies, the dust content in the interstellar medium (ISM) is expected to increase during the first billion years, while generations of stars die and pollute their surrounding environment. The amount of dust absorption also limits our understanding of how star-formation evolves in this cosmic epoch. 
While the easiest approach to infer the dust attenuation for large samples of high-$z$, UV detected galaxies relies on their UV continuum slopes, an alternative approach adopting directly the stellar masses has also become common in recent years. Indeed, a correlation between $\beta$ (or A$_{V}$) and M$_\ast$ has been found up to $z\sim3.5$ \citep[e.g.,][]{buat12,heinis14,pannella15,hathi16,mclure18a}, with a $1\sigma$ scatter in $\beta$ of the order of $0.5$. It is thus interesting to investigate with our large VANDELS sample how these quantites are related. 

\begin{figure*}[t!]
    \centering
    \includegraphics[angle=0,width=\linewidth,trim={0.cm 0.cm 0.cm 0cm},clip]{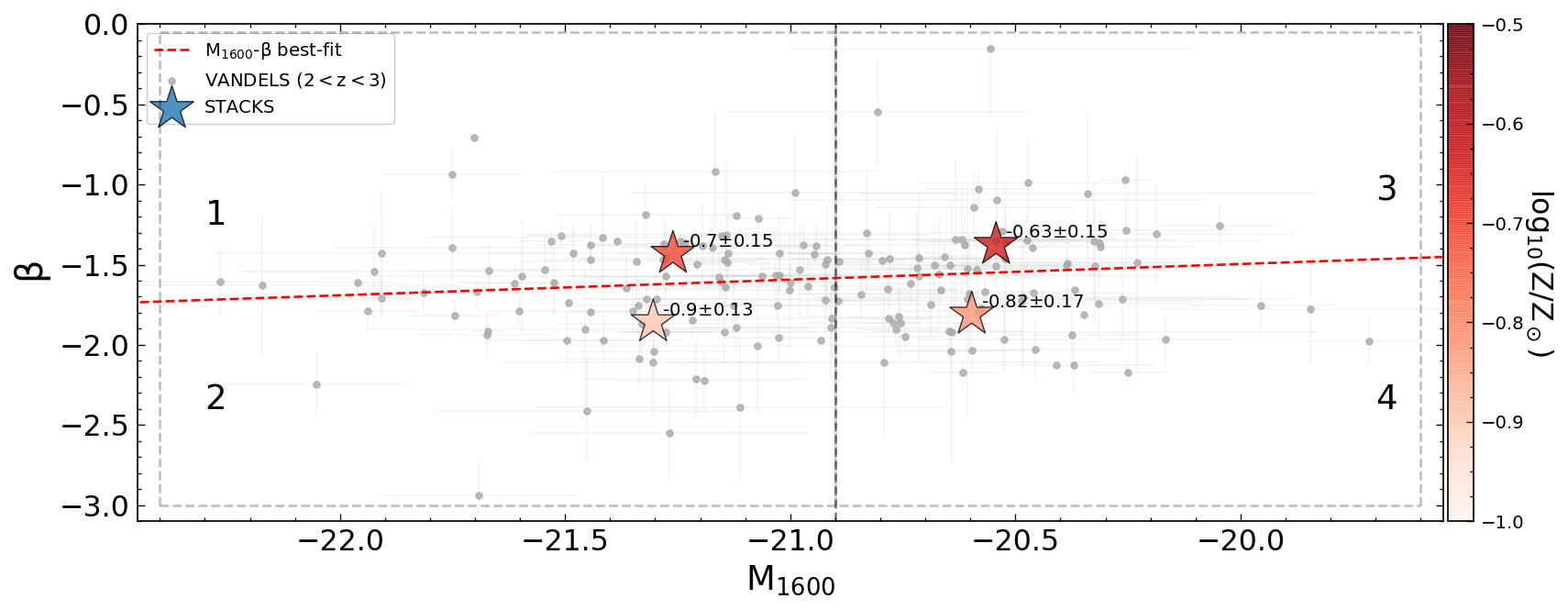}
    \includegraphics[angle=0,width=\linewidth,trim={0.cm 0.cm 0.cm 0cm},clip]{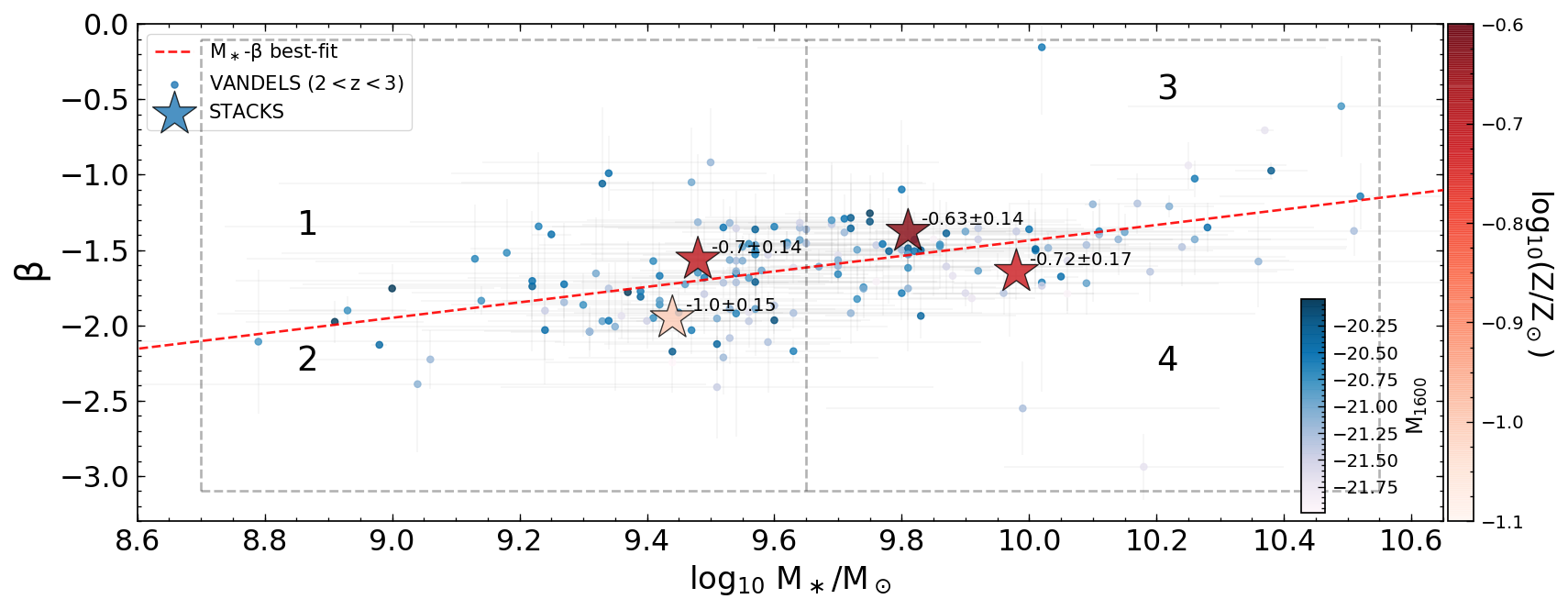}
    \caption{\small \textit{Top}: Metallicity dependence of the M$_{1600}$-$\beta$ relation for VANDELS star-forming galaxies at redshifts $2<z<3$. The position of the stars (color-coded in metallicity) is representative of the median M$_{UV}$ and $\beta$ of the galaxies in each bin. 
    The metallicities ($\pm 1\sigma$ uncertainty) from the stacked spectra of galaxies in the same bin are written in black on the right side of each star. Single galaxies are drawn with squares. The best-fit relation (red dashed line) is used to divide the bins in $\beta$, while mass bins are shown with gray dashed lines. The resulting four bins are marked by 1,2,3,4. \textit{Bottom}: Metallicity dependence of the M$_\ast$-$\beta$ relation for the same subset of galaxies analyzed in the upper panel. The metallicity is calculated in four bins of stellar mass and UV slope, defined by the dashed lines. 
    }\label{metallicity_coding}
\end{figure*}

In Fig. \ref{mass-beta} we display, for our VANDELS selected subset $\#2$, the UV slope as a function of stellar mass. We remark that we are taking here only those galaxies for which $\beta$ is well constrained from the power-law fit to the available photometric data (see Section \ref{uncertainties}). 
Given the rapid evolution of $\beta$ expected with cosmic time, we divided the main sample in two redshift bins, above and below the median redshift of the sample. Then we constructed additional subsamples of different stellar masses by requiring an equal number of objects for each subset, and we calculated the median $\beta$ and the median absolute deviation (MAD) in all of these bins. 

We find that the median UV slopes in our stellar-mass bins range from $-2.2$ to $-1.4$, which correspond to an attenuation at $1600$ \AA\ (A$_{1600}$) from $\sim1$ to $\sim2.5$ mag, assuming the \citet{meurer99} relation. 
We can see that, calculating the median $\beta$ in $5$ equal sized bins of stellar mass, more distant galaxies at $3<z<5$ (blue squares) are bluer than the low-$z$ subset (red squares) by $\sim0.3$ on average, even though the dispersion of the points around the median relations (larger for the high-$z$ subset because of the higher uncertainties of individual measurements) is typically greater than the average difference between the two subsets ($0.2$ to $0.5$). 
However, we also notice that such difference is more evident in the third and fourth bins of the lower-$z$ sample, whose $\beta$ values are significantly ($>1\sigma$) redder compared to the second subset at $z>3$. 
This result also indicates that dust attenuation is less relevant as redshift increases, as suggested by \citet{bouwens07}. 

The horizontal dashed line in Fig. \ref{mass-beta} highlights the dust-free level of $\beta$. Despite some galaxies lying below this limit, we notice that the majority of them are still consistent within $1$ or $2 \sigma$ uncertainty with very little or zero attenuation. We warn the reader that the intrinsic value of $\beta$ in absence of dust attenuation has also a second order dependence on the stellar metallicity, stellar population age and IMF, hence the above line should be considered as an approximation. 

We determine the dust-free value of $\beta$ from the same Starburst 99 models we use to calibrate the metallicity with absorption lines. Varying the stellar age and the stellar metallicity of the models, respectively in the range 50-500 Myr and 0.05-2.5 times Z$_\odot$ has a minor effect on the intrinsic $\beta$, in all cases no larger than $0.1$. Given our small metallicity range, the contribution of Z$_\ast$ would be even smaller, of the order of $\sim 1 \%$, thus largely negligible. As a result, the dust attenuation has by far the biggest effect on shaping the UV slope, as already claimed by \citet{bouwens12}. In the following, we consider an intrinsic $\beta$ of $-2.67$, found for an age of $100$ Myr (assumed for the indexes calibrations) and metallicity $-1.2 < $ log$_{10}$ (Z/Z$_\odot$) $< - 0.5$. This value is also consistent with the extrapolation from the $\beta$-$A_{1600}$ relation of LBGs at $z\sim4$ by \citet{castellano14} ($\beta_\text{dust free}=-2.67_{-0.2}^{+0.18}$) and from a similar analysis performed by \citet{debarros14}.

Moreover, we also remark that assuming different dust attenuation laws and different dust geometries than the foreground screen may lead to a different conversion between A$_{1600}$ and $\beta$ \citep[for more details see, e.g.,][]{reddy18,salim18,mclure18a}. However, this would only produce a constant rescaling of the absolute attenuation axis (A$_{1600}$), if all the galaxies obey the same law. In order to better constrain the A$_{1600}$ - $\beta$ relation, we would need an independent estimate of dust attenuation for these galaxies, which could come from analyzing their far-infrared emission. 

In Fig. \ref{mass-beta} we also show for comparison the results found by \citet{pannella15}, \citet{hathi16} and \citet{mclure18a} for star-forming galaxies in the redshift range between $2$ and $3.3$, with colored diamonds, gray empty circles and a green line, respectively. In particular, we notice that the study of \citet{pannella15} at $z\sim3.3$ (compatible with our median redshift) comprises only galaxies more massive than $10^{10.5}$ M$_\odot$, thus they lie outside the mass range where we have robust statistics in VANDELS.
Secondly, the median photometric based UV slopes of \citet{hathi16} are slightly above our estimates, which is likely due to the lower redshift range they probe in their study. Nevertheless, their results are in agreement within $1\sigma$ with our median $\beta$ at $z_\text{median}=2.58$.
Finally, despite the large dispersion of our points, the slope of our M$_\ast$ - $\beta$ relations is in reasonable agreement in all redshift bins with that found by \citet{mclure18a}, even though our $\beta$ measurements are systematically lower by $\sim 0.3$, depending on the stellar mass range. We remark that the two analyses are performed with different methods: galaxies from \citet{mclure18a} were indeed derived by fitting pure power-law SEDs to the photometry and requiring only central bandpasses not to lie outside the Calzetti ranges when estimating $\beta$. This effect was already discussed in Section \ref{uncertainties}, and accounts for an offset of $\sim 0.15$-$0.2$ dex.  
An additional offset of $0.05$-$0.1$ in $\beta$ (lower than the typical $\beta$ uncertainty) can come from the sample selection, because we are considering here only high quality spectroscopic determinations (flags 3 and 4). In any case, this slight offset does not imply a significant physical difference compared to lower quality flags or to the full parent sample, as both of them are representative of the star-forming Main Sequence \citep{mclure18b}.

\subsection{Metallicity dependence of the M$_\ast$-$\beta$ and M$_{UV}$-$\beta$ relations}\label{met_dependence}

We have seen that the UV magnitude and the stellar mass can be used as proxies to infer the UV slope or the dust attenuation level in the UV, providing useful corrections to derive dust-unbiased luminosity functions and total SFRs of high-redshift galaxies. 
We analyze in the following whether the stellar metallicity plays any role in these conversions, and whether it can improve our estimate of dust attenuation. 

In Fig. \ref{metallicity_coding}-\textit{top} we study the metallicity dependence of the M$_{UV}$-$\beta$ relation shown in Fig. \ref{MUV-beta}. Given the evolution of $\beta$ with cosmic time, to test variations of metallicity we have to focus on a limited range of redshifts throughout the analysis. Because of the higher SNR available, we considered the lowest redshift bin between $z=2$ and $z=3$. This allows us to explore a larger portion of the M$_{UV}$-$\beta$ plane and define four bins of galaxies with similar median $z$, using for separation both the M$_{1600}$-$\beta$ relation from the whole sample at $z<3$ and the median UV magnitude of this subset. The stacked spectra in all the four bins have a SNR above $25$, ideal for our metallicity estimation method based on the absorption indexes described in Section \ref{calibrations}. 

The result in Fig. \ref{metallicity_coding}-\textit{top} indicates a metallicity dependence of the M$_\text{UV}$-$\beta$ relation: galaxies with redder UV slope (i.e., higher attenuation) have an enhanced metallicity at fixed UV absolute magnitude compared to less attenuated objects by $\sim + 0.2$ dex on average. The difference is larger than the typical $1 \sigma$ uncertainties of the metallicity estimates, hence it is significant both for UV bright and faint sources. This also indicates a probability of less than $\sim 2 \%$ to obtain this configuration if there is no dependence of the $\beta$-M$_\text{UV}$ relation on the stellar metallicity. 

We also applied the same approach explained above for the same galaxies to the M$_\ast$-$\beta$ relation (Fig. \ref{metallicity_coding}-\textit{bottom}). As in the previous case, we use for the separation the median stellar mass of the subset ($10^{9.7}$ M$_\odot$) and its best-fit M$_\ast$-$\beta$ relation, constructed by fitting a linear relation to the median $\beta$ values in five bins of M$_\ast$. This way we are able to study the stellar metallicity in each stellar mass regime. 

We can see in Fig. \ref{metallicity_coding}-\textit{bottom} that the largest increase in metallicity occurs in the direction of increasing stellar mass, which is a further evidence of the tight relation between these two quantities already seen in Section \ref{MZR_results}.
At fixed M$_\ast$, a statistically significant difference (at $>2\sigma$) of $0.3$ dex in metallicity is found in the lower stellar mass range (M$_\ast <$ M$_\text{median}$) between galaxies with UV slope lower and higher than the best-fit M$_\ast$-$\beta$ relation. In contrast, in the highest mass bin (M$_\ast$ $>10^{9.65}$ M$_\odot$), while the metallicity of redder galaxies is still higher than less attenuated objects, the difference is smaller ($\sim0.1$ dex), and the two measurements are consistent within their $1\sigma$ errors. 

Overall, Fig. \ref{metallicity_coding} indicates that the stellar metallicity can explain part of the scatter of the M$_{UV}$-$\beta$ and M$_\ast$-$\beta$ relations. In particular, galaxies show a spread of metallicity with dust attenuation at fixed M$_{UV}$ and M$_\ast$, even though more massive and evolved systems ((M$_\ast$ $>10^{9.65}$ M$_\odot$) tend to have more homogeneous metallicity values compared to lower mass systems.

\subsection{The attenuation - metallicity relation}\label{metallicity_attenuation}

\begin{figure}[t!]
    \centering
    \includegraphics[angle=0,width=\linewidth,trim={0.cm 0.cm 0.cm 0cm},clip]{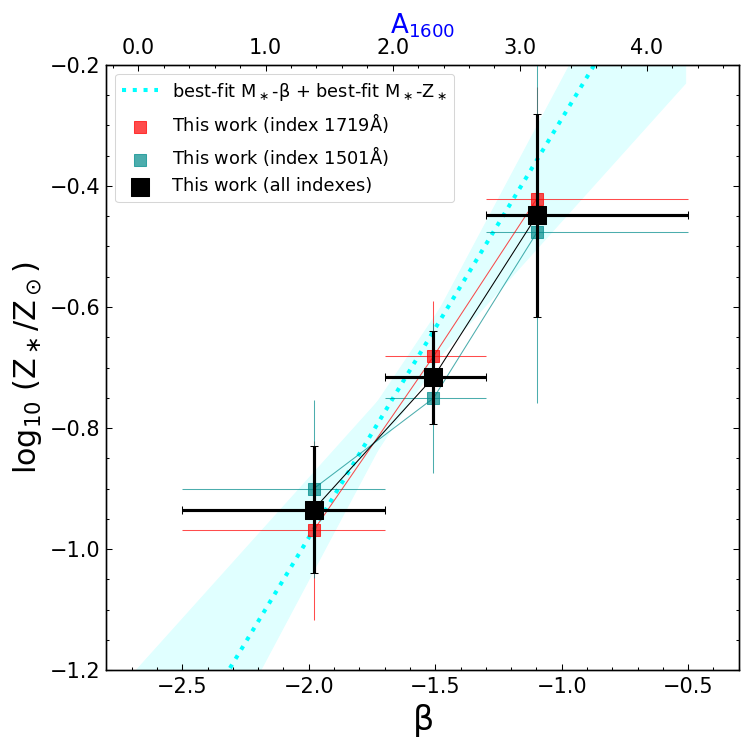}
    \caption{\small Relation between the UV slope and the stellar metallicity for VANDELS galaxies. The average relation from the two indexes ($1501$ and $1719$ \AA) is drawn with black squares, while the pale colored squares represent the underlying metallicities from single absorption features. Horizontal bars associated to each of the three data points indicate the range of $\beta$ slopes of galaxies in the same bin. The relation obtained from combining the best-fit MZR and the M$_\ast$-$\beta$ relation is highlighted with a cyan dashed line and cyan $1\sigma$ uncertainty area. On the top x-axis, A$_{1600}$ is shown for comparison purposes, as determined from the $\beta$-A$_{1600}$ conversion of \citet{meurer99}. We warn that this conversion may also depend at a second order on the metallicity itself. 
    }\label{beta_metallicity}
\end{figure}

\begin{figure*}[h!]
    \centering
    \vspace{-0.2cm}
    \includegraphics[angle=0,width=0.46\linewidth,trim={0.cm 0.cm 0.1cm 0cm},clip]{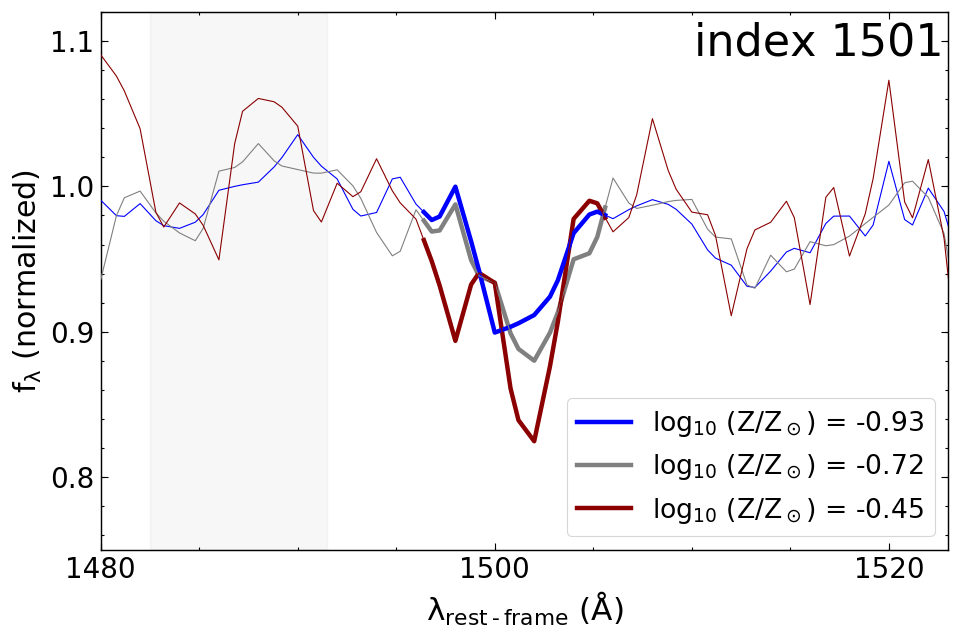} 
    \includegraphics[angle=0,width=0.46\linewidth,trim={0.cm 0.cm 0.3cm 0cm},clip]{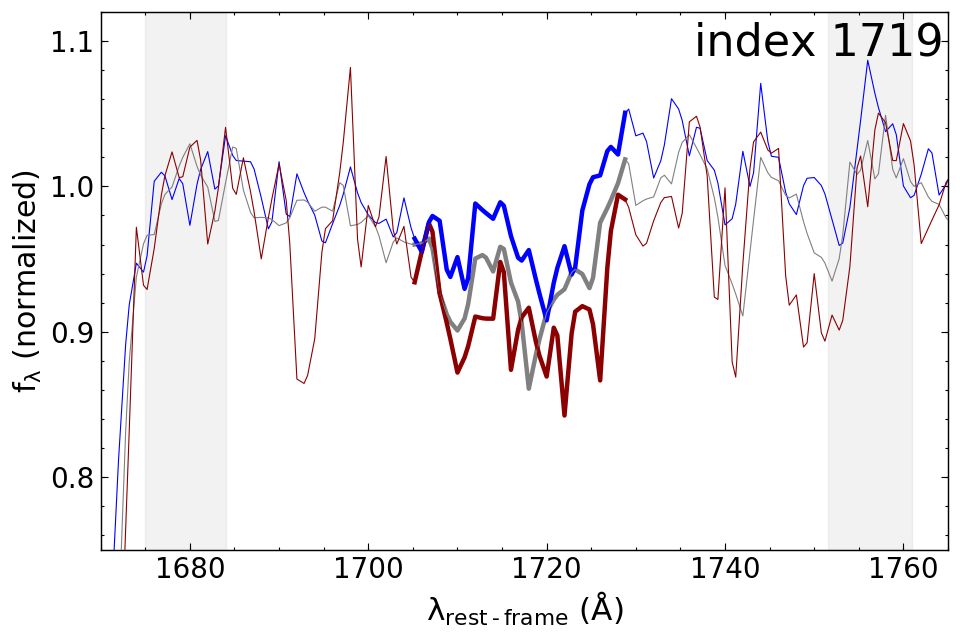}
    \vspace{-0.33cm}
    \includegraphics[angle=0,width=0.91\linewidth,trim={0.cm 0.cm 0.1cm 0cm},clip]{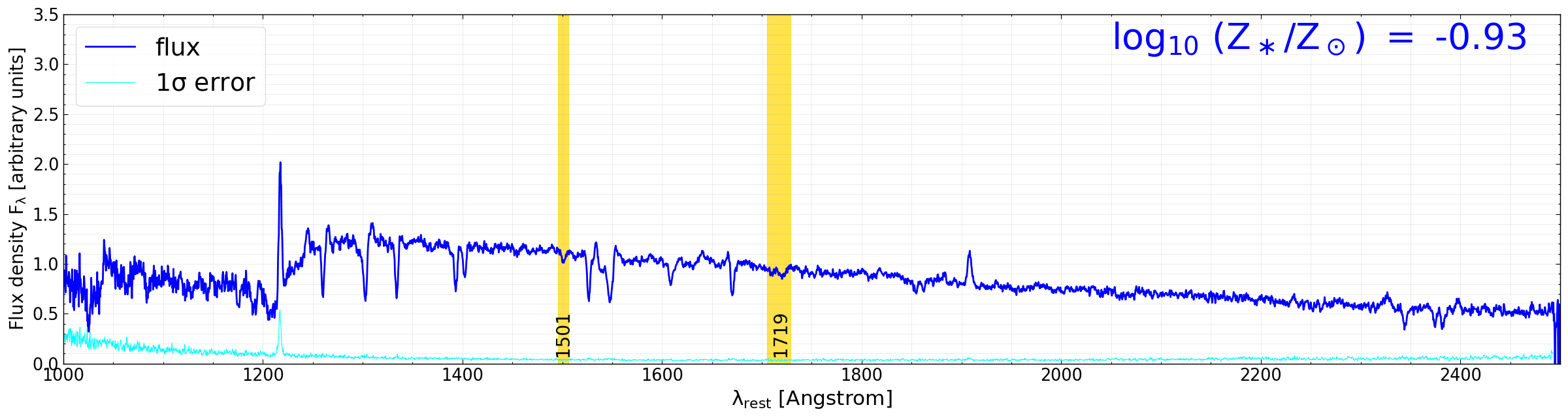}
    \vspace{-0.33cm}
    \includegraphics[angle=0,width=0.91\linewidth,trim={0.cm 0.cm 0.1cm 0cm},clip]{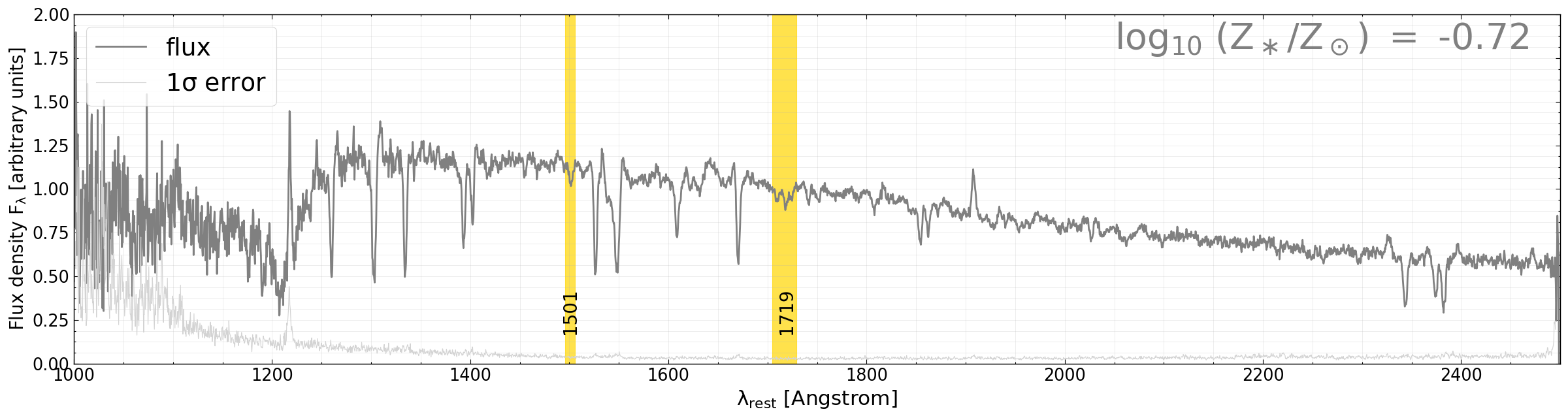}
    \vspace{-0.33cm}
    \includegraphics[angle=0,width=0.91\linewidth,trim={0.cm 0.cm 0.1cm 0cm},clip]{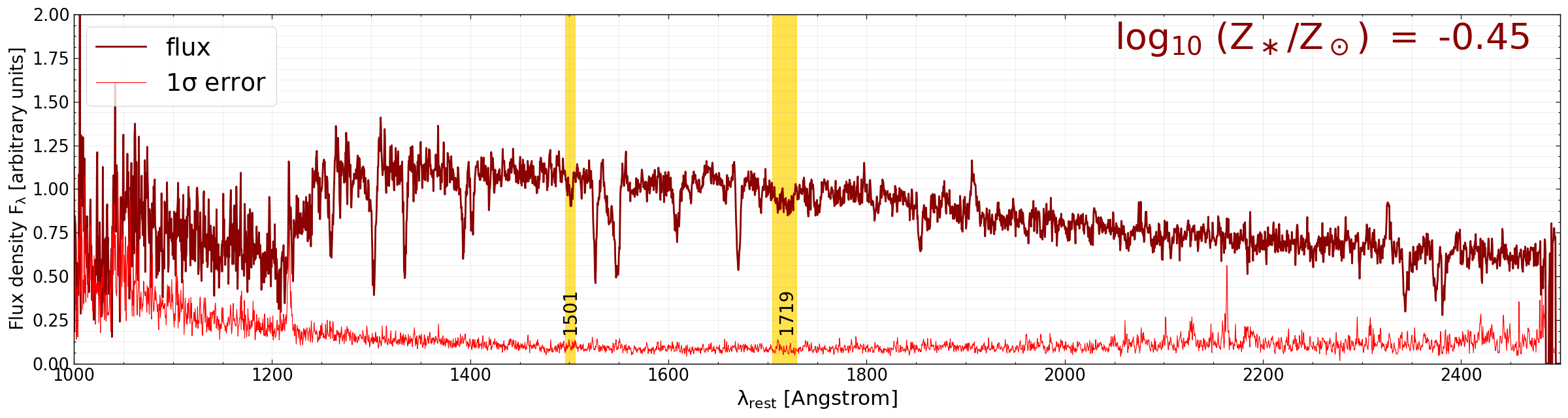}
    \vspace{-0.02cm}
    \caption{\footnotesize Stacked spectra between $1000$ and $2500$ \AA\ in three bins of increasing $\beta_\text{median}$: $-1.98$ (blue line), $-1.51$ (gray line), and $-1.10$ (red line) (see also Table \ref{table2}). The absorption complexes used in this work to estimate the metallicity, located at $1501$ and $1719$ \AA, are highlighted with yellow vertical bands.
    In the two top panels, the three stacks are plotted together and zoomed around the two metallicity indexes. A thicker line specifies both the width of each absorption complex, while grey shaded areas indicate the spectral ranges for the estimation of the pseudocontinuum.
    Spectral stacks of galaxies with redder UV slopes have deeper absorption features, hence higher metallicity. 
    }\label{fitlineprofile}
\end{figure*}

As we have shown in previous sections, both the metallicity and the UV slope increase with the stellar mass. Therefore, also Z$_\ast$ and $\beta$ should be tightly related each other. Given that $\beta$ is mostly influenced by the level of dust attenuation in the galaxy (as discussed in Section \ref{beta_mass}), this also means that the light emitted by less or more chemically enriched stellar populations is subject to different levels of dimming.
Analyzing VANDELS galaxies in bins of Ly$\alpha$ EW, \citet{cullen20} suggest an increase of Z$_\ast$ with UV slope. It is thus interesting to directly compare these two quantities, as done with the stellar metallicity and the stellar mass. From the physical point of view, the exact dependence between Z$_\ast$ and $\beta$ is influenced by several, often simultaneous phenomena, including dust formation mechanisms in metal-rich or metal-poor environments, grain growth from capture of heavy metal particles produced inside stars, destruction by SNe explosions, and ejection through AGN or stellar driven winds. 

To investigate this relation, we consider the subset $\#2$ selected in Section \ref{selection}. Because of the larger uncertainties of $\beta$ estimations compared to the stellar masses, and given the lower number of objects than those used for the MZR, we defined here three bins of galaxies, representative of a bluer, an intermediate, and a redder slope population, with median $\beta$ values of $\sim -2$, $\sim -1.5$, and $-1$, respectively (see Table \ref{table2}). In each bin we stacked all the spectra with the same procedure illustrated in Section \ref{stacks}, and we measured the metallicity from the $1501$ and $1719$ \AA\ absorption features. 
The resulting trend is shown with black squares and error bars in Fig. \ref{beta_metallicity}. As for the MZR in Fig. \ref{MZR_figure}, these represent the median $\beta$ in each bin as a function of the stellar metallicity estimated for each stacked spectrum (from the average of the two indexes), with the $\beta$ range and Z$_\ast$ uncertainty highlighted with horizontal and vertical error bars, respectively. 
We observe an increasing trend of metallicity toward redder spectra: $Z$ rises by $\sim 0.5$ dex between $\beta=-2$ and $\beta=-1.1$. Even though the metallicities from single indexes (drawn with pale red and dark-cyan smaller squares) show a larger uncertainty, they are remarkably in agreement each other within $1 \sigma$ and with the global relation, displaying a similar increase in metallicity from bluer to redder galaxy spectra. 

Since it was possible to model with a first order polynomial both the MZR (see Fig. \ref{MZR_figure}-\textit{top}) and the M$_\ast$-$\beta$ relation without redshift binning (derived from the same VANDELS data, even though with a slightly larger sample in the first case), it is interesting to combine these two best-fit lines, removing the dependency on stellar mass. This could provide a consistency check of the results obtained with different approaches. The outcome of this exercise is shown in Fig. \ref{beta_metallicity} with a cyan dashed line and cyan shaded region (representing $1\sigma$ confidence limits). We can see that it reproduces qualitatively the general trend established by our direct measurements of stellar metallicity in stacks of $\beta$ (black squares).

Despite the relatively large uncertainties of our average data points, we also fitted them with a linear relation, in order to quantitatively compare these results to the above mentioned analytical calculation, and with the predictions of semi-analytic models in the following section. This exercise yields the following equation : 
\begin{equation}\label{Zbetarelation}
\mathrm{log}_{10} (Z/Z_\odot) = (0.53 \pm 0.07) \times \beta + (0.11 \pm 0.11)
\end{equation}
We remark again that this is the simplest approximation, and we do not attempt to extrapolate or constrain more complex dependecies between Z$_\ast$ and $\beta$, especially in the bluer and redder tails. However, it is important and reassuring to find a consistency between our best-fit $Z_\ast$-$\beta$ relation in Eq. \ref{Zbetarelation} and that derived independently from two underlying trends of the stellar mass with the UV slope and the metallicity.

Finally, to highlight the variation of absorption line depth with increasing $\beta$ for each of our two metallicity indexes, we plot together the spectral portions of the stacks close to the $1501$ and $1719$ \AA\ absorption complexes. In Fig. \ref{fitlineprofile} we draw the spectra obtained for the three stacks in bins of increasing UV slope, which also correspond to increasing levels of stellar metallicity. 
We find that the stacked spectrum derived in the first bin at lower $\beta$ has a lower absorption EW in all the two above indexes. As we move to bins of higher UV slope, the depths of the absorption features increase in a visible manner. This visual inspection thus confirms the tight relation between $\beta$ and $Z$ already shown in Fig. \ref{beta_metallicity}. 
In the next section we will compare this observational trend with predictions of theoretical galaxy evolution models.


\section{Discussion}\label{discussion}

Our findings show that the stellar metallicity is tightly related not only to the stellar mass of the galaxies, but also to the UV continuum slope, which is a proxy of the dust attenuation. In the following we compare our results to theoretical predictions from semi-analytic models (SAM) of galaxy formation and evolution. 

\subsection{Comparison to semi-analytic models}\label{GAEA}

\begin{figure}[ht!]
    \centering
    \includegraphics[angle=0,width=\linewidth,trim={0cm 0cm 0.1cm 0cm},clip]{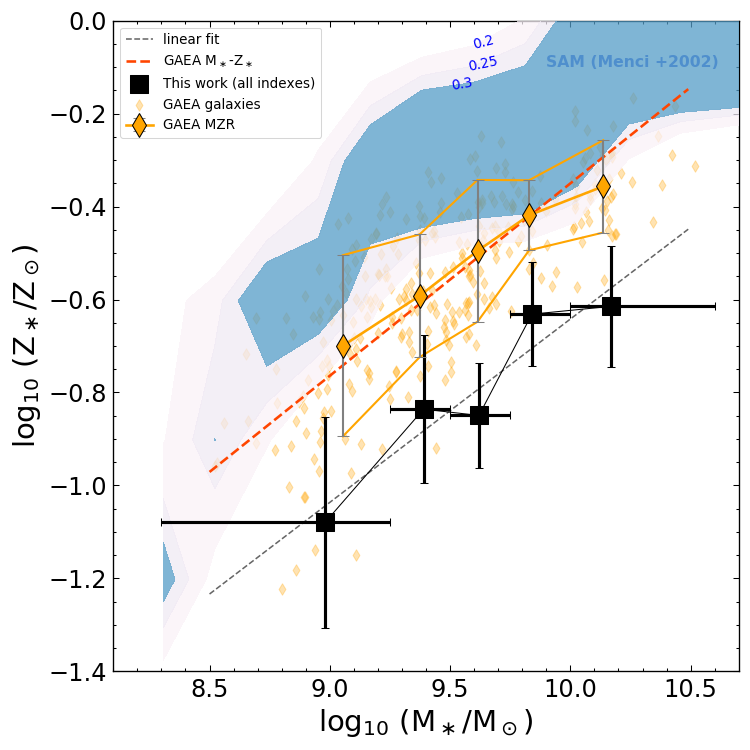}
    \includegraphics[angle=0,width=\linewidth,trim={0cm 0cm 0.1cm 0cm},clip]{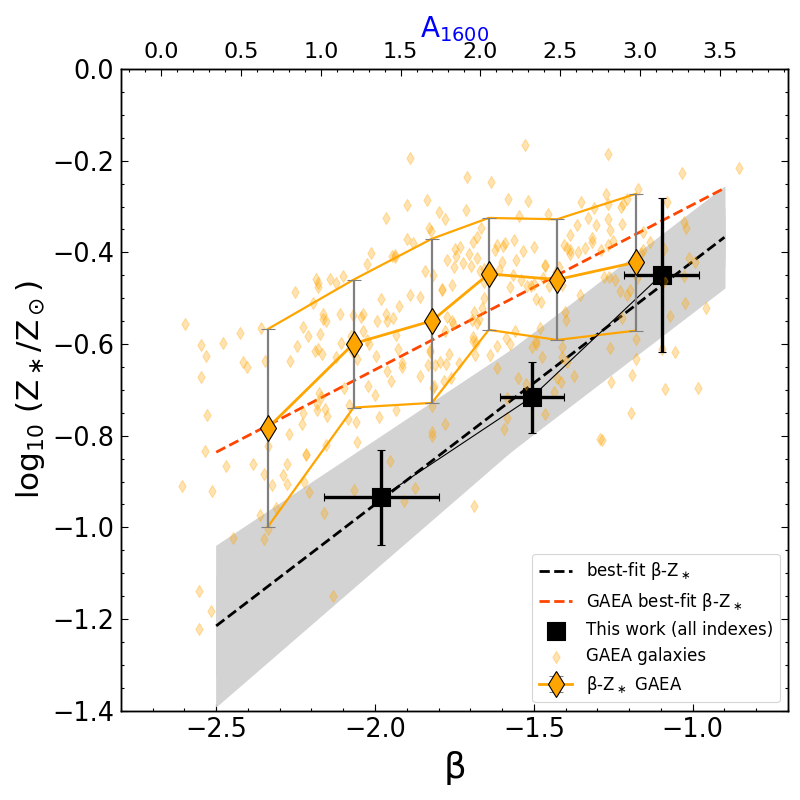}
    \caption{\small \textit{Top:} Comparison between the MZR derived from VANDELS (black squares and error bars) and from predictions of two semi-analytic models: orange diamonds represent the median masses and metallicities of GAEA galaxies in five bins of stellar mass defined as for the observations. Blue contour regions indicate the fraction of galaxies in the SAM of \citet{menci14} with different Z$_\ast$ at a given mass, according to the 2D bins defined in the text. \textit{Bottom:} Comparison of VANDELS results to the UV slope - metallicity relation predicted by the GAEA models. The orange big diamonds were derived with the same method of the upper panel. We highlight with a black dashed line the best-fit relation for the VANDELS sample, while the gray shaded area encloses its $1\sigma$ uncertainty region.}\label{comparison_GAEA}
\end{figure}

The first model that we consider has been developed by Menci and collaborators \footnote{More details are found on the following webpage: https://lbc.oa-roma.inaf.it/menci/}, and first presented in \citet{menci02,menci04}. Here we use the most updated version of the SAM, which is described in \citet{menci14}. Summarizing its important features, it creates a subset of Dark Matter (DM) merging trees following the extended \citet{press74} statistics. The code generates then the merging history of DM galactic subhalos and describes the evolution of the baryonic component following the main physical mechanisms such as gas condensation and cooling, star-formation, black-hole growth, and feedback processes, both of AGN or stellar origin. 
This allows the computation of the properties of the galaxies associated to each DM subhalo (which could eventually merge together) including, among all, their stellar mass, gas mass, and metallicity of both the gas and stellar components. The effects of dust extinction are not included. 
As far as the chemical abundance is concerned, it is calculated considering the whole star-formation history of each model galaxy, adopting a yield (i.e., the fraction of metals in stars that returns to the ISM during their lifetime) of $0.02$ and a recycled gas fraction of $0.45$, appropriate for the Chabrier IMF. The model also adopts the instantaneous recycling approximation (IRA), and it does not distinguish between SNII and SNIa chemical enrichment. 
This SAM has been shown to accurately predict the observed luminosity function of galaxies from the Local Universe to high-redshift \citep{menci02}, the quasar luminosity function up to $z\sim4$ \citep{menci06}, and the color bimodality of galaxies \citep{menci05}.

In order to have a more manageable dataset, especially for the visualization, we studied the distribution of galaxies in the M$_\ast$-Z$_\ast$ plane, using a grid with 20 bins in mass (from log$_{10}$ (M$_\ast$/M$_\odot$) $=8.3$ in steps of $0.2$) and 20 bins also in metallicity (from log$_{10}$ (Z$_\ast$) $=-3.35$ in increasing steps of $0.3$). Then, at a given mass, the fraction of galaxies residing in each bin of metallicity was calculated. 

As an alternative approach, we also consider the GAlaxy Evolution and Assembly model (GAEA) for the formation and evolution of galaxies across cosmic time. 
This model represents an evolution with respect to the earlier \citet{delucia07} code. GAEA traces the evolution of the multi-phase baryonic gas (i.e. hot gas, cold gas, stars) in the different galaxy components (i.e. disc, bulge and halo). The mass and energy exchanges between the different reservoirs are followed by solving a system of approximated differential equations, that account for the physical mechanism acting on the baryonic component, such as gas cooling, star formation, stellar and AGN feedback. In detail, the main improvements in GAEA include an improved treatment of chemical enrichment and stellar feedback. \citet{delucia14} relax the instantaneous recycling approximation of stellar ejected metals assumed in the original version. They instead consider the different lifetimes of stars with varying initial masses \citep{padovani93} and track the enrichment of single chemical elements at various stages of stellar lives. 
Moreover, \citet{hirschmann16} propose an improved feedback scheme aimed at reproducing the evolution of the galaxy stellar mass function up to $z\sim 3$. This stellar feedback scheme is inspired by the results of hydro-dynamical simulations and includes both stellar-driven winds able to efficiently eject the hot gas and a mass-dependent reincorporation mechanism for the ejected material. These new prescriptions provide an explanation for the 'anti-hierarchical' galaxy evolution scenario, with low-mass galaxies increasing in number density towards lower redshifts at a faster pace than more massive counterparts. In detail, GAEA is able to reproduce the evolution of the GSMF up to $z\sim 7$ and of the cosmic SFR up to $z\sim10$ \citep{fontanot17}, the $z\sim0$ gas-phase MZR (De Lucia et al. 2020) and its redshift evolution (Fontanot et al., in preparation). In the following we consider GAEA prediction corresponding to a realization based on the merger trees extracted from the Millennium Simulation \citet{springel05}, and corresponding to a WMAP1 cosmology (i.e., $\Omega_\Lambda$ $= 0.75$, $\Omega_\text{m} = 0.25$, $\Omega_\text{b} = 0.045$, n $= 1$, $\sigma_8 = 0.9$, H$_0 = 73$ km/s/Mpc). 

For a fair comparison with our results, we use a GAEA light cone produced inside the collaboration to mimick the VANDELS survey. This cone covers the same area of VANDELS and was generated following the procedure of \citet{zoldan17}, with a stellar mass limit of $10^{8.5}$ M$_\ast$ to match the lower limit of our galaxies. For each object identified in the cone, luminosities are calculated assuming a Chabrier IMF and \citet{bruzual03} stellar population models (see \citet{delucia14} for more details), while SFR$_{UV}$ were derived from the unobscured UV luminosity L$_{\nu,1600}$, ensuring they have similar timescales to those estimated from SED fitting. 
The models also predict the magnitudes observed in common broad photometric bands ranging from U to H. 
Effects of dust attenuation are included in the observed magnitudes, assuming the double screen model of \citet{charlot00}, with the light of younger stellar populations experiencing an additional effective absorption inside the birth clouds compared to older stars affected only by the ambient ISM attenuation. 
In the cone, we consider a filter set corresponding to those available in the framework of the VANDELS survey, therefore we could derive an estimate of the $\beta$ slopes in the range $1250$-$1750$ \AA\ using similar techniques as the real data explained in Section \ref{beta}.
Finally, for each galaxy, the stellar metallicity is computed as the mass fraction of metal elements in the stellar component, normalized to Z$_\odot=0.0142$. Before comparing to the models, we also matched the 3D distribution in redshift, stellar mass and SFR of galaxies in the GAEA light-cone to that of VANDELS objects selected in this work. 

The resulting M$_\ast$-Z$_\ast$ diagrams from the two models are presented in Fig. \ref{comparison_GAEA}-top. We can see that individual galaxies in GAEA span stellar metallicites log$_{10} (Z/Z_\odot)$ between $-1.2$ and $-0.2$. A linear fit to individual galaxies yields a best-fit slope as $0.41 \pm 0.02$ and a normalization $+0.27$ dex higher compared to VANDELS observations. 
Considering the five stellar mass bins in the $10^9$-$10^{10.2}$ M$_\odot$ range, the median metallicities in the bins increase mothonically from $-0.7$ to $-0.4$.
For the galaxies modeled with the approach of \citet{menci14}, we draw a contour plot of their distribution in the M$_\ast$-Z$_\ast$ diagram. We notice that for stellar masses ranging $10^9<M_\ast<10^{10.2}$ M$_\odot$, the stellar metallicities of the average star-forming galaxy population are also $0.2$ dex higher than GAEA predictions, with log$_{10} (Z/Z_\odot)$ varying from $-0.8$ to above solar. 
Overall, despite the different metallicity normalizations, the two relations have slopes that are remarkably in agreement with the observed MZR from VANDELS data. This suggests that models connect the natural evolution from low-mass, more metal-poor galaxies to high-mass, more chemically enriched systems in a way that is consistent with our observations.  

In Fig. \ref{comparison_GAEA}-bottom we show the comparison between VANDELS data and GAEA models for the $\beta$ - Z$_\ast$ relation. First, we notice that the range of UV slopes predicted by GAEA is consistent with the values found in our VANDELS selected sample. 
Nonetheless, over this range we confirm that the typical metallicities of GAEA galaxies are systematically higher than our estimates, although the tension is reduced to a 1-$\sigma$ level (even less toward redder slopes, corresponding to $\beta > -1.5 $). In order to understand if the models is able to reproduce the dependence of $Z_\ast$ from $\beta$, we perform a linear fit of individual mock galaxies, which provides the following best fit relation: Z$_\ast$ $=$ ($0.36 \pm 0.03$)  $\times \beta$ $+$ ($0.06 \pm 0.05$) (red dashed line in Fig. \ref{comparison_GAEA}). While this relation is slightly flatter than VANDELS data, the two slopes are consistent at $2 \sigma$ level, and they agree even more if we consider bluer galaxies with $\beta < -1.5$. Overall, we remark that the existence of a well defined relation between the UV slope and the stellar metallicity is a success for this model. 

Finally, it is interesting to ask where do these constant metallicity offsets in Fig. \ref{comparison_GAEA}-top come from. 
It is worth stressing that our derivation of metallicity in GAEA represents, by construction, a mass-weighted metallicity. 
\citet{cullen19}, using cosmological simulations at $z\sim5$, showed that mass-weighted metallicities are generally higher than FUV-weighted observed values estimated from the UV rest-frame spectra, by an amount of $0.1$-$0.2$ dex, depending on the simulations adopted. This difference is due to the fact that younger and more metal rich stellar populations are typically affected by a higher level of dust attenuation inside their birthclouds, while older (hence more metal poor) stars are less attenuated and thus contribute more to the observed FUV emission. 
Unfortunately, the correct assessment of FUV-weighted metallicities in GAEA is beyond the current capabilities of the model, partly because of the simplified assumptions for dust obscuration (a screen model), and in part for the lack of spatial resolution in the treatment of star forming discs, which does not allow a detailed treatment of individual star forming regions. However, we notice that the results obtained in the framework of hydro-simulations \citep[see, e.g., Fig.8 in ][]{cullen19} go into the right direction to reduce the tension between GAEA and VANDELS data. 
On the other hand, the discrepancy with the \citet{menci14} SAM is larger than what we can recover with FUV-weighting. In this case, we remind that additional metallicity offsets can come from the treatment of the metal yield: if we decrease the effective total yield, we would obtain lower metallicities, more consistent with the observations. However, this approach cannot be used in GAEA, as this model do not treat yields as a free parameter.

Finally, we also warn the reader that, if we use BPASS models, the calibration function for the $1719$ \AA\ index gives metallicities that are $+0.25$ dex higher, but no offsets with the Starburst99 results are found when using the $1501$ \AA\ index alone. 
While it is beyond the goals of this paper to discuss the origin of this discrepancy (which might be related to the different chemical composition and/or physics adopted by the two stellar models), it is true that applying the BPASS calibration on the $1719$ \AA\ index would make the observed relations more consistent at least with the GAEA predictions. However, we think this is unlikely, because the positive offset that we see for the $1719$ \AA\ absorption complex is not found for the other lines, as it is shown in more detail in the Appendix \ref{appendix3}. Moreover, our result remains unaltered for the $1501$ index and is consistent with the previous work of \citet{cullen19}, based on fitting the entire FUV spectrum.
To conclude, we stress again that, most importantly, the shape of the theoretical relations analyzed in this section appears to be consistent with our data.

\subsection{Future developments}\label{future}

The UV slope remains a fundamental quantity to constrain the properties of galaxies at all redshifts, and search for more extreme candidates resembling to even higher redshift systems.
For example, we find in our sample a significant population of galaxies ($\sim 33\%$) with a UV slope bluer than $\sim -2$. From Fig. \ref{mass-beta}, these systems also have preferentially a lower stellar mass ($\lesssim 10^{9.5}$ M$_\odot$) and a small dust attenuation (A$_{1600} \lesssim 1.5$) according to the standard assumptions of the \citet{meurer99} calibration. 
As claimed in other works \citep[e.g.,][]{steidel99,ouchi04,bouwens06,bouwens09,hathi08,erb10,bouwens16b,shivaei18}, a very blue UV slope indicates the presence of very young, metal-poor stellar populations, and it is suggestive of a higher ionizing photons production efficiency and escape of ionizing radiation from such galaxies, which are the typical conditions in the reionization epoch. 

We have also seen how the measurement of stellar metallicity remains not easy, requiring simultaneously a high sensitivity (or many hours of integration) to enhance the SNR, and possibly a high spectral resolution. In the Appendix \ref{appendix-resolution}, we will discuss the effect of the resolution on the metallicity calibration functions and on the reliability with which Z$_\ast$ can be constrained. 
On the other hand, Fig. \ref{beta_metallicity} suggests that the UV slope can be used to derive a first, approximate estimation of the metal abundance in stars and pre-select extremely metal poor galaxy candidates with Z$_\ast$ around $1/10$ Z$_\odot$ or below, allowing to save observational time and to better focus on specific targets.
In the near future, new spectrographs like NIRSpec on board of JWST or HIRES mounted at the ELT, thanks to their higher spectral resolution compared to VANDELS ($R>1000$), will allow focused follow-ups of hundreds of galaxies to constrain their stellar metallicity from faint photospheric absorption lines, also at much higher redshifts and sensitivity (in case of ELT) than those reached in this study. 
Thanks also to the improved imaging sensitivity in the infrared compared to current instruments, we can aim at measuring the UV slope of the faintest systems at redshifts $> 6$.
An important goal that remains is then to observe and characterize statistical samples of pristine (supposedly PopIII stars dominated) galaxies, in order to constrain our Z$_\ast$-$\beta$ relation at even lower metallicities than our study and shed light on galaxy evolution in the earliest phases of our Universe. 

\section{Summary and conclusions}\label{conclusions}

We have selected a representative sample of star-forming Main Sequence galaxies in the redshift range $2<z<5$ and stellar mass $8.3$ $<$ M$_\ast$/M$_\odot$ $<$ $10.6$ with FUV rest-frame spectra obtained by the VANDELS survey. Measuring stellar metallicities from two UV rest-frame absorption lines, and estimating stellar masses and UV spectral slopes from photometric data, we study the stellar mass - stellar metallicity diagram, and the metallicity dependence of the stellar mass - $\beta$ and UV magnitude - $\beta$ relations. This sheds light on the role of metallicity on scaling relations that are typically adopted to infer dust corrections for optical and near-IR detected galaxies at high-$z$.
We summarize our main findings in the following :
\begin{itemize}
\item Using Starburst99 synthetic spectra adapted to VANDELS resolution, we calibrated the stellar metallicity with two absorption complexes located at $1501$ and $1719$ \AA\ rest-frame, which are largely independent on the IMF, age, dust content, and nebular continuum emission (Section \ref{calibrations}). The EWs of these two indexes, measured in VANDELS stacked spectra, are well in agreement with Starburst99 model predictions, yielding fully consistent metallicity values in all the range explored with our data. 
\item The mass-metallicity relation (MZR) estimated from the above far-UV absorption indexes is consistent with the previous determination based on VANDELS data \citep{cullen19}, obtained through a global fitting to the full FUV spectrum with similar stellar models (Fig. \ref{MZR_figure}).  
\item The MZR does not significantly evolve from $z\sim2$ to $z\sim3.5$, in agreement with \citet{cullen19}: even though a slightly lower metallicity (by $0.05$ dex) is measured for typical star-forming galaxies at redshift $z\sim3.5$ compared to samples at lower $z_\text{median} \sim 2.5$ (from our VANDELS sample) and $\sim 2$ \citep[from ][]{halliday08}, the difference is lower than the typical uncertainties currently associated to stellar metallicity estimates. 
\item The stellar metallicity partly explains the scatter of the M$_\text{UV}$-$\beta$ and M$_\ast$-$\beta$ relations, hence Z$_\ast$ could discriminate between different dust attenuation correction levels (Fig. \ref{metallicity_coding}). 
Using a subset at $2<z<3$ with higher SNR, we find a difference in metallicity of $0.2$ dex on average between galaxies redder and bluer than the best-fit M$_\text{UV}$-$\beta$ and M$_\ast$-$\beta$ relations, even though more massive and evolved systems tend to have more uniform metallicity values. 
\item The remarkable correlation between Z$_\ast$ and $\beta$ seen in Fig. \ref{beta_metallicity} suggests that the UV slope can be used as a proxy for the stellar metallicity. This would be of particular importance to constrain properties of galaxies at redshifts $2 < z < 5$, and search for candidates with more extreme stellar populations in terms of their metal content, more resembling to those typical of the reionization epoch.
\item We contrast our findings with the predictions of two different SAMs, namely GAEA \citep{hirschmann16} and \citet{menci14}. We first construct two mock galaxy catalogs with a redshift, stellar mass and SFR distributions matched to our VANDELS data. We show that these models predict a slope of the mass-metallicity relation that is in agreement with that found from VANDELS star-forming galaxies, even though their metallicity normalizations are higher by $\sim 0.2$ and $\sim 0.4$ dex, respectively. In addition, the slope of the Z$_\ast$-$\beta$ relation predicted by GAEA is consistent at $2 \sigma$ level with our result.
An offset of $\sim0.1$-$0.2$ dex between mass-weighted and FUV-weighted stellar metallicities, or a lower yield in \citet{menci14} SAM, could account for the different normalization between our data and model predictions.
\end{itemize}

\noindent
Several instruments coming with future telescopes like JWST and ELT, thanks to the improved sensitivity, spectral coverage and spectral resolution, will enable to push the investigation of metal and dust enrichment down to the earliest epochs of stellar and galaxy formation for a large number of systems, and assess the physical properties of the first systems born in the Universe. Similarly, also Euclid will allow to measure the UV slope and the dust properties of young galaxies in large volume surveys, with significantly higher statistics than today.
In this context, the UV slope remains a powerful diagnostic tool to distinguish between galaxies of different metal content and, possibly, evolutionary stages. 

\begin{acknowledgements}
AC and MT acknowledge the support from grant PRIN MIUR 2017 20173ML3WW$\_$001. RA acknowledges support from ANID FONDECYT Regular 1202007.
\end{acknowledgements}

\clearpage
\appendix

\section{Metallicity calibrations}\label{appendix}

\subsection{Derivation of the UV slope from VANDELS spectra}\label{appendix0}

The VANDELS spectra used in this paper offer an opportunity to compare our photometric-based UV slopes to those derived from the entire spectrum fitting. Following the original definition of $\beta$, we calculated the median flux in the $20$ rest-frame spectral windows introduced by \citet{calzetti94} as free of absorption or emission lines in the range $1250$-$2750$ \AA. We then fitted these points with a power slope function, as done with the photometric data. 
We found that the spectroscopic measurements tend to give redder slopes compared to photometry-based determinations, and this is due primarily to the small range of rest-frame wavelengths spanned by our spectra, which in most cases do not probe $\lambda > 2000$ \AA. Limiting the comparison to galaxies with most reliable $\beta$ ($\sigma_\beta <0.5$) at redshifts $\leq2.6$, thus probing at least up to $\lambda_{rest}\sim2600$ \AA, we recover a better agreement.
However, we note that the shape correction in the bluest part of VANDELS spectra, explained in Section \ref{observations}, was determined in a statistical way. Since object by object variations are still possible, it is not desirable to completely rely on the spectra for the derivation of UV slopes. 
In addition, the continuum in some spectral windows in VANDELS spectra is not always detected, especially in the redder part, reducing the available range for proper measurements of colors.
Because of these limitations, we preferred to adopt the UV-slopes obtained from photometric data. This also enables a comparison to many other works at similar redshifts that are based, because of their simplicity, on photometric datasets.

\subsection{Dependence of the metallicity uncertainty on the spectrum SNR}\label{appendix1}

\begin{figure}[t!]
    \centering
    \includegraphics[angle=0,width=\linewidth,trim={0cm 0cm 0cm 0cm},clip]{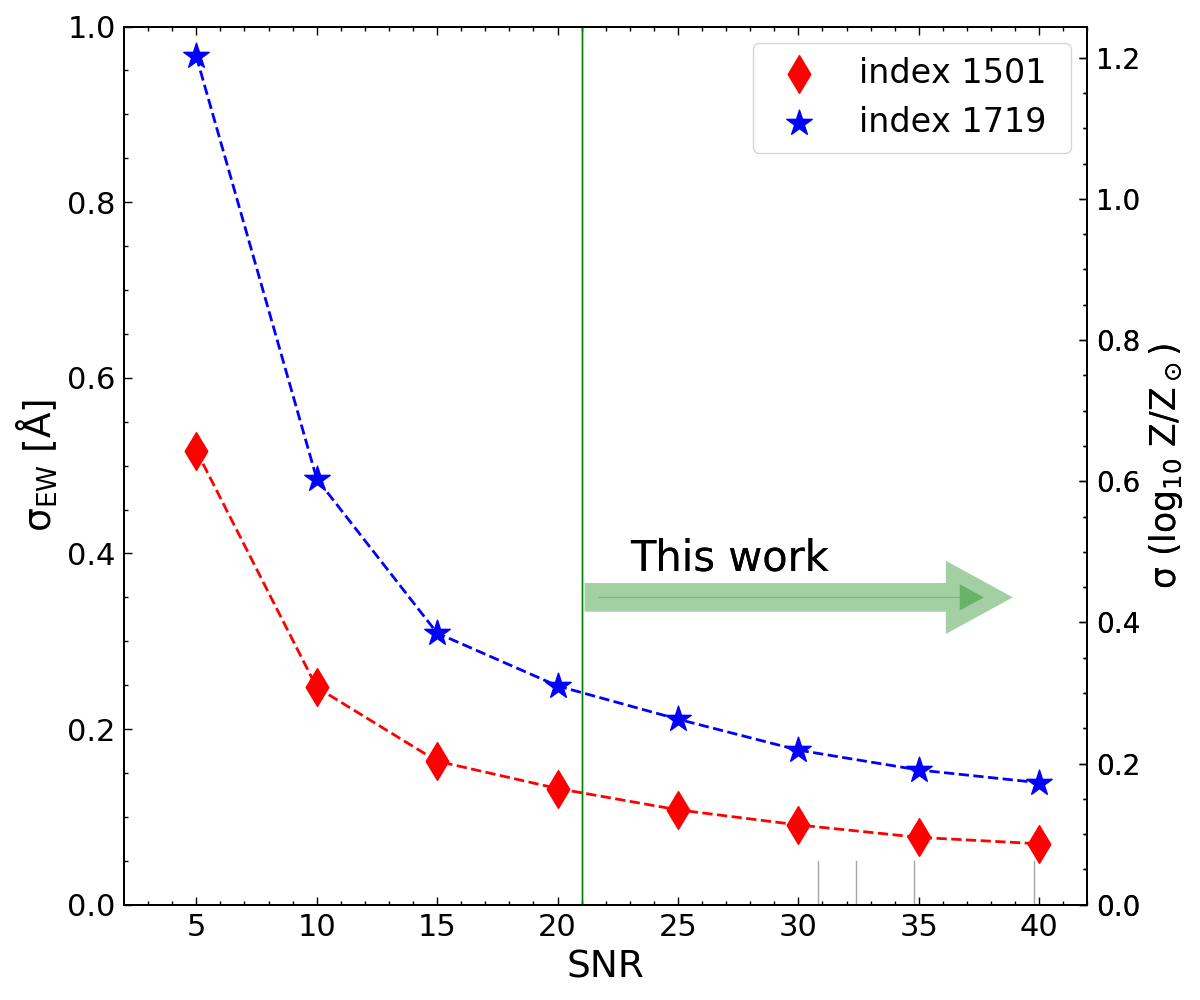}   
    \caption{\small Relation between the uncertainty of the EW (or stellar metallicity on the left y-axis) and the SNR of the input spectrum. This is obtained through Monte Carlo simulations where gaussian noise is added to Starburst99 templates with different metallicities. In this figure, different markers and colors refer to different metallicity indexes used in this work. For this example we use the template with a metallicity of Z$_\ast = 0.2 \times$ Z$_\odot$, close to the value of our selected VANDELS galaxies. The green vertical line highlights the range of SNR of the stacks that we analyze in this paper, while the smaller gray vertical bars indicate the SNR reached by the stacks in stellar mass from which the mass-metallicity relation is derived in Section \ref{MZR_results}.}\label{Zerror_SNR}
\end{figure}

The signal-to-noise of the spectra strongly affects the uncertainty that we get on the equivalent width of the absorption lines, hence on the accuracy of the metallicity estimation. The diagram in Fig. \ref{Zerror_SNR} was derived from $500$ simulations, adding to the theoretical Starburst99 templates an increasing random gaussian noise, and then measuring the equivalent-width uncertainty as the standard deviation of all these different realizations. As a result, for all the two metallicity indexes adopted in this work, $\sigma_{EW}$ follows an exponential decline as we increase the SNR of the stack from $5$ to $40$. In the same range, the uncertainty on the metallicity also decreases. Remarkably, these trends do not depend significantly on the metallicity of the model, while we can have different normalizations according to the index considered. The results of this paper are derived from spectral stacks with a minimum SNR of $\sim21$, therefore ensuring an uncertainty on log$_{10} (Z/Z_\odot)$ lower than $\sim 0.4$ from single indicators taken separately. For the mass-metallicity relation presented in Section \ref{MZR_results}, the minimum SNR is $\sim32$, which yields $\sigma_{log_{10} (Z/Z_\odot)} \lesssim 0.2$ in all the two cases.

\subsection{Metallicity calibrations for other indexes and with the BPASS model}\label{appendix3}


\begin{figure*}[t!]
    \centering
    \vspace{-0.15cm}
    \includegraphics[angle=0,width=0.44\linewidth,trim={0cm 0cm 0cm 0cm},clip]{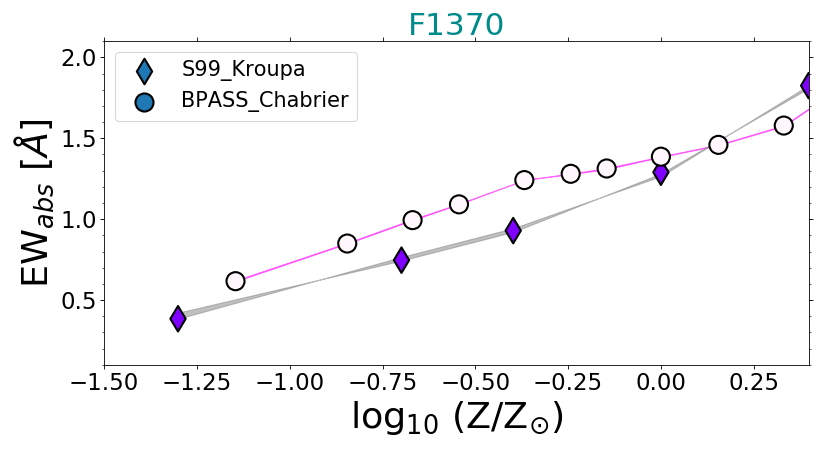}
    \includegraphics[angle=0,width=0.44\linewidth,trim={0cm 0cm 0cm 0cm},clip]{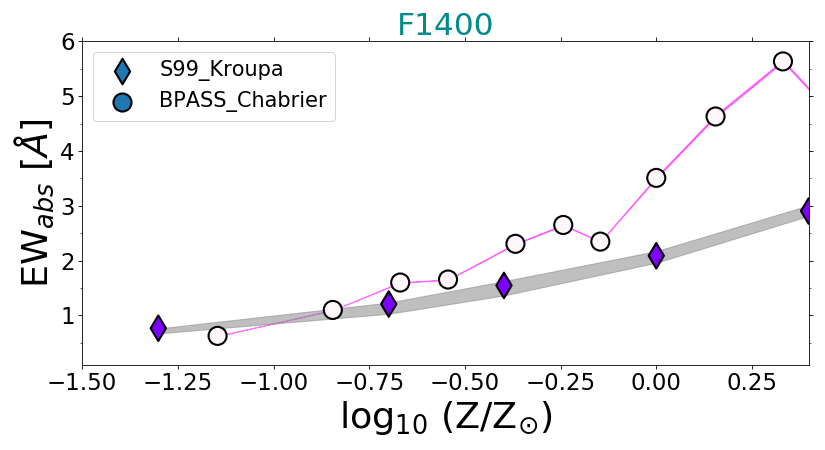}

    \vspace{-0.15cm}
    \includegraphics[angle=0,width=0.44\linewidth,trim={0cm 0cm 0cm 0cm},clip]{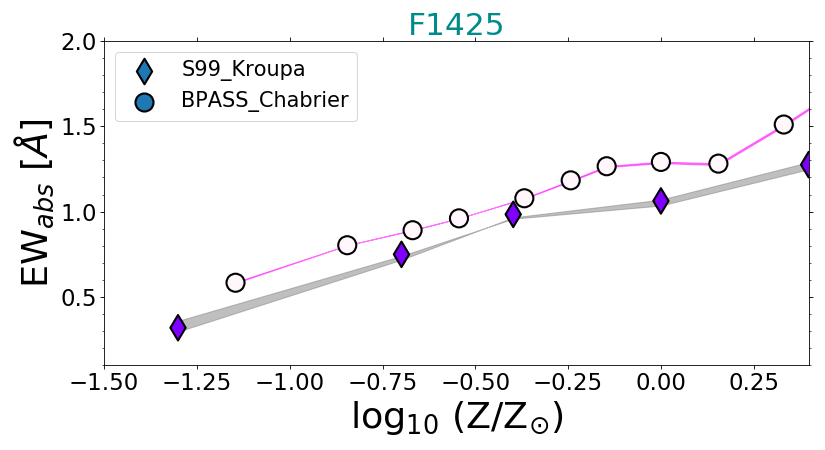}
    \includegraphics[angle=0,width=0.44\linewidth,trim={0cm 0cm 0cm 0cm},clip]{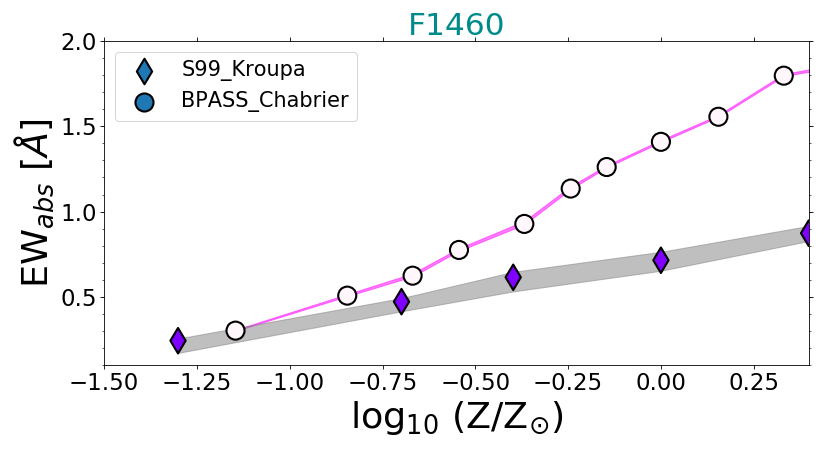}
    
    \vspace{-0.15cm}
    \includegraphics[angle=0,width=0.44\linewidth,trim={0cm 0cm 0cm 0cm},clip]{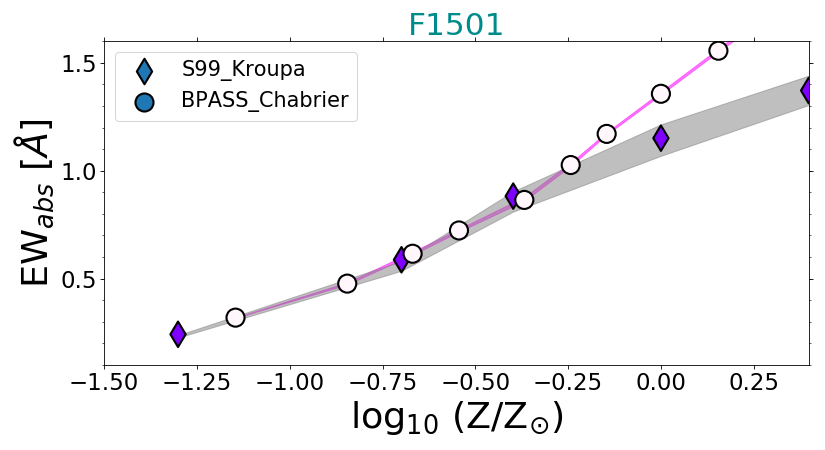}
    \includegraphics[angle=0,width=0.44\linewidth,trim={0cm 0cm 0cm 0cm},clip]{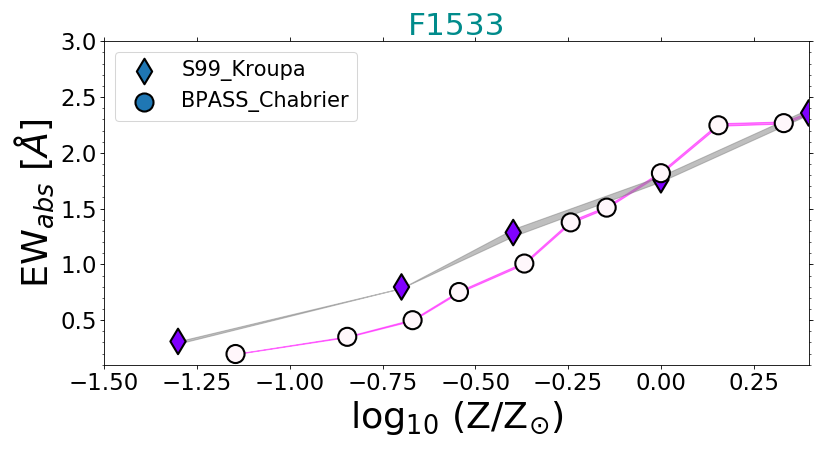}
    
    \vspace{-0.15cm}
    \includegraphics[angle=0,width=0.44\linewidth,trim={0cm 0cm 0cm 0cm},clip]{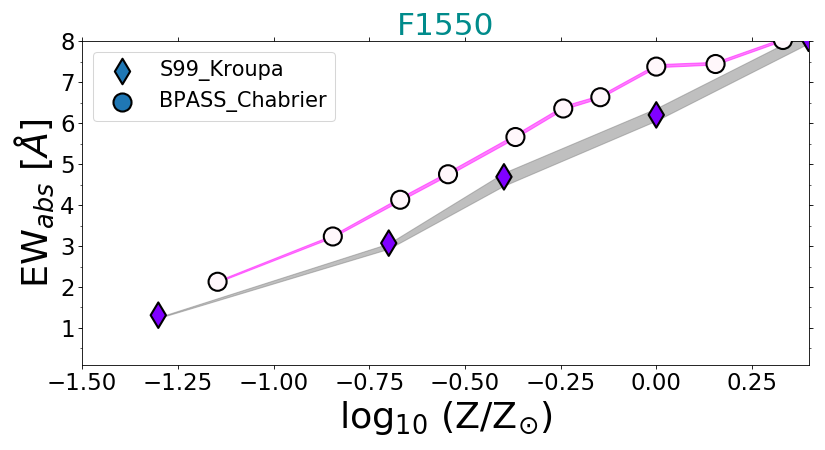}
    \includegraphics[angle=0,width=0.44\linewidth,trim={0cm 0cm 0cm 0cm},clip]{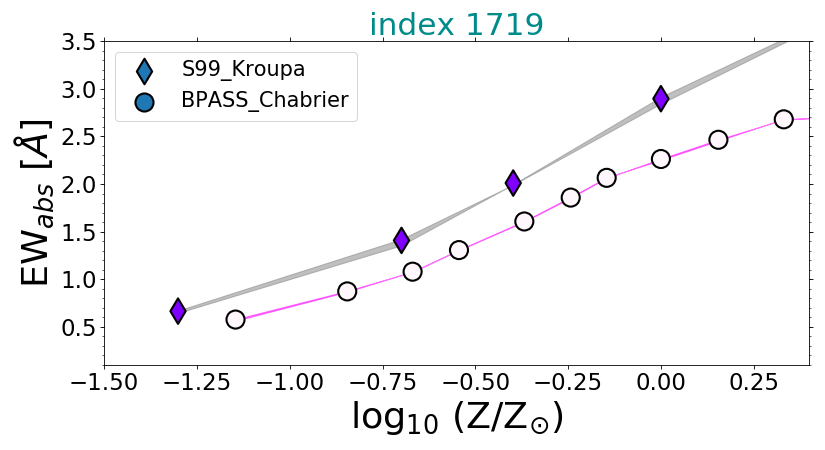}
    
    \vspace{-0.15cm}
    \includegraphics[angle=0,width=0.44\linewidth,trim={0cm 0cm 0cm 0cm},clip]{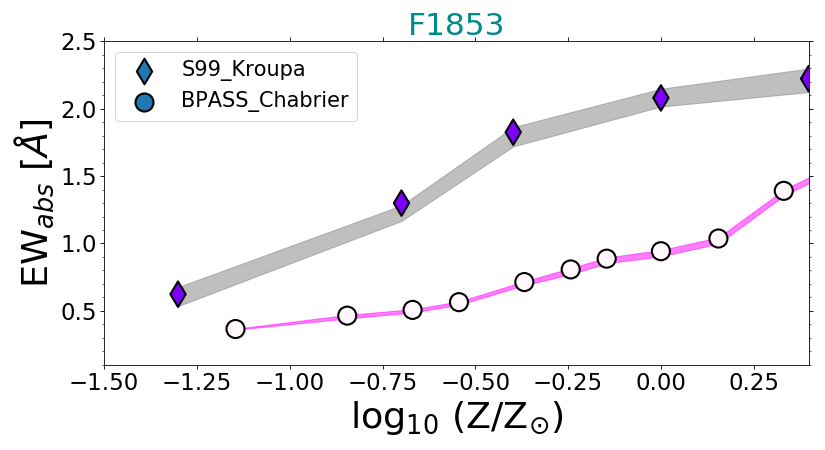}
    \includegraphics[angle=0,width=0.44\linewidth,trim={0cm 0cm 0cm 0cm},clip]{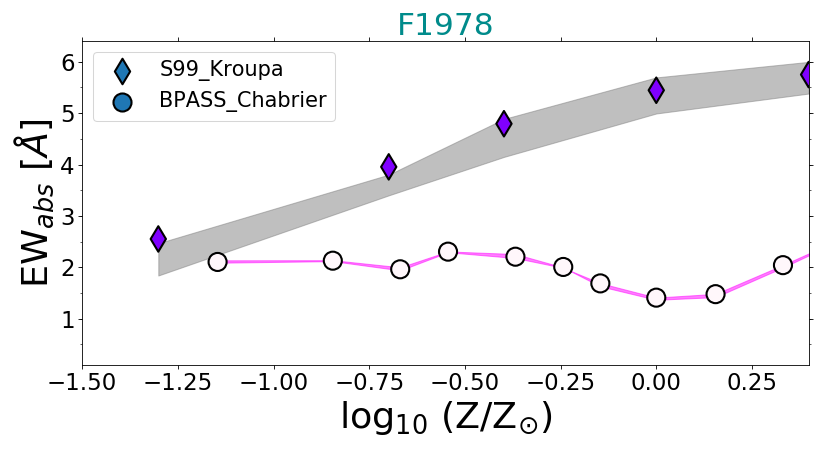}
    
    \vspace{-0.15cm}   
    \caption{\small Comparison between the input stellar metallicity and the EW predicted by Starburts99 and BPASS synthetic spectra (violet diamonds and white filled circles, respectively), for $9$ absorption line indexes typically adopted in the literature to calibrate the metallicity. We considered a continuous SFH for an age varying between $50$ Myr and $2$ Gyr (shaded region). The markers in each panel refer instead to an age of $100$ Myr, which is taken for the final calibration of the $1501$ and $1719$ indexes (see Fig. \ref {calibration_figure1}).
    }\label{other_models1}
\end{figure*}

\begin{figure*}[t!]
    \centering
    \includegraphics[angle=0,width=0.48\linewidth,trim={0cm 0cm 0cm 0cm},clip]{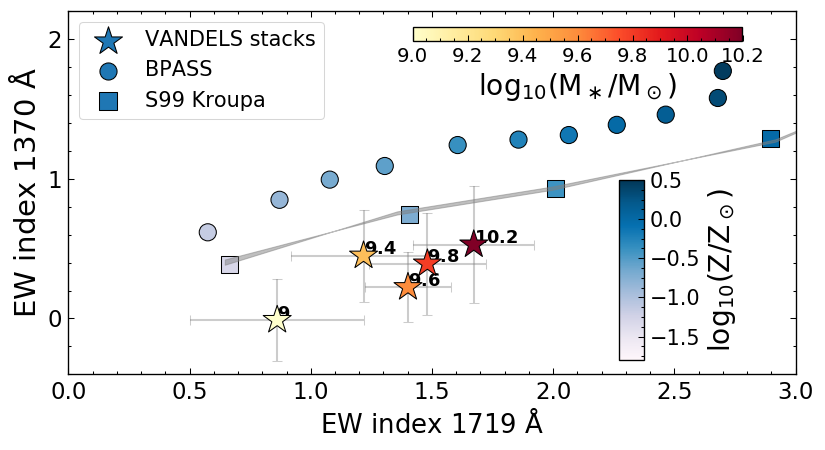}
    \includegraphics[angle=0,width=0.48\linewidth,trim={0cm 0cm 0cm 0cm},clip]{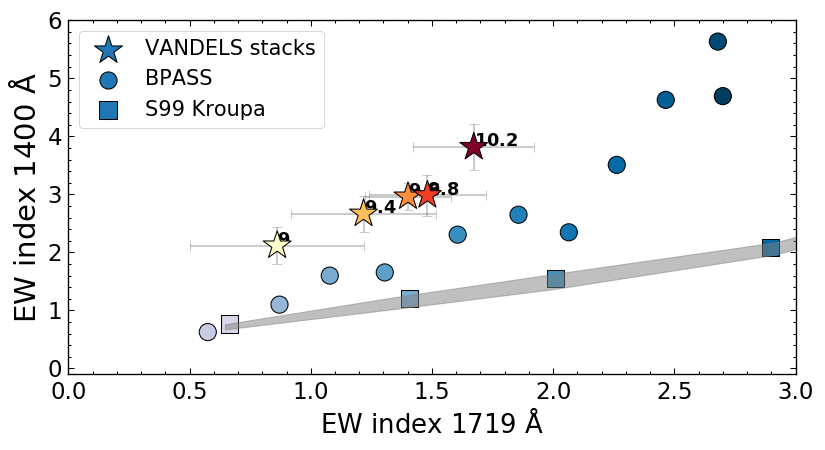}
    \includegraphics[angle=0,width=0.48\linewidth,trim={0cm 0cm 0cm 0cm},clip]{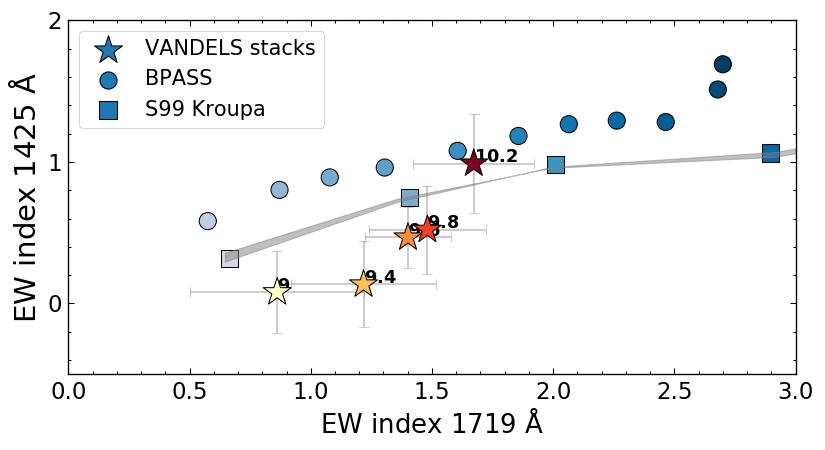}
    \includegraphics[angle=0,width=0.48\linewidth,trim={0cm 0cm 0cm 0cm},clip]{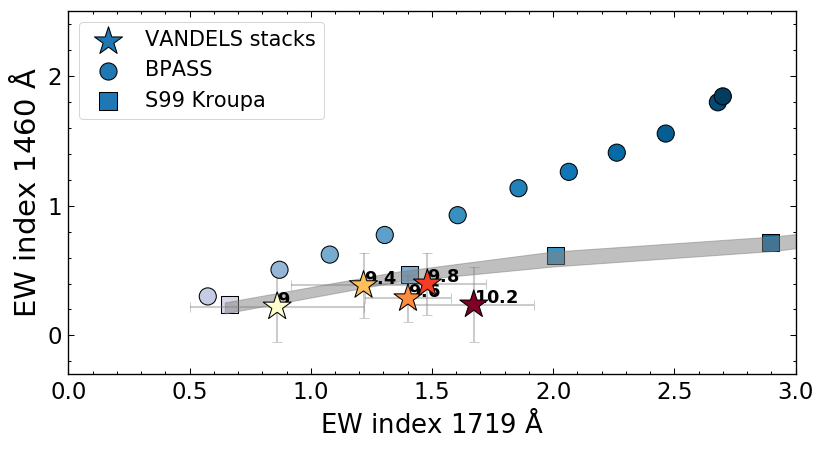}
    \includegraphics[angle=0,width=0.48\linewidth,trim={0cm 0cm 0cm 0cm},clip]{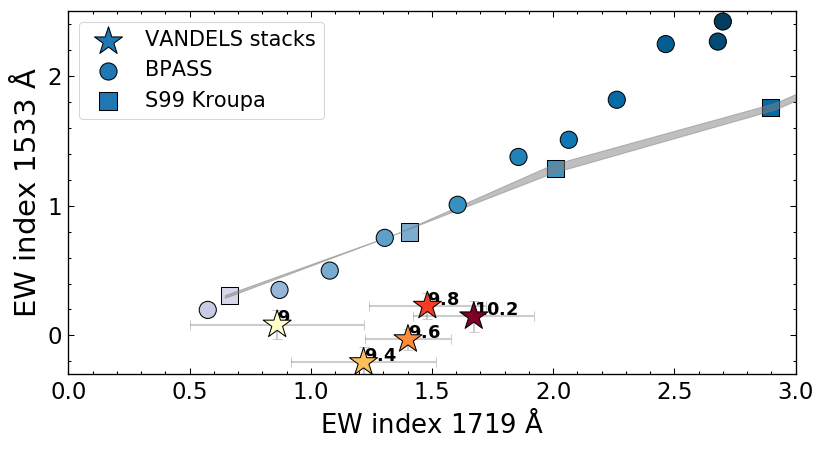}
    \includegraphics[angle=0,width=0.48\linewidth,trim={0cm 0cm 0cm 0cm},clip]{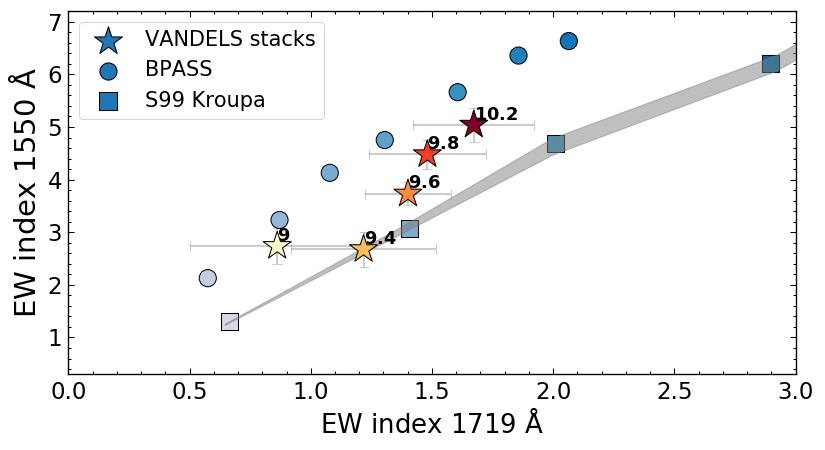}
    \includegraphics[angle=0,width=0.48\linewidth,trim={0cm 0cm 0cm 0cm},clip]{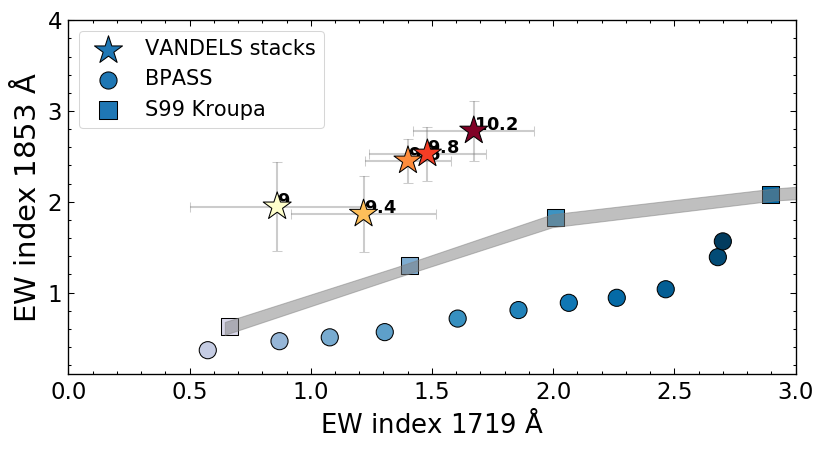}
    \includegraphics[angle=0,width=0.48\linewidth,trim={0cm 0cm 0cm 0cm},clip]{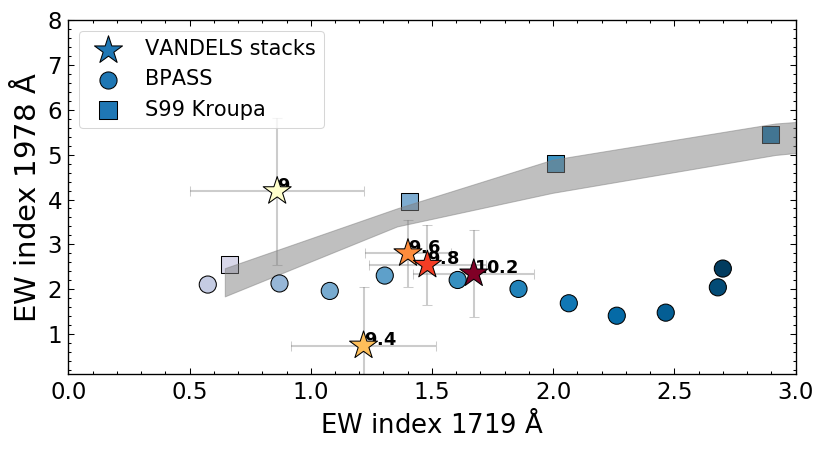}
   
    \caption{\small The six panels compare the EW of the $1719$ metallicity indicator to the EW of other absorption indexes in the FUV rest-frame that were not considered in this work, either because no correlation was found with Z, or because the observational data are not in agreement with the predictions of any models (as in the case of significant ISM contamination). In each panel we show the EW measurements from the five VANDELS stacks in stellar mass bins with colored stars (the median stellar mass of each stack is annotated in black). In addition, we superimpose EW-EW relations predicted by Starburst99 models (square symbols, with shaded gray area indicating different stellar ages from $50$ Myr to $2$ Gyr), and BPASS models (big circles).}\label{other_calibrations}
\end{figure*}

In addition to Starburst99, we also tested other models among those most used in the literature. The Binary Population and Spectral Synthesis code (BPASS, \citet{eldridge17}) has established as one of the most powerful to predict the emission of simple stellar populations and galaxies from UV to mid-infrared wavelengths. The main advantage of these models is that they include also binary stars evolution, while the important limitation is related to the lower native spectral resolution ($1$ \AA) compared to Starburst99. 

In order to test the metallicity calibrations, we ran a series of BPASS models with both single stars and binary stars, upper stellar mass limit of $100$ and $300$ M$_\odot$, Chabrier IMF and constant SFH from $50$ Myr to $2$ Gyr, as done for Starburst99 templates. The stellar metallicity was varied among all possible values allowed by the models: $0.001$, $0.002$, $0.003$, $0.004$, $0.006$, $0.008$, $0.01$, $0.02$, $0.03$, and $0.04$, that we rescaled to a solar metallicity of $0.0142$ to be consistent with the assumption throughout this paper. We restricted the output spectra to the FUV rest-frame range between $1000$ and $2500$ \AA\, and smoothed them to reach the same spectral resolution of VANDELS. Then, we measured the equivalent widths of the same absorption indexes introduced in Table \ref{table_features}, which are commonly used as metallicity tracers in the literature. 
In Fig. \ref{other_models1}, we show the relation between the model metallicity (both Starburst99 and BPASS) and the equivalent width of the following absorption complexes located at $1370$, $1400$, $1425$, $1460$, $1501$, $1533$, $1550$, $1719$, $1853$, and $1978$ \AA. 
In Fig. \ref{other_calibrations}, we compare instead the EWs estimated for different indexes to the $1719$ metallicity tracer, and we overplot the predictions of the two models we are analyzing in this appendix.

First we focus on the indexes adopted in this paper to estimate the stellar metallicity.
It is important to recognize that the metallicity calibrations for the index at $1501$ \AA\ are tightly in agreement between Starburst99 and BPASS models (Fig. \ref{other_models1}). 
On the other hand, for the $1719$ index, they give slightly different predictions, with metallicities differing by $\sim 0.25$ dex at fixed EW. 

\begin{figure*}[t!]
    \centering
    \includegraphics[angle=0,width=0.24\linewidth,trim={0cm 0cm 0cm 0cm},clip]{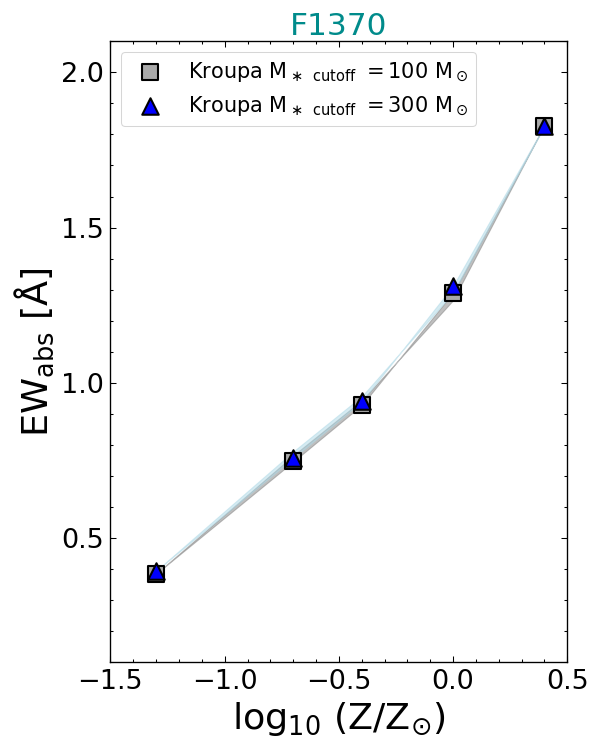}  
    \includegraphics[angle=0,width=0.24\linewidth,trim={0cm 0cm 0cm 0cm},clip]{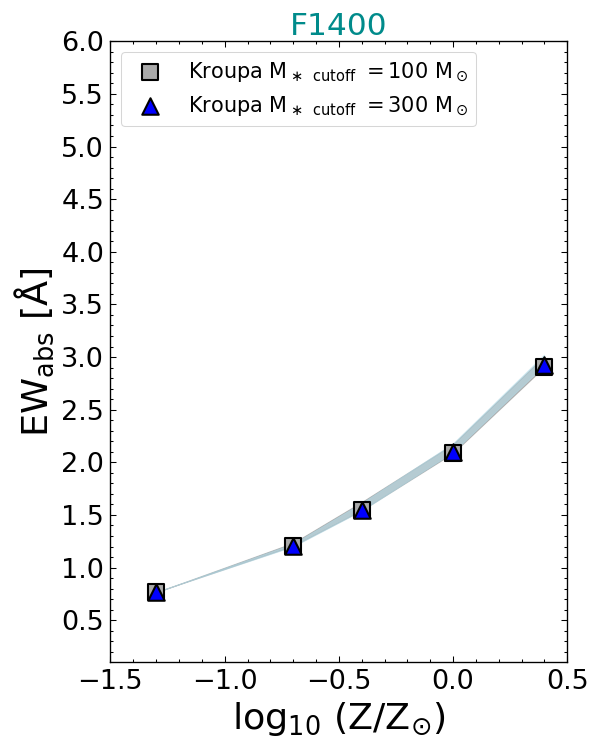} 
    \includegraphics[angle=0,width=0.24\linewidth,trim={0cm 0cm 0cm 0cm},clip]{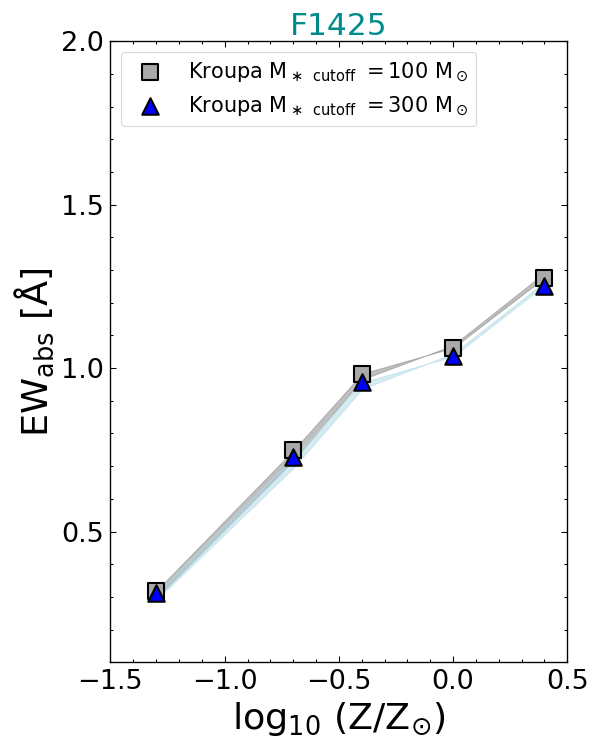} 
    \includegraphics[angle=0,width=0.24\linewidth,trim={0cm 0cm 0cm 0cm},clip]{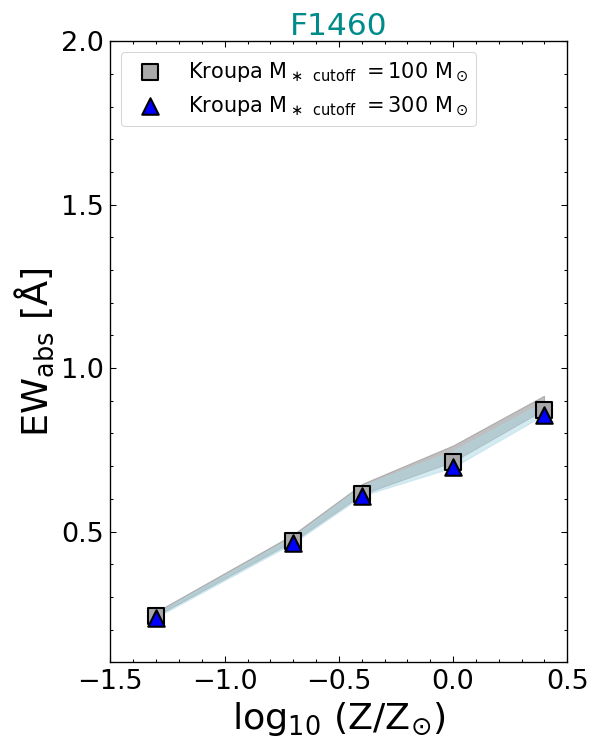} 
    \includegraphics[angle=0,width=0.24\linewidth,trim={0cm 0cm 0cm 0cm},clip]{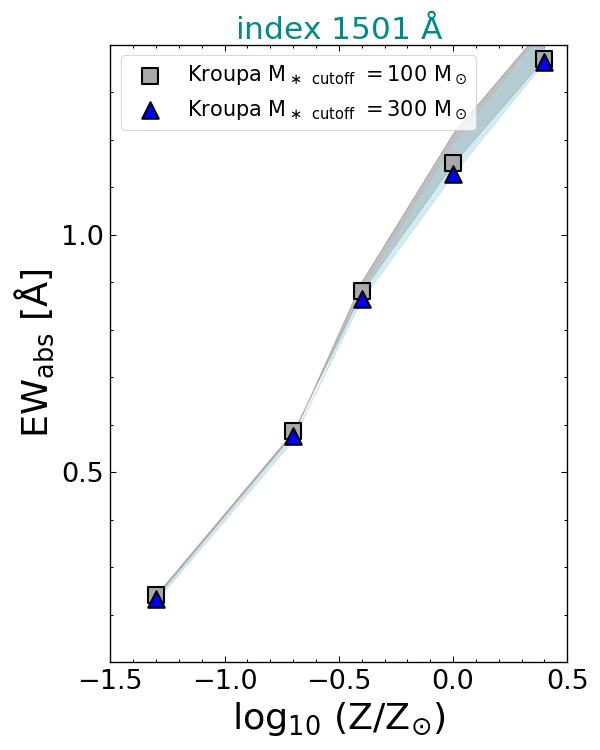} 
    \includegraphics[angle=0,width=0.24\linewidth,trim={0cm 0cm 0cm 0cm},clip]{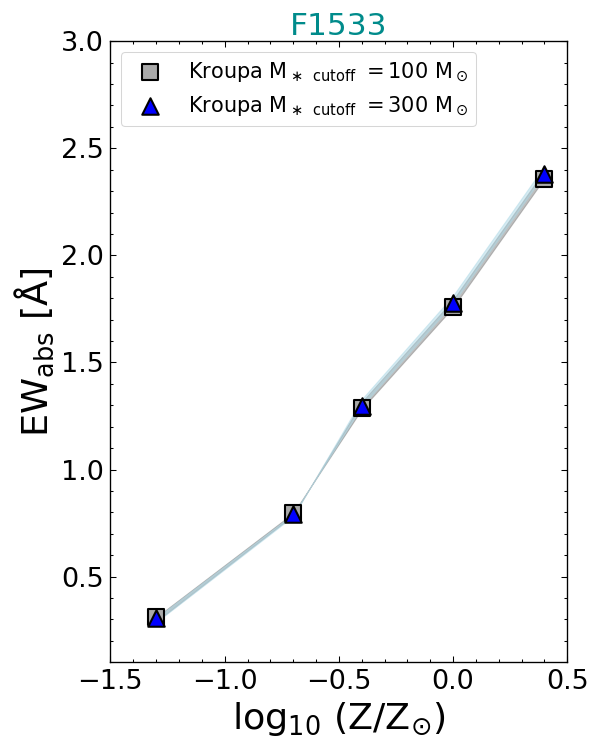} 
    \includegraphics[angle=0,width=0.24\linewidth,trim={0cm 0cm 0cm 0cm},clip]{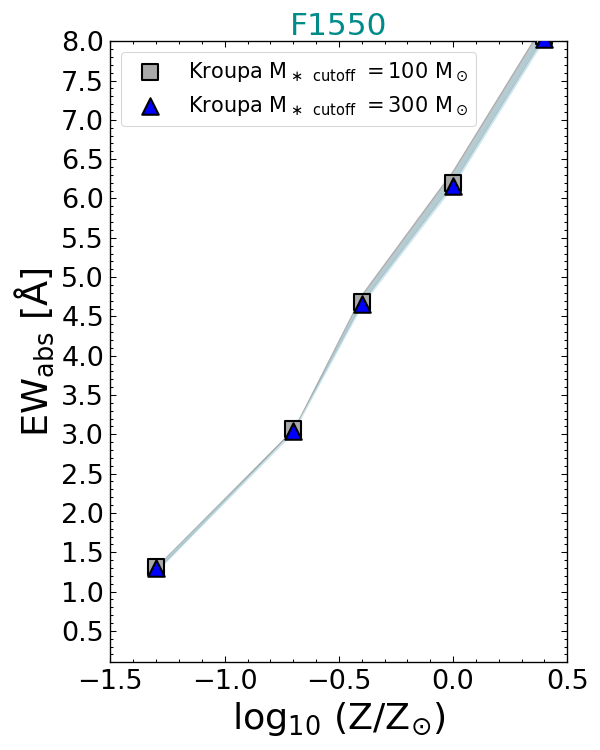} 
    \includegraphics[angle=0,width=0.24\linewidth,trim={0cm 0cm 0cm 0cm},clip]{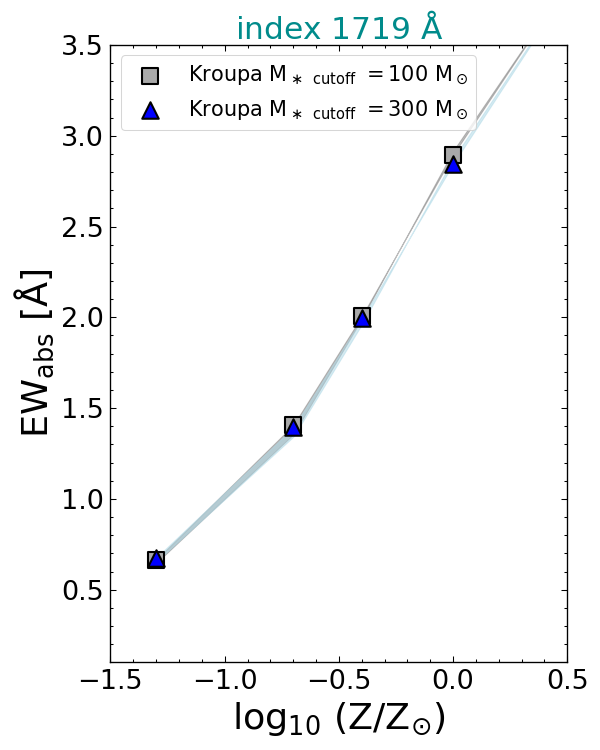} 
    \includegraphics[angle=0,width=0.24\linewidth,trim={0cm 0cm 0cm 0cm},clip]{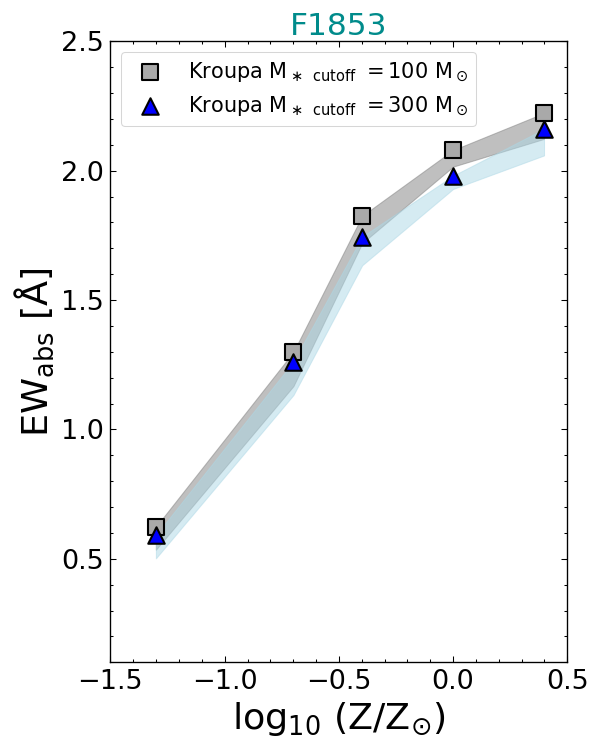} 
    \includegraphics[angle=0,width=0.24\linewidth,trim={0cm 0cm 0cm 0cm},clip]{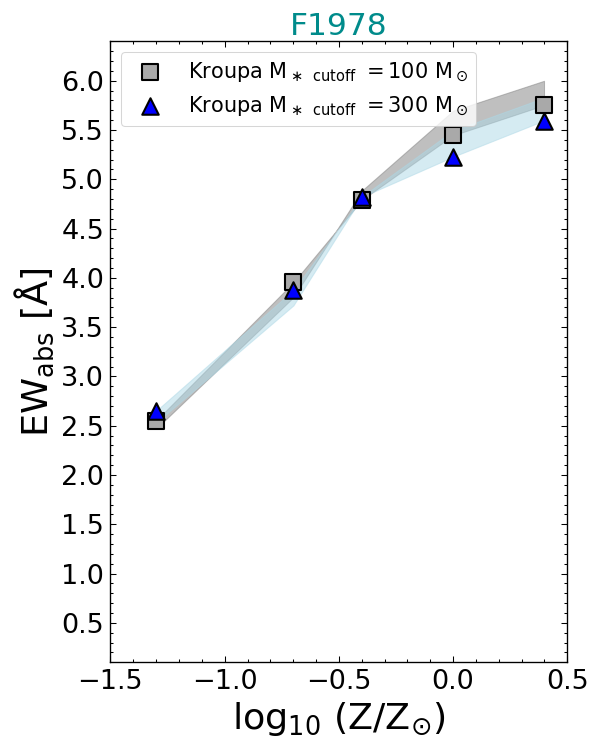} 
    \caption{\small Stellar metallicity vs equivalent width predicted by Starburts99 models for the ten absorption lines studied in this paper. The gray squares are derived assuming a standard upper-mass cutoff of $100$ M$_\odot$, while the blue triangles are calculated assuming a cutoff mass of $300$ M$_\odot$. All the symbols are representative of a constant SFH and a stellar age of $100$ Myr. The shaded regions allow the stellar age to vary between $50$ and $100$ Myr. The other details of the models are the same as in Fig. \ref{other_models1}. }\label{IMF_variations_test1}
\end{figure*}

The index at $1425$ \AA\, already adopted in \citet{sommariva12}, spans a small dynamic range according to Starburst99 models, varying from $0$ to $1$ \AA\ in absorption in the entire metallicity range possible. In contrast with the $1501$ index, the stellar models (converted to the VANDELS resolution) predict here a flattening of the EW at Z/Z$_\odot$ $> 0.5$, so that it is nearly impossible to constrain the metallicity above this value. Observationally, the EWs measured in the five bins are systematically below the model predictions by more than $1\sigma$, unless the last bin at higher stellar masses, for which the EW from $1719$ and $1425$ are in agreement with Starburst99. 

The $1370$ index, as the previous indicator, is also very faint and does not allow an accurate derivation of log$_{10}$ (Z/Z$_\odot$). The observations show an EW systematically lower compared to model predictions at the same EW$_{1719}$, even though three of them are marginally consistent within $1\sigma$ to the theoretical expectation. 

For the $1460$ absorption feature, the slope in the EW$_{1460}$-Z diagram is very shallow, with the EW ranging from $0.3$ to $0.8$ \AA, so that even a small error on the EW would leave the metallicity basically unconstrained. The values measured from the stacks, even though they are consistent with models within $2\sigma$, tend to place systematically below, and we do not observe any increase of absorption depth at higher stellar masses. A similar behavior hold for the $1533$ index introduced by \citet{sommariva12}. 

The similar behavior of these three indexes could be due to the relatively low spectral resolution of our spectra, suggesting that much higher resolving power (and sensitivity) are needed to increase the contrast between absorption feature and pseudocontinuum, eventually identifying the real, unabsorbed continuum level. We thus exclude them from our analysis, and report any further test of such indexes to the availability of higher SNR and resolution spectra.

On the other hand, the $1400$, $1550$, and $1853$ \AA\ absorption features have EWs well above the value expected from a pure stellar component (based on Starburst99 templates). In the first and third case, the observed EWs are between double and three times higher than pure stellar absorption expectations, which is likely due to ISM absorption contamination. Similarly, for the $1550$ feature we measure EWs $1$-$1.5$ \AA\ higher than S99 model predictions (but lower by the same amount if compared to BPASS). A more detailed discussion of the $1400$ and $1550$ \AA\ absorption lines can be found in Talia et al. 2020 (in preparation). 
As a result of this analysis, also these three lines were excluded from our work. 

\begin{figure*}[t!]
    \centering
    \includegraphics[angle=0,width=0.24\linewidth,trim={0cm 0cm 0cm 0cm},clip]{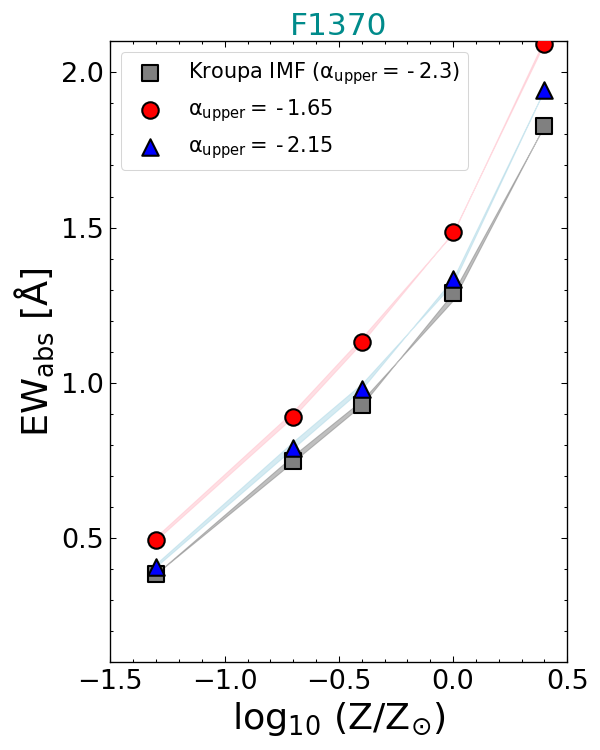}  
    \includegraphics[angle=0,width=0.24\linewidth,trim={0cm 0cm 0cm 0cm},clip]{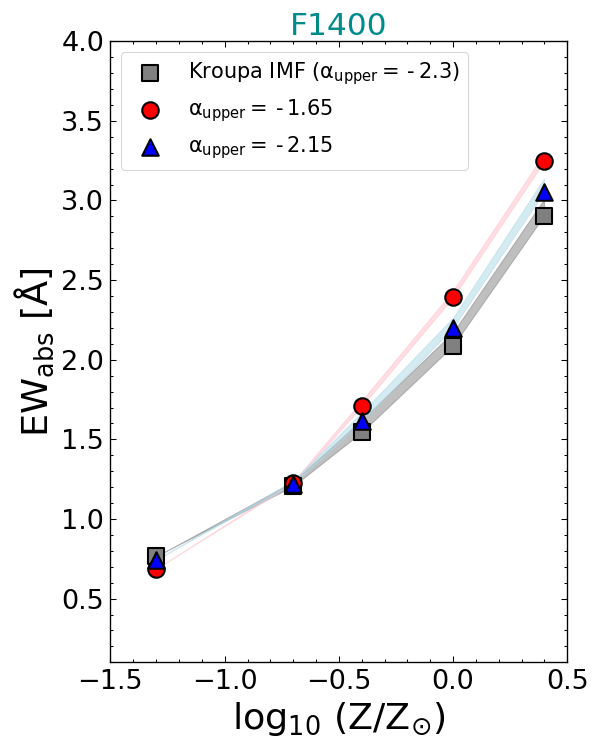} 
    \includegraphics[angle=0,width=0.24\linewidth,trim={0cm 0cm 0cm 0cm},clip]{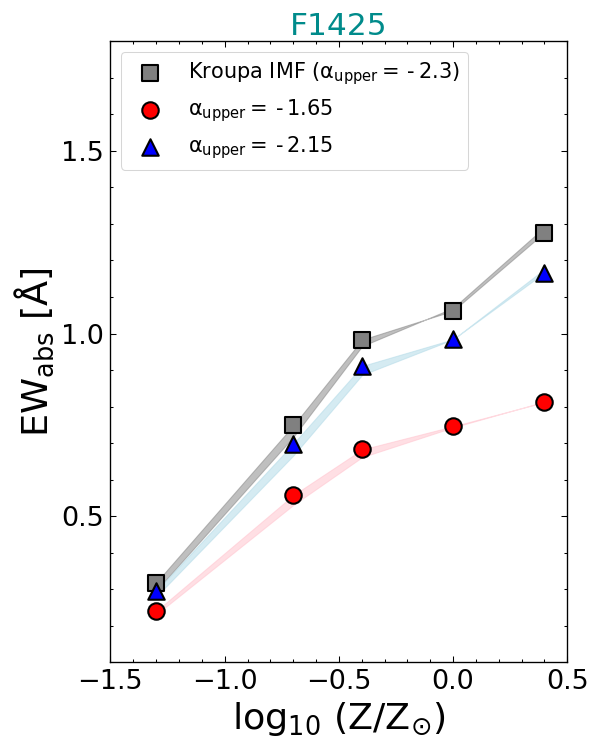} 
    \includegraphics[angle=0,width=0.24\linewidth,trim={0cm 0cm 0cm 0cm},clip]{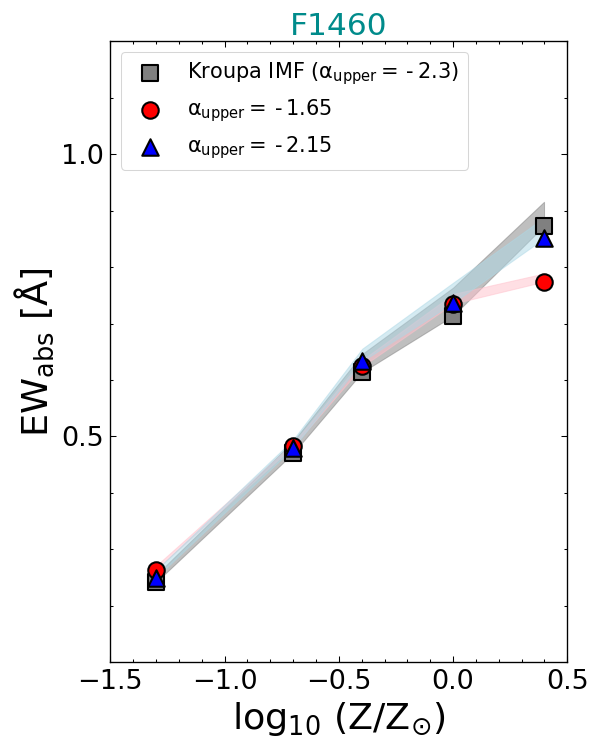} 
    \includegraphics[angle=0,width=0.24\linewidth,trim={0cm 0cm 0cm 0cm},clip]{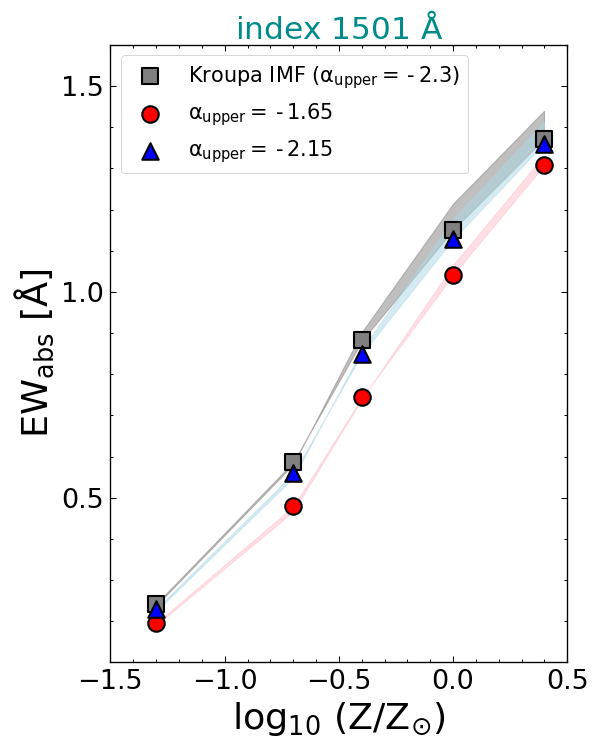} 
    \includegraphics[angle=0,width=0.24\linewidth,trim={0cm 0cm 0cm 0cm},clip]{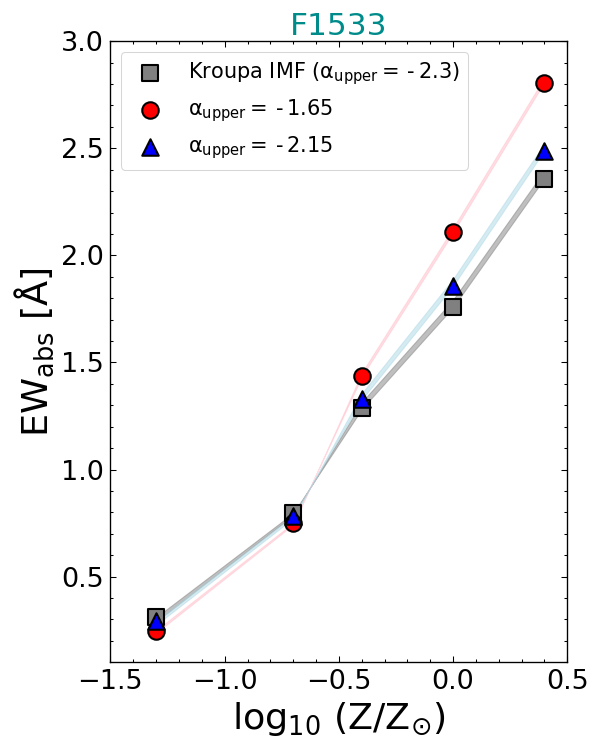} 
    \includegraphics[angle=0,width=0.24\linewidth,trim={0cm 0cm 0cm 0cm},clip]{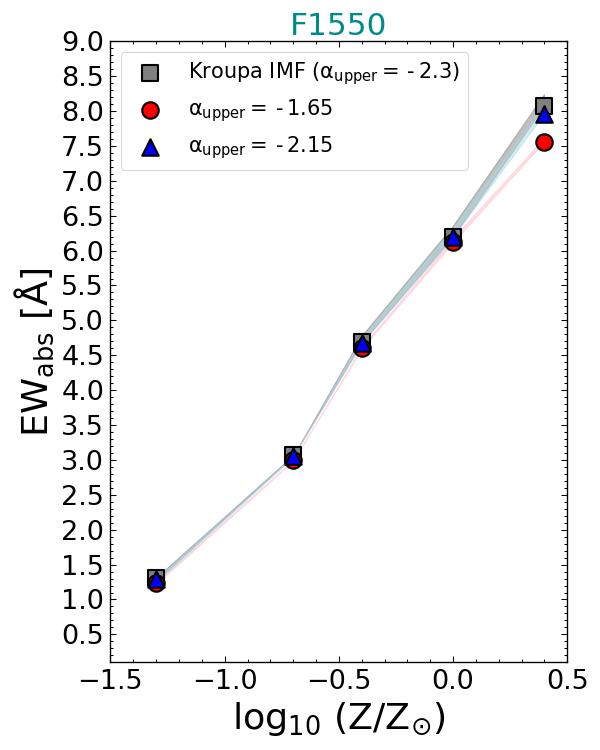} 
    \includegraphics[angle=0,width=0.24\linewidth,trim={0cm 0cm 0cm 0cm},clip]{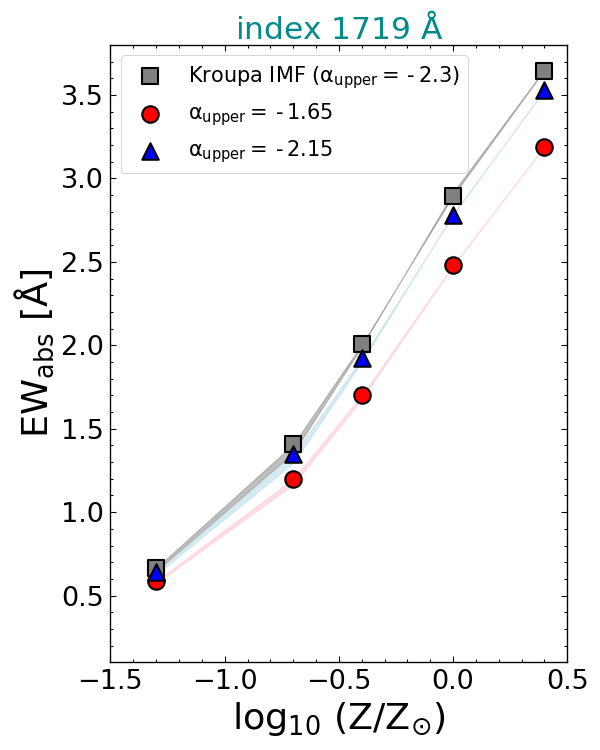} 
    \includegraphics[angle=0,width=0.24\linewidth,trim={0cm 0cm 0cm 0cm},clip]{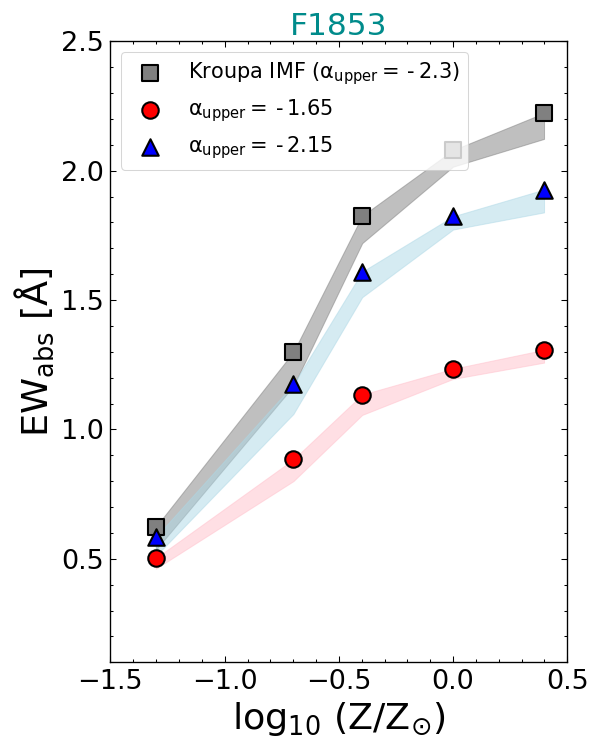} 
    \includegraphics[angle=0,width=0.24\linewidth,trim={0cm 0cm 0cm 0cm},clip]{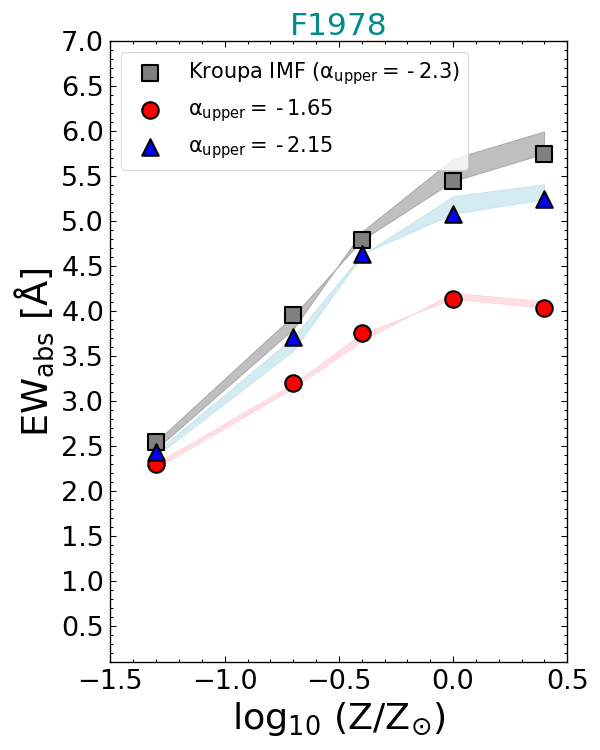} 
   
    \caption{\small Stellar metallicity vs equivalent width predicted by Starburts99 models for the ten absorption lines studied in this paper. The gray squares are derived assuming a Kroupa IMF, the blue triangles are built by changing the high-mass end slope to $\alpha$ $_{upper}=-2.15$ \citep[as in][]{baldry03}, while the red circles assume an extremely flat slope $\alpha$ $_{upper}=-1.65$. All the symbols are representative of a constant SFH and a stellar age of $100$ Myr. The shaded regions allow the stellar age to vary between $50$ and $100$ Myr. The other details of the models are the same as in Fig. \ref{other_models1}. }\label{IMF_variations_test2}
\end{figure*}

Overall, for the large majority of the indexes considered, the metallicity calibrations might vary with the model considered, just because of the slightly different spectral shapes of the templates. 
The different EW - metallicity relations predicted by BPASS and Starburst99 could be mostly related to a combination effect of different stellar atmosphere composition, stellar evolutionary tracks, stellar wind modeling and composition, while the different native resolution of the models could also play a role. However, despite such differences, the $1501$ \AA\ tracer seem the most stable between the two models analyzed here, thus supporting the conclusions of this paper. Deeper and higher resolution spectra, available with the next generation telescopes, will allow to enhance the accuracy of the metallicity measurements from the faintest absorption features, and better analyze the comparison with different models.

\subsection{IMF variations of the absorption lines}\label{IMF_variations}

We explored the effect of different IMFs on all the indexes studied in this paper. In particular, we varied both the upper-mass cutoff M$_\text{max,IMF}$ and the slope in the high-mass end ($\alpha_{upper}$) compared to the standard assumptions adopted so far, that is, a Salpeter or Kroupa functions, and M$_\text{max,IMF}=100$ M$_\odot$. 

First, we increased the upper-mass cut to $300$ M$_\odot$, which results only in a slight variation of the equivalent width (at fixed metallicity) for the majority of the indexes (see Fig. \ref{IMF_variations_test1}). The largest variation happens with the $1853$ \AA\ index, while for the others it is always smaller than $1\%$.

More interesting is the analysis of different slopes. \citet{bouwens12} assume a Salpeter stellar IMF, and claim there is no clear evolution from low to high redshift ($z>4$). 
However, several papers have investigated possible deviations from standard IMFs in high redshift galaxies.
\citet{elmegreen06} finds that galaxy-averaged IMFs are in general not steeper than a Salpeter slope in the high-mass end. Exploring different cosmic star-formation histories (SFH), \citet{baldry03} find a best slope $\alpha_{upper} = -2.15$ ($\pm0.2$), slightly lower but still consistent with the Kroupa or Salpeter slope ($\sim -2.3$). 
\citet{fontanot18} show that $\alpha_{upper}$ of the galaxy-wide IMF can vary with the SFR and the cosmic-ray (CR) density in the system. They find it can be in general top-heavy and shallower than the Kroupa IMF for SFR $>1$ M$_\odot$/yr at all CR densitites, which in any case occurs for the highest, 'starburst-like' SFRs $\geq 100$ M$_\odot$/yr, which are not present in our VANDELS selection. 
Additionally, \citet{vandokkum08} investigates the stellar IMF at the epoch of reionization as a function of the ISM clumping factor and the escape fraction of ionizing radiation. For different reionization histories, they find that galaxy-averaged IMFs could be flatter than a Salpeter slope, with $\alpha_{upper}=-2.1$ if it extends to $200$ M$_\odot$, or $\alpha_{upper}=-1.65$ if it extends to $50$ M$_\odot$.

Given these previous studies, we tested the effect of a Kroupa IMF with two different slope in the high-mass end regime: a moderately shallower IMF $\alpha_{upper}=-2.1$, as in \citet{baldry03}, and an extreme value of $-1.65$ (Fig. \ref{IMF_variations_test2}). 
In the first case, we observe only a slight change of the metallicity calibration function for most of the indexes, with variations of the equivalent width of less than $\sim5\%$ in the metallicity range of our work. Typically, the largest variations (up to $8\%$) occur at solar and super-solar Z$_\ast$.
The only exceptions are the indexes at $1853$ and $1978$ \AA, which can show a decrease of EW up to $0.25$ \AA. However, these indexes are not considered for our metallicity calculation.

Finally, for an extremely flat IMF slope ($\alpha_{upper}=-1.65$), the effects on the EW are not negligible for the majority of the absorption lines. In particular, for those adopted in this paper to derive the stellar metallicity, we observe a variation of EW (at fixed Z$_\ast$) of $0.15$ and $0.3$ \AA\ for the $1501$ and $1719$ \AA\ indexes, respectively, in the Z$_\ast$ range of our dataset. In other words, this IMF would produce a constant upward shift of stellar metallicity of $0.1$-$0.13$ dex. However, we remark that this is a very unlikely IMF for typical star-forming galaxies at $z\sim3$ in the stellar mass range explored by VANDELS.

\subsection{The effect of spectral resolution on the metallicity calibrations}\label{appendix-resolution}

The spectral resolution of the observations is important to determine the stellar metallicity from absorption lines. Indeed, it affects the calibration functions that are applied to the various indexes. This behavior can be understood if we think that degrading the resolution tends to smooth the few narrow spectral regions that are intrinsically free of absorption and to decrease the overall pseudo-continuum level, lowering the contrast with the absorption features. As a result, the resolution affects the dynamic range of the EWs of the lines, setting a lower limit on the metallicities that can be reliably constrained. 
For each specific index, the minimum metallicity measurable can be defined as the value for which the corresponding equivalent width in absorption would be equal to the typical uncertainty of such measurement (i.e., the relative error is $\sim 100\%$). In this case, the EW estimation would be consistent with $Z_\ast=0$ within $1\sigma$, and only an upper limit can be eventually determined. 

\begin{figure*}[t!]
    \centering
    \includegraphics[angle=0,width=0.24\linewidth,trim={0cm 0cm 0cm 0cm},clip]{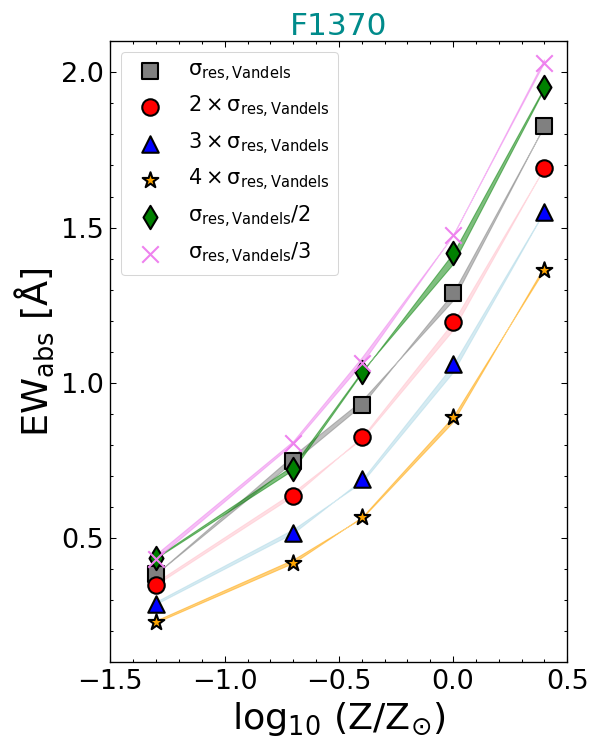}  
    \includegraphics[angle=0,width=0.24\linewidth,trim={0cm 0cm 0cm 0cm},clip]{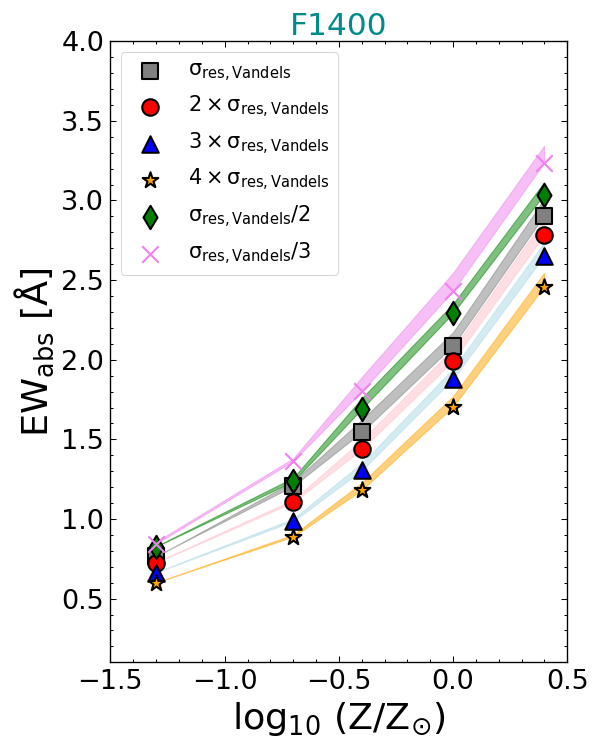} 
    \includegraphics[angle=0,width=0.24\linewidth,trim={0cm 0cm 0cm 0cm},clip]{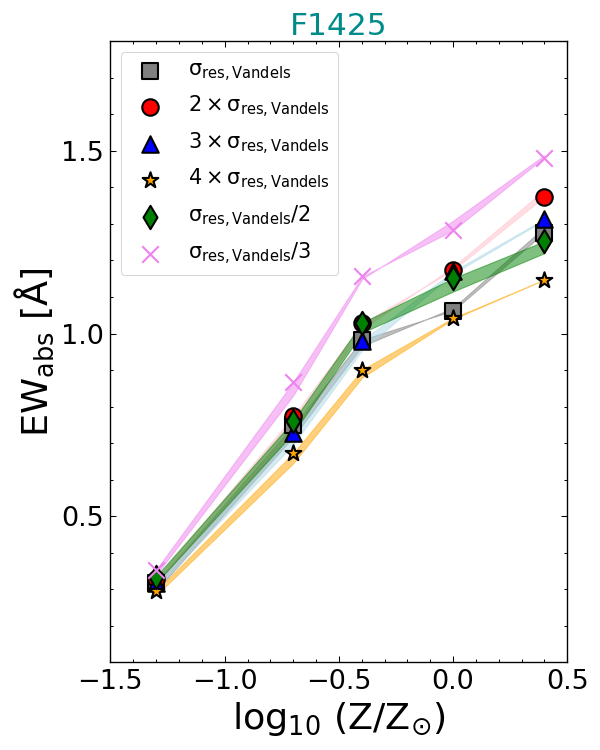} 
    \includegraphics[angle=0,width=0.24\linewidth,trim={0cm 0cm 0cm 0cm},clip]{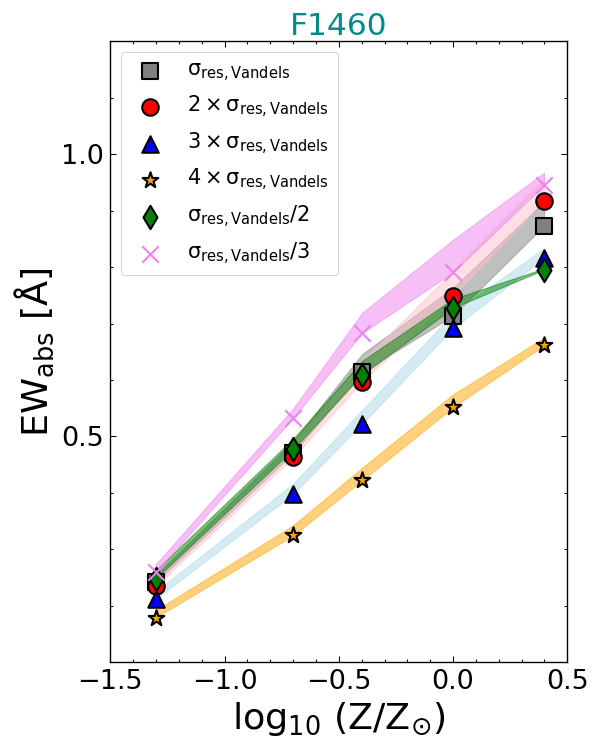} 
    \includegraphics[angle=0,width=0.24\linewidth,trim={0cm 0cm 0cm 0cm},clip]{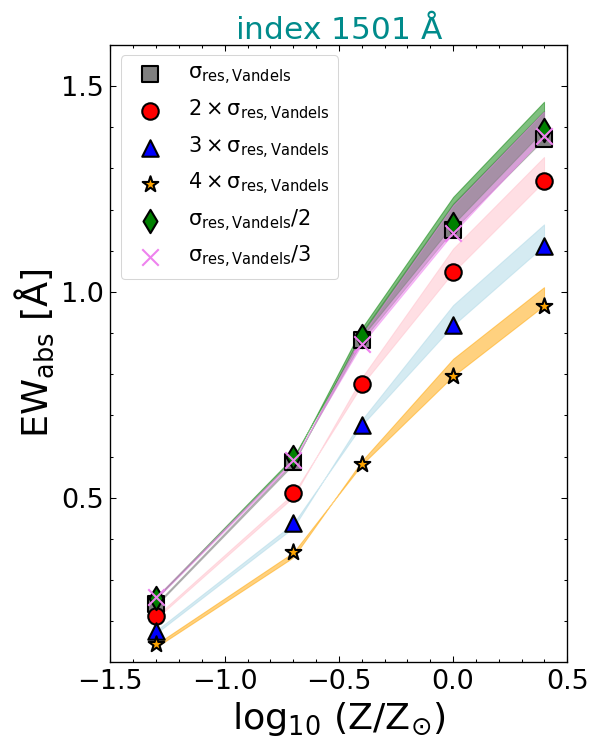} 
    \includegraphics[angle=0,width=0.24\linewidth,trim={0cm 0cm 0cm 0cm},clip]{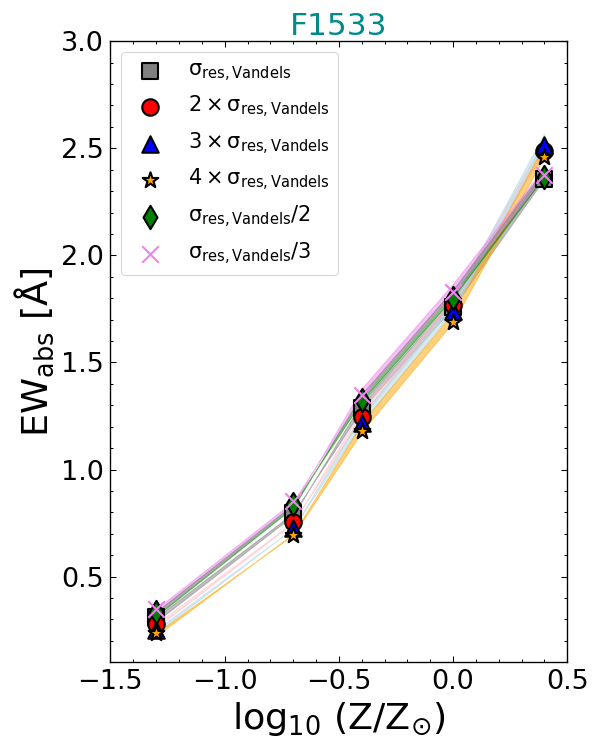} 
    \includegraphics[angle=0,width=0.24\linewidth,trim={0cm 0cm 0cm 0cm},clip]{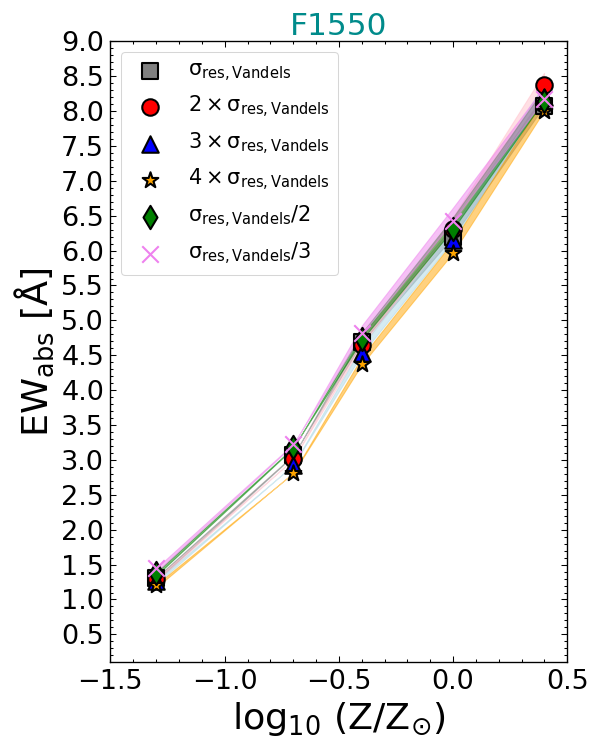} 
    \includegraphics[angle=0,width=0.24\linewidth,trim={0cm 0cm 0cm 0cm},clip]{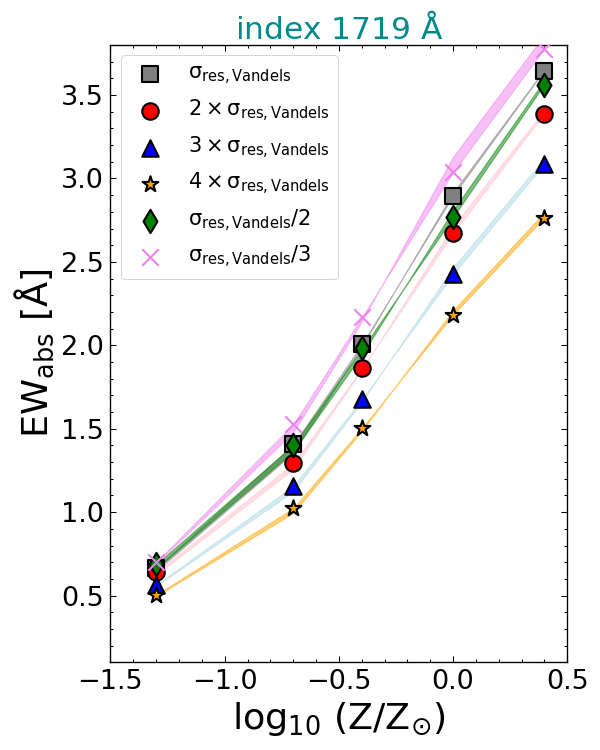} 
    \includegraphics[angle=0,width=0.24\linewidth,trim={0cm 0cm 0cm 0cm},clip]{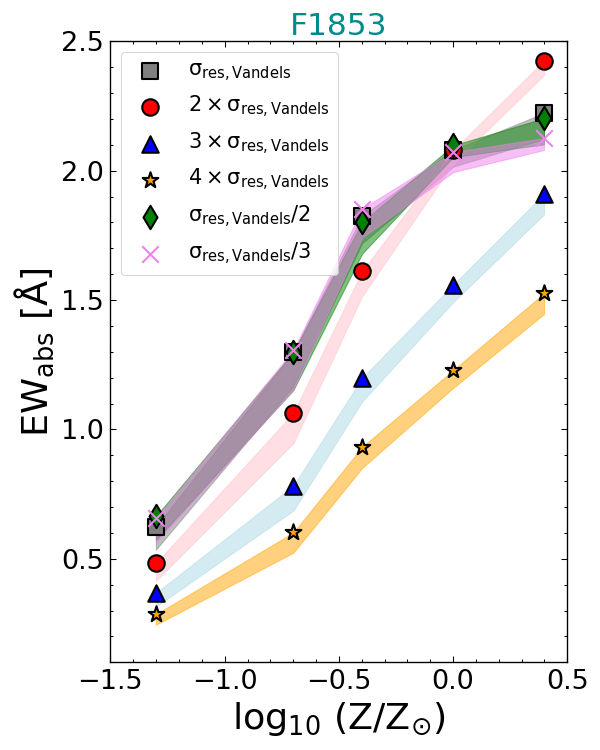} 
    \includegraphics[angle=0,width=0.24\linewidth,trim={0cm 0cm 0cm 0cm},clip]{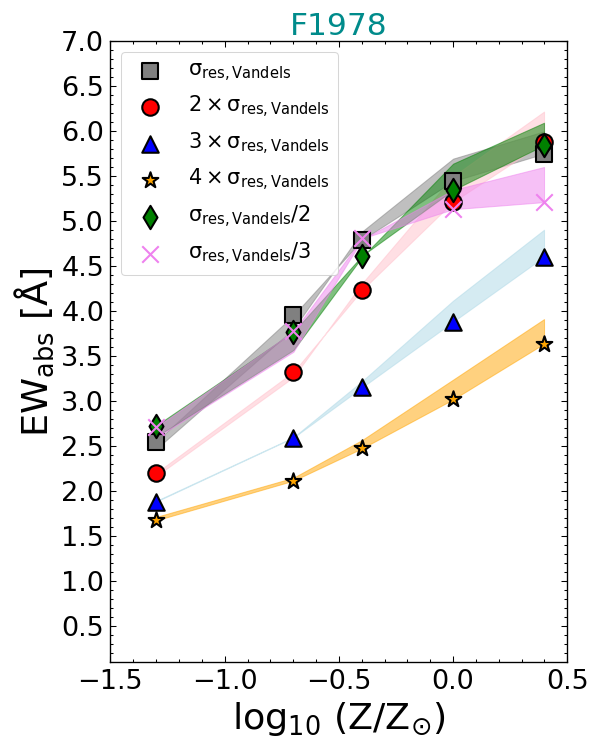} 
  
    \caption{\small Stellar metallicity vs equivalent width predicted by Starburts99 models for the ten absorption lines studied in this paper and for different spectral resolutions. The resolution elements $\sigma_\text{res}$ tested are $1/2$, $1/3$, $1$, $2$, $3$, and $4$ times the VANDELS value. All the symbols are derived with a constant SFH and a stellar age of $100$ Myr. The shaded regions allow the stellar age to vary between $50$ and $100$ Myr.}\label{figure-resolution}
\end{figure*}

We analyzed the effect of the resolution for all the ten indexes presented in this work. In particular, we tested FWHM resolution elements two, three and four times larger than VANDELS, where the latter value was set to ensure that it is still lower than the majority of absorption line widths. We also tested higher spectral resolutions, with $\sigma_\text{resolution}$ equal to $1/2$ and $1/3$ of VANDELS. The results, displayed in Fig. \ref{figure-resolution}, show that degrading the resolution systematically decreases the EW of most of the absorption lines. 
It is also interesting to notice that the calibrations obtained for VANDELS and for the two improved resolutions are essentially the same for the majority of the lines, including those at $1501$, $1533$, $1550$ and $1583$ \AA. For the $1719$ \AA\ absorption complex, the highest resolution spectra of our analysis would give higher equivalent widths than in the VANDELS case by $\sim 0.1$ - $0.15$ \AA\ in the range $0.2<$ Z$_\odot$ $<1$ (thus always below the typical uncertainties), while they are consistent at Z$_\ast <0.2$.

In general, lower metallicities are more difficult to properly constrain as we increase $\sigma_\text{resolution}$.
For example, with the worst resolution tested and the $1501$ \AA\ index, given its typical uncertainties of $\sigma_{EW}$ $\sim0.15$-$0.2$ \AA\ (see also Appendix \ref{appendix1} for the dependence on the SNR), it is basically impossible to constrain within $1\sigma$ the metallicity better than the entire range of our study ($-1.1<$ log$_{10}$(Z/Z$_\odot$) $<-0.6$). 
Overall, this analysis can provide a useful reference to evaluate the calibration functions for the ten indexes when observing at different spectral resolutions. 

\end{document}